%% file: OledTheoryPaper.r0.19.tex
\documentclass[aip,letterpaper,11pt,nofootinbib]{revtex4}

\usepackage{amsmath,mathtools}    
\usepackage{amssymb}    
\usepackage{mathrsfs}
\usepackage{graphicx}   
\usepackage{float}
\usepackage{chemfig}
\usepackage{array}
\usepackage{mdframed}
\usepackage{multirow}
\usepackage{pdflscape}
\usepackage{verbatim}   
\usepackage{subfigure}  
\def\app#1#2{%
  \mathrel{%
    \setbox0=\hbox{$#1\sim$}%
    \setbox2=\hbox{%
      \rlap{\hbox{$#1\propto$}}%
      \lower1.1\ht0\box0%
    }%
    \raise0.25\ht2\box2%
  }%
}

\def\FkPrime[#1]{f_{k'_{#1}}}
\def\TauPrime[#1]{\tau'_{#1}}
\def\TauSP[#1]{\tau_{sp_{#1}}}
\def\LLambda[#1]{K_{d_{#1}} + {1\over\tau'_{#1}}}
\def\LY[#1]{{f_{k'_{#1}}\over\tau_{sp_{#1}}}}
\def\LP[#1]{{P_{k'_{#1}}\over\tau_{k'_{#1}}}}
\def\Lk[#1]{L_{k'_{#1}}}
\def\lLambda[#1]{K_{d_{#1}} + 1/\tau'_{#1}}
\def\lY[#1]{{f_{k'_{#1}}/\tau_{sp_{#1}}}}
\def\lP[#1]{{P_{k'_{#1}}/\tau_{k'_{#1}}}}
\def\Alpha[#1]{\alpha_{#1}}

\def\OTauPrime[#1]{{1\over\tau'_{#1}}}
\def\oTauPrime[#1]{{1/\tau'_{#1}}}

\def\OTauSP[#1]{{1\over\tau_{sp_{#1}}}}
\def\oTauSP[#1]{{1\over\tau_{sp_{#1}}}}

\def\Rnr[#1]{{1\over\tau_{nr_#1}}}
\def\rnr[#1]{{1/\tau_{nr_#1}}}

\def\Rsp[#1]{{1\over\tau_{sp_#1}}}
\def\rsp[#1]{{1/\tau_{sp_#1}}}

\def\Rprime[#1]{{1\over\tau'_{#1}}}
\def\rprime[#1]{{1/\tau'_{#1}}}

\def\geV{{\gamma \over e V_a}}
\def\gev{\left(\gamma/ e V_a\right)}

\def\gE{{\gamma \over e}}
\def\ge{(\gamma/ e)}

\def\Keh{K_{eh}}
\def\Kd{K_d}

\def\tsp{\tau_{sp}}
\def\tnr{\tau_{nr}}
\def\tcav{\tau_{cav}}

\def\Pcav{{P\over\tcav}}

\def\nlim{n_{lim}}
\def\nnlim{n^2_{lim}}

\def\nse{n_{SE}}
\def\NSE{N_{SE}}
\def\Pse{P_{SE}}

\def\nsp{n_{sp}}
\def\Nsp{N_{sp}}
\def\Psp{P_{sp}}

\def\Kn{K_n}

\def\V{\mathbb{V}}

\def\KehN{K_{eh(2)}}
\def\t2{\tau'_{(2)}}

\def\n{[n]}
\def\Sa{[S_0]}
\def\Sb{[S^*]}
\def\Tb{[T^*]}
\def\nST{\eta_{S/T}}
\def\nTS{\eta_{T/S}}

\def\Sbb{[S^{**}]}
\def\Tbb{[T^{**}]}
\def\KehS{K_{eh(S^*)}}
\def\KehT{K_{eh(T^*)}}
\def\tSbb{\tau'_{(S^{**})}}
\def\tTbb{\tau'_{(T^{**})}}

\def\tnrS{\tau_{nr_S}}
\def\tspS{\tau_{sp_S}}
\def\tS{\tau'_S}

\def\fkS{f_{k_S}}
\def\gkS{g_{k_S}}
\def\PkS{[P_{k_S}]}
\def\tkS{\tau_{k_S}}

\def\fkSout{f_{k'_S}}

\def\LkSout{L_{k'_S}}

\def\tnrT{\tau_{nr_T}}
\def\tspT{\tau_{sp_T}}
\def\tT{\tau'_T}

\def\fkT{f_{k_T}}
\def\gkT{g_{k_T}}
\def\PkT{[P_{k_T}]}
\def\tkT{\tau_{k_T}}

\def\fkTout{f_{k'_T}}

\def\LkTout{L_{k'_T}}

\def\KdS{K_{d_S}}
\def\KdT{K_{d_T}}

\def\Kisc{K_{isc}}
\def\Krisc{K_{risc}}

\def\KTS{K_{TS}}
\def\KTn{K_{Tn}}
\def\KSn{K_{Sn}}

\def\KSS{K_{SS}}
\def\KTT{K_{TT}}
\def\KTTT{K_{TT,T}}
\def\KTTS{K_{TT,S}}

\begin{document}

\title{Efficiency and Stimulated Emission in Quarter Wave OLEDS}
\author{Mitchell C. Nelson}
\email[Correspondence:\ ]{drmcnelsonm@gmail.com}

\date{10 October 2015}

\begin{abstract}
  Quarter-wave OLEDS are microcavity devices that can operate in the low finesse limit and achieve high efficiency ($> 300$ lm/W) by using interference to reduce the onset current for the transition to stimulated emission.
  In this work we study the transition to stimulated emission and compare the kinetics and electrical properties of conventional and quarter-wave devices.
  We show that
  suppression of spontaneous emission into the vertical mode can result in a sharp transition to stimulated emission at $\gev I \sim \NSE/\tsp$, where $\NSE/\tsp$ is determined by optical parameters,
  and we find a previously observed electrical signature for the transition where the excited state population becomes fixed at low current density.
  We then study the role of loss mechanisms in the quarter-wave configuration
  and conclude with some requirements for practical devices.
\end{abstract}

\maketitle

\section{Introduction}
Light production in organic light emitting diodes (OLEDS) comprises a broad set of topics spanning material properties, electrical, spectral and photochemical processes, and classical and quantum optics.\cite{GasparBook2015}
OLEDS are typically formed by layers of organic and metallo-organic materials deposited between two electrodes. 
Charge is injected as holes and electrons which then migrate inwards and recombine to form excited states that can then be quenched, undergo intersystem crossing, migrate, or relax and produce light, some of which exits from a vertical mode as useful light and some of which may go into other modes in the device and substrate and exit as wasted light and heat.
There is often a wavelength scale separation between two parallel reflective interfaces so that the device formally meets the definition of a microcavity.\cite{KavokinBook2007}
While developments in materials and device architecture have contributed to improvements in internal efficiency and outcoupling, important issues remain in efficiency, roll-off and device lifetime.\cite{Reineke2013,Lee2015}.
An avenue for improvement in these areas is the role of device architecture in manipulating the kinetics of light production.

As is well known, emission of light results when an electronic transition is able to couple to an allowed mode.\cite{Fox2014,Miller2004,Berman1994}
A cavity is thus able to enhance or suppress emission\cite{Purcell1946,Kleppner81} and alter the kinetics and behavior of the device.\cite{Yokoyama89,Yamamoto92}
An emitter located at an odd multiple of 1/4 wavelength (QW) from the back mirror in a resonant cavity, for example, has zero amplitude for spontaneous emission into the vertical mode\cite{Deppe1994,Dutra1996}
and inhibited spontaneous emission is associated with stimulated emission without inversion (SWI) at low current density.\cite{Yablonovitch87,Yablonovitch88}
Once in stimulated emission, excited state populations and losses can be locked to fixed values determined by optical parameters, and efficiency can approach ideal limits.\cite{Noda06}
But getting to stimulated emission can be difficult because of losses or material limits.\cite{Bulovic1998,Yamamoto2005,GiebinkForrest2009}
SWI has been reported in devices with high finesse cavities and materials with large Stokes shifts.\cite{Yokoyama92,Mompart2000}
The QW architecture offers a way to achieve SWI in a low finesse cavity, and at low current density, by reducing the photon threshold for the transition.

High efficiency at 318 lm/W with linear output was first observed in an OLED with an emitter located at one quarter-wave from a highly reflective cathode and with a distributed Bragg reflector (DBR) interposed between the ITO anode and substrate.\cite{Nelson2015,Dodabalapur1996}
In a second device, a design that was previously reported to produce 55 lm/W\cite{Hung2012} was altered to locate the recombination zone at the quarter-wave position in a low-finesse cavity formed between a Ag top mirror and a simple ITO-glass interface. The modifications resulted in a device that produces 340 lm/W.\cite{Nelson2015}
It was thought that orientation of the emitter\cite{Kim2013} in the second device would obviate the need for the DBR.
Later work indicated that the emitter is isotropic\cite{Graf2014} and revisiting earlier data we found that the central wavelength emitted by the DBR fitted QW device matches a cavity formed by the interface after the transparent anode before the DBR.
Thus it appears that the effect does not depend on a constrained mode space and does not require an elaborate or high finesse cavity.

Since the proposed effect operates in the interaction between optical boundary conditions and excited state kinetics, and towards the goal of designing devices that exploit the effect,
we have undertaken a study of QW and conventional OLED behaviors in spontaneous and stimulated emission as limiting cases and in the transition to stimulated emission.
We report here the results of our analyses and describe some requirements for achieving stimulated emission in a QW device.

\section{Theory of OLED devices}
\begin{figure}[h]
  \includegraphics[scale=1.]{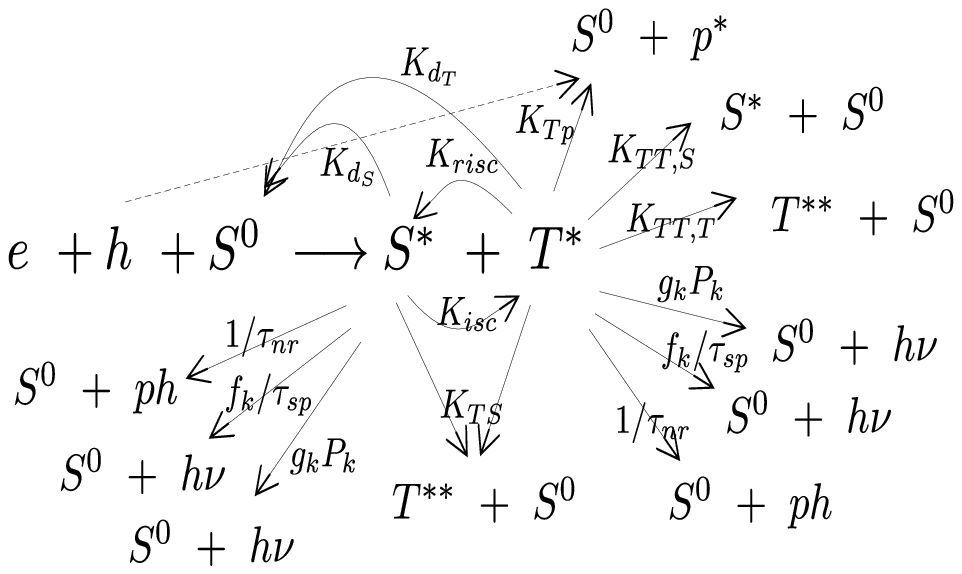}
  \caption{\label{processes} Some of the processes in a generic OLED with singlet and triplet states.}
\end{figure}

Processes in a generic single component emitter in an OLED\cite{Peng2012} are shown in FIG.~\ref{processes}.
Charge carriers recombine to form a statistical mixture of excited state species, which then relax, undergo intersystem crossing, or interact with each other or with charge. Losses in this model include non-radiative relaxation, singlet-triplet quenching, triplet-triplet quenching, singlet-singlet quenching and charge quenching.
Light production processes include spontaneous emission which is proportional to terms of form $N/\tsp$ and stimulated emission which is proportional to terms of form $g P N$, where $N$ is an excited state population and $P$ is photon density and where there is not significant overlap in the absorption and emission spectra.
The stimulated process can be much faster and alter the kinetic description of the system.
Our task is to understand performance in these two regimes and the conditions for achieving stimulated emission controlled kinetics.
Empirically, we expect that the requirements include that any asymptote or local maximum in the system occurs above this transition, and that current and charge density remain small through the transition.


In the following,
Section \ref{ssec:twolevel} describes limiting case behaviors and the transition to stimulated emission in a two level model,
Section \ref{ssec:electrical} derives an electrical behavior at the transition that was observed in QW devices,
Section \ref{ssec:quenching} describes the contribution to onset current and performance by second order and charge quenching losses,
Section \ref{ssec:threelevel} treats intersystem crossings and losses in a three level model, and
Section \ref{ssec:numerical} describes numerical evaluations for a realistic system in three configurations and with different phosphorescent lifetimes.

\subsection{Two level emitter} \label{ssec:twolevel}
Our analysis begins with an idealized electroluminescent microcavity device with a single excited state and well separated emission and absorption lines.
The initial processes will include charge recombination and dissociation, non-radiative relaxation, spontaneous emission and stimulated emission, represented schematically as,
\begin{equation}
    \begin{aligned}
      e + h + N_0 & \xrightarrow{\Keh} N \\
      N & \xrightarrow{\Kd} N_0 + e + h\\
      N & \xrightarrow{1/\tau_{nr}} N_0 + {\rm phonon} \\
      N & \xrightarrow{1/\tau_{sp}} N_0 + P \\
      N & \xrightarrow{g_k P_k} N_0 + 2  P_{k}
    \end{aligned}
\end{equation}
where the charge species $e$ and $h$ are supplied by charge injection and transport across the adjacent layers.
Rate equations for charge $n$, excited state $N$ and photon $P$ densities can be written as,\cite{AgrawalBook2002}
\begin{eqnarray}
  {d n \over dt} & = & \geV I - K_{eh} n^2 (1 - N ) + \Kd N \label{dnBasic} \\
  {d N \over dt} & = & K_{eh} n^2 (1 - N) - \left(\Kd + {1 \over \tsp} + {1 \over \tnr }\right) N - g P N \label{dNBasic} \\
  {d P \over dt} & = & g N P + \chi {N \over \tsp} - {P \over \tcav}\label{dPBasic}
\end{eqnarray}
where $I$ is current,
$\gamma$ is the fraction of current that recombines to form excited states,
$e$ is the unit charge,
$V_a$ is the volume and $\Keh$ is the rate constant for charge recombination
with units chosen such that the ground state population $N_0$ can be written as $N_0 = 1 - N$,
$\Kd$ is the rate constant for charge dissociation from the excited state molecule,
$\tsp$ is the spontaneous emission lifetime,
$\tnr$ is the non radiative lifetime,
$g$ is the rate coefficient for stimulated emission,
$\chi$ is the fraction of photons from spontaneous emission that enter the vertical mode,
and $\tcav$ is the lifetime in the vertical mode of the cavity.
Considering all of the modes of the device, $\chi = f_{k'}/\sum_k f_k$ where $f_k$ is the cavity enhancement factor for mode $k$ and $k'$ is the vertical mode.
Setting $\sum_k f_k = 1$, means that $1/\tau_{sp}$ is the spontaneous emission rate in the cavity rather than the free space value.\cite{Purcell1946}
Adding the rate equations we obtain
\begin{equation}
  \label{ILPbasic}
  \geV I = {N\over\tau_{nr}} + (1 - \chi){N\over\tau_{sp}} + {P\over\tau_{cav}}
\end{equation}
The term on the left represents current injection, the first term on the right represents non-radiative first order losses, the second represents radiative losses to modes outside of the vertical mode, and the last term is the light output from the device from its vertical mode.

\begin{figure}[h]
  \includegraphics[scale=0.5]{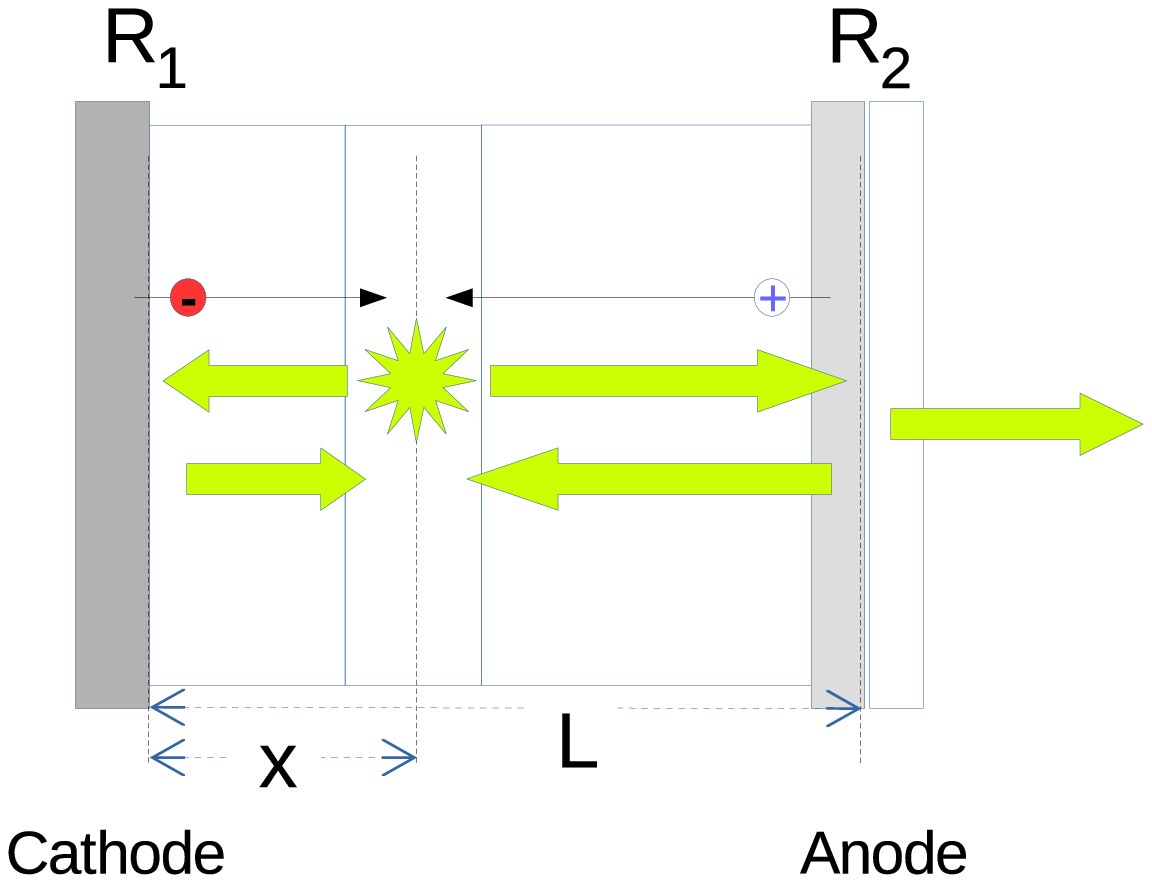}
  \caption{
    \label{microcavity}
    Microcavity device with optical length $L$.
    The back mirror is at the cathode-transport layer interface.
    The exit mirror is at the semi-transparent anode-glass interface.
    A thin emitting region is located at optical length $x$ from the back mirror.
    For an emitter at $x = (2n+1)\lambda/4$, emission to the left is reflected from the mirror and cancels emission to the right.
  }
\end{figure}

The geometry for a planar microcavity device with a thin emitter layer is illustrated in FIG.~\ref{microcavity}.
In a resonant vertical mode $k'$ where one end of the cavity is totally reflective ($R_1 = 1$), $f_{k'}$ depends on position as,\cite{Deppe1994,Dutra1996}
\begin{equation}
  \label{QWcondition}
  f_{k'}(x) \propto 1 + \cos\left[4\pi {x(\lambda_{k'})\over\lambda_{k'}}\right]
\end{equation}
where $x(\lambda_{k'})$ is the optical distance between the emitting region and mirror.
The amplitude for spontaneous emission into mode $k'$ is enhanced at $x(\lambda_{k'}) = \lambda_{k'}/2$ and suppressed at $x(\lambda_{k'}) = \lambda_{k'}/4$.
Stimulated emission is not attenuated at the quarter wave position because the emitted photon is in phase with the stimulating photon.\cite{Einstein1917}
In a standing wave mode, the stimulated emission rate follows the local field so that,\cite{Martini1992,DuanAgrawal1993,Gupta2007}
\begin{equation}
  g_{k'}(x) \propto \sin^2\left[2 \pi {x(\lambda_{k'})\over L(\lambda_{k'})}\right]
\end{equation}

In typical emitter materials, recombination occurs in a region of width $ w_r/d \sim 4 \mu_e \mu_h / (\mu_e +\mu_h)^2$ centered at $x_r/d\sim \mu_h/ (\mu_h + \mu_e)$ from the hole injection side within the emitter layer, where $\mu_q$ with $q \in e,h$ are the carrier mobilities and $d$ is the thickness of the emitter layer.\cite{Kalinowski} 
Emission occurs within exciton diffusion length of the recombination event.

Two further properties of the cavity are of interest: the lifetime for a photon to exit the vertical mode  $\tcav = 2 L(\lambda) / c ( 1 - R )$,
and the linewidth $\Delta \lambda/\lambda = (1 - R)/2\pi$.
For a one wavelength low finesse cavity (small $R$), $\tcav \sim 4\times 10^{-15} \mathrm{sec}$ for $\lambda \sim 500\ \mathrm{nm}$.

In equation \eqref{dPBasic} we find two limiting case behaviors.
The first is when $\chi/\tau_{sp} >> g P$, for example when the emitter is located at a half wave point in a high finesse cavity or when $P$ is small.
Light output is then proportional to the excited state population
\begin{equation}
  \label{Psp}
  {P_{SP}\over\tcav} = {\chi \over \tsp} N
\end{equation}
the excited state population is proportional to current,
\begin{equation}
  \geV I = {N \over \tau'}
\end{equation}
where $\tau' = 1/\tnr + 1/\tsp$, and the external quantum efficiency equation is
\begin{equation}
  \label{LIsp}
  L_{sp} =  \gamma \chi \phi  {I\over e} 
\end{equation}
where $L_{sp} = V_a \Psp/\tcav$ is the light output from the vertical mode and $\phi = \tau'/\tsp$ is the internal quantum yield for radiative relaxation. This is the classic formula for efficiency sans the singlet-triplet factor.\cite{Tsutsui13}

The second limiting case is when $g P >> \chi/\tau_{sp}$, either at high output or with the emitter at the quarter wave position where $\chi \sim 0$.  Then, the excited state population is constant,
\begin{equation}
  \label{Nse}
  N_{SE} = {1\over g\ \tcav}
\end{equation}
and light output is equal to electrical input minus a constant.
\begin{equation}
  \label{qwEQEbasic}
  L_{SE} = {\gamma \over e} I\ - \alpha_{SE}
\end{equation}
where $\alpha_{SE}/V_a = \ N_{SE}/\tau'$.
As power is increased, efficiency approaches an asymptote at $\gamma/e$.
If there is a small overlap in the absorption and emission spectra within the bandwidth of the cavity,  $N_{SE}$ is replaced by $\NSE + N'_0$ and the asymptotic efficiency is still $\gamma/e$.

The transition region between the two limiting cases provides some important insights.
In the general case, when $\NSE = (g\ \tcav)^{-1} \leq 1$ the relationship between light and the excited state population is,
\begin{equation}
  \label{PNinter}
  {P\over\tcav} = {\chi\over\tsp}\ {N \over 1 - N / \NSE}
\end{equation}
the relationship between current and excited state population becomes,
\begin{equation}
  \label{INinter}
  {\geV I} =
       {N\over\tau'}
       \left( 1 + \chi \phi { N/N_{SE} \over 1 - N/N_{SE}} \right)
\end{equation}
and the relationship between current and light is
\begin{equation}
  \label{IPinter}
  \geV I = \left({\NSE\over\tau'} + \Pcav\right)\ { P/\tcav \over \chi\ \NSE/\tsp\ +\ P/\tcav}
\end{equation}
We learn from equation \eqref{PNinter} that the physical solution is where $N$ approaches $\NSE$ from below as current goes to infinity.
Similarly, from equation \eqref{INinter}, the transition to stimulated emission is controlled by $\chi$ and $\NSE/\tau'$.
As $\chi \rightarrow 0$, the transition becomes sharp with an onset current at $\gev I \sim \NSE/\tau'$.
The Einstein A/B relationship gives us
$g \propto c \lambda^2 / 4 \tsp$\cite{Hilborn1982} and therefore $\NSE/\tau' \propto 1/\tcav$.
The transition and onset current are then controlled primarily by optical parameters.

\begin{figure}[h]
  \subfigure[Transition to stimulated emission] {
    \includegraphics[scale=0.35]{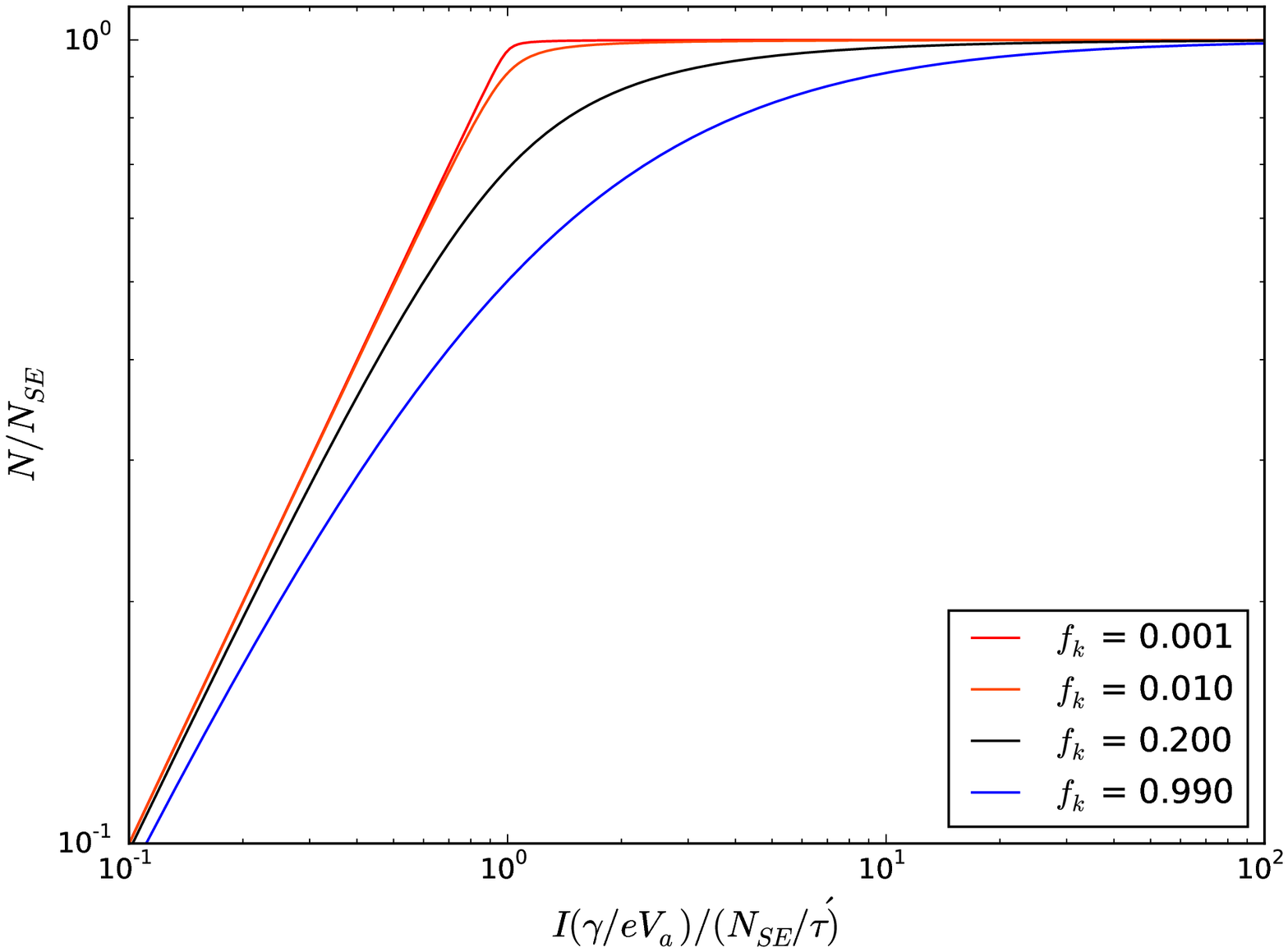}
    \label{fig:spontstimN}
    }
  \subfigure[Charge density in the transition] {
    \includegraphics[scale=0.35]{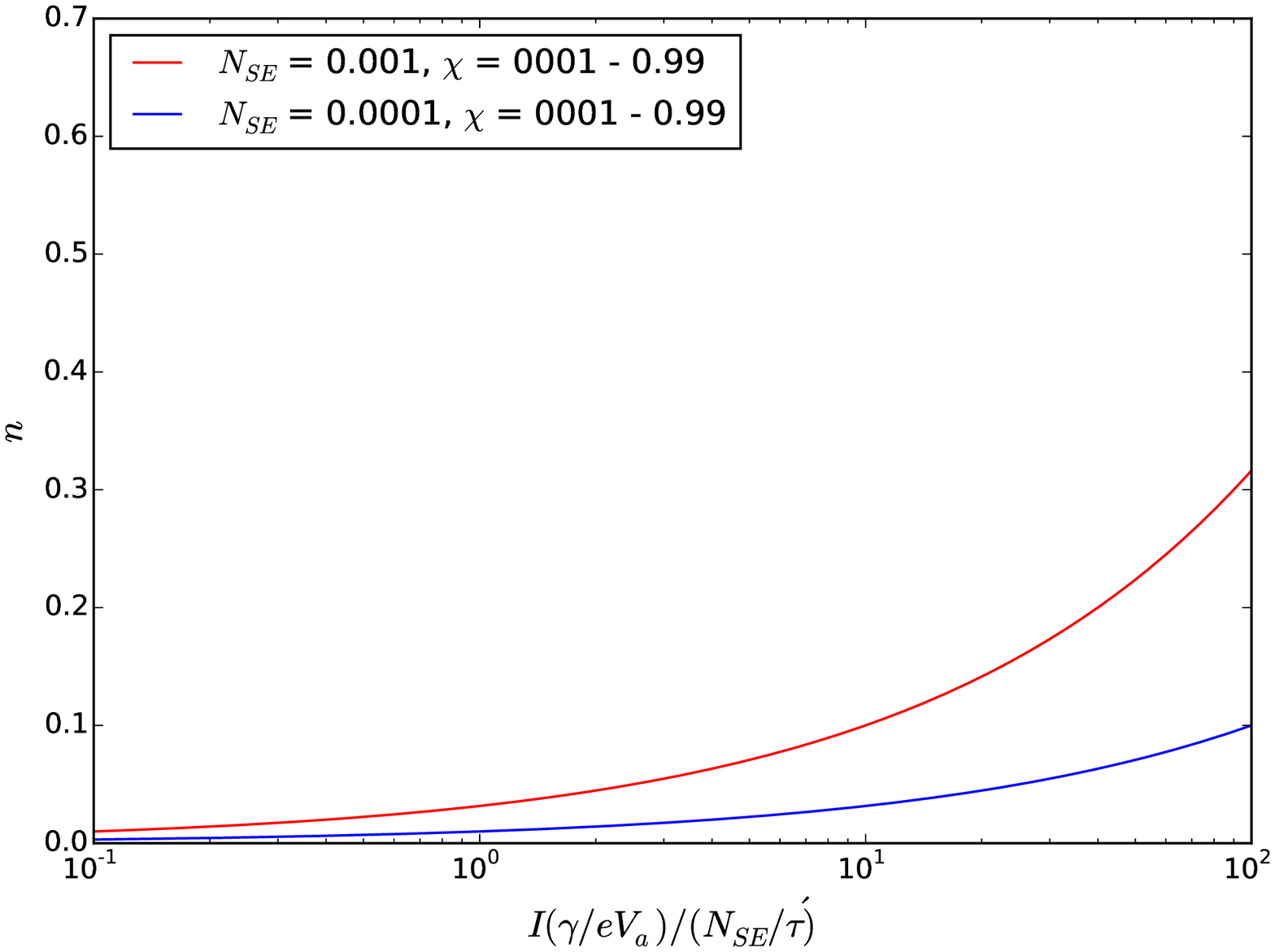}
    \label{fig:spontstimNSE}
    }
  \caption{\label{fig:twoleveltransition}
    Excited state population approaching $\NSE$ for QW (red, red-orange), normal (black) and $f_k \sim 1$ (blue) devices, and charge density in the transition region for different values of $\NSE$.}
\end{figure}

The behavior in the transition region is summarized in FIG.\ref{fig:twoleveltransition}.  The graphs are in unitless coordinates with excited state population as $N/\NSE$ and current as $\gev {\rm I}/(\NSE/\tau')$.
The excited state population is graphed in FIG.~\ref{fig:spontstimN} for different values of $\chi$. 
The transition moves to low current and becomes sharp as $\chi \rightarrow 0$.
Charge density is graphed in FIG.~\ref{fig:spontstimNSE} using equation \eqref{dnBasic} and equation \eqref{INinter}.
Charge density is not sensitive to $\chi$, but in these units, decreases with $\NSE$.
For a QW device with $\tcav \sim 10^{-15}$ the onset current is order of $\mathrm{1\ mA/mm^2}$.
In a device with $\chi \gtrsim 0.2$, the
current required to approach within 99\% of $\NSE$ can be higher by a factor of $10^4$ in real units.\cite{Yamamoto2005}
Since charge density increases with current, charge quenching losses may become significant and push the approach to $\NSE$ to still higher current.

Charge density in the spontaneous emission case is found from equation \eqref{dNBasic} in terms of the excited state population as,
\begin{equation}
  \label{nsp}
  \nsp = \sqrt{ {\Kd + 1/\tau'\over \Keh}\ \left({N\over 1 - N}\right) }
\end{equation}
and in terms of current as,
\begin{equation}
  \label{nspcurrent}
  \nsp = \sqrt{{\Kd\ \tau' + 1\over \Keh}\ { \gev I \over 1/\tau' - \gev I} }
\end{equation}
where $1/\tau' = 1/\tsp + 1/\tnr$.
The infinity is avoided for $N < \NSE < 1$ or by including loss terms.
In stimulated emission, charge density  can be related to light as
\begin{equation}
  \label{nselight}
  \nse \sim \sqrt{ { 1\over\Keh } \left(\Pcav + (\Kd + 1/\tau') \NSE\right) }
\end{equation}
or to current as
\begin{equation}
  \label{nsecurrent}
  \nse \sim \sqrt{ { 1\over\Keh } \left(\geV I + \Kd \NSE\right) }
\end{equation}
Voltage in all cases is proportional to charge density through an effective capacitance,
\begin{equation}
  q = \mathrm{C}\ \V
\end{equation}
where $n = q/{\rm volume}$, and $\V$ is the voltage across the emitter layer.\cite{Pitarch2006}
Maximum output at this stage in our analysis might be defined in terms of a limiting charge $\nlim$ or voltage above which the device might suffer an irreversible change.
In the spontaneous emission region,
\begin{equation}
    {P_{sp}\over\tcav}\ \lesssim\ \chi\phi\ {\Keh \nnlim \over  1 + \Keh \tau' \nnlim }
\end{equation}
and for stimulated emission,
\begin{equation}
  {P_{SE}\over\tcav}\ \leq\ \Keh\ \nnlim
\end{equation}
Electrical characteristics are discussed in more detail in section \ref{ssec:electrical}.

Properties in the limiting case regimes of spontaneous emission and stimulated emission with losses to first order, are summarized in Table~\ref{O1properties}.
We can characterize spontaneous emission as having output proportional to the excited state population and efficiency limited by outcoupling and internal losses.
Stimulated emission is characterized by the excited state population being constant while efficiency approaches $\gamma/e$.
\begin{table}[h]
  {
    \setlength{\tabcolsep}{1.0em}
    \renewcommand{\arraystretch}{2.0}%
    \begin{tabular}{l|c|c}
      \centering
      Property & Spontaneous Em. & Stimulated Em.
      \\
      \hline
      $\displaystyle N$
      &
      $\displaystyle N \propto \mathrm{I}, L$
      &
      $N = {\rm constant}$
      \\
      \hline
      $n, \V$
      &
      $\displaystyle \sqrt{ {I\over 1/\tau'-I}}$
      &
      $\displaystyle \sqrt{\mathrm{I}}$ \\
      \hline
      EQE (L/I) & $\displaystyle \gamma\ \chi\ \phi$
      &
      $\displaystyle \gamma\ (\sim 1)$ \\
      \hline
      $\displaystyle L_{max} (fl)$
      &
      $\displaystyle \chi\ \phi\ \Keh\ {\nnlim\over 1 + \nnlim}$
      &
      $\displaystyle \Keh\ \nnlim$
    \end{tabular}
  }
  \caption{\label{O1properties} Comparison of properties of spontaneous emission and stimulated emission with loss terms to first order in $N$.}
\end{table}

\subsection{Electrical behavior at the transition}\label{ssec:electrical}

\begin{figure}[h]
  \subfigure[\ Light and current versus voltage (measured).]{
    \includegraphics[scale=0.5]{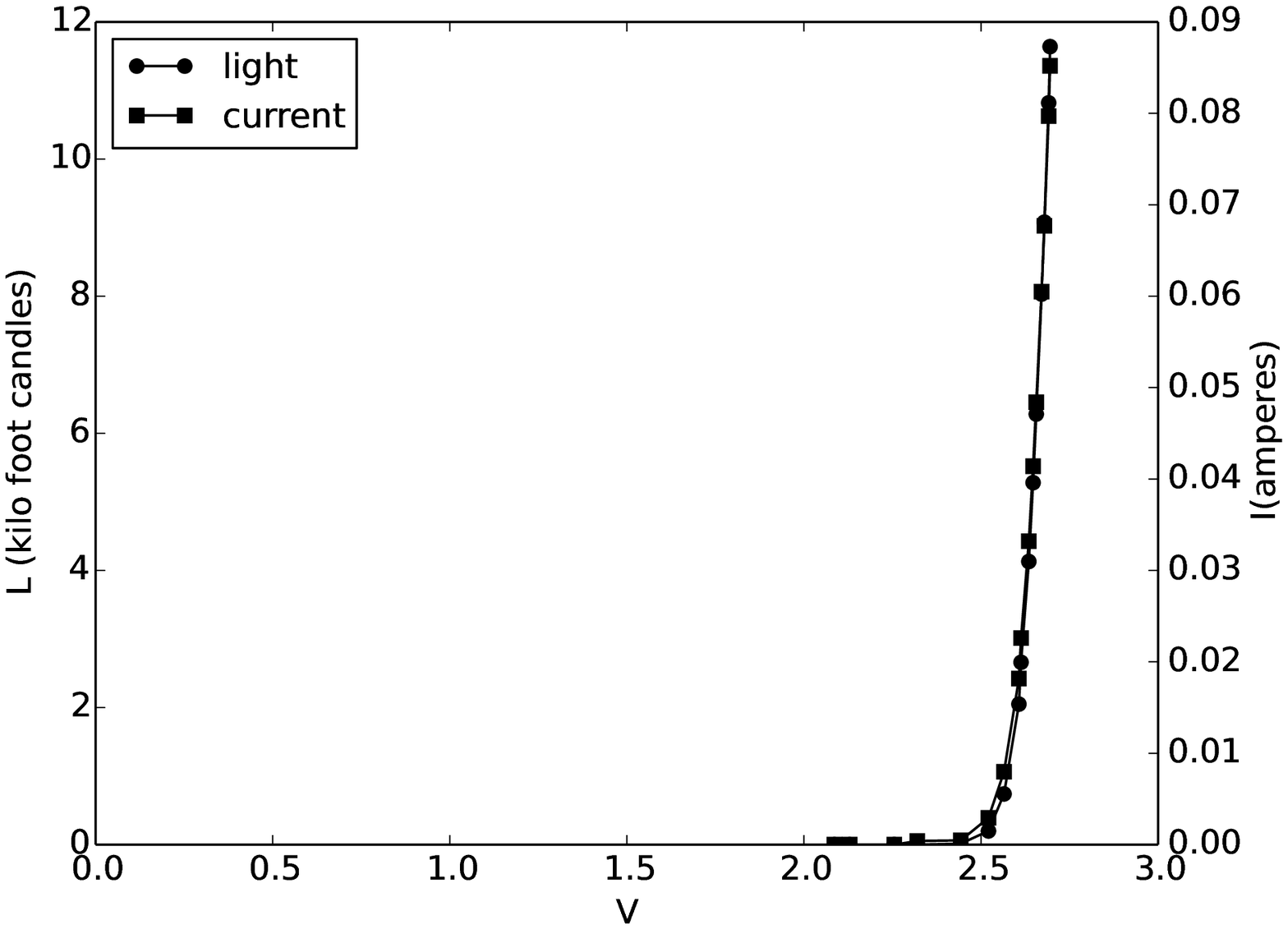}
    \label{LIVdata}
  }
  \subfigure[\ Light versus current (measured).]{
    \includegraphics[scale=0.5]{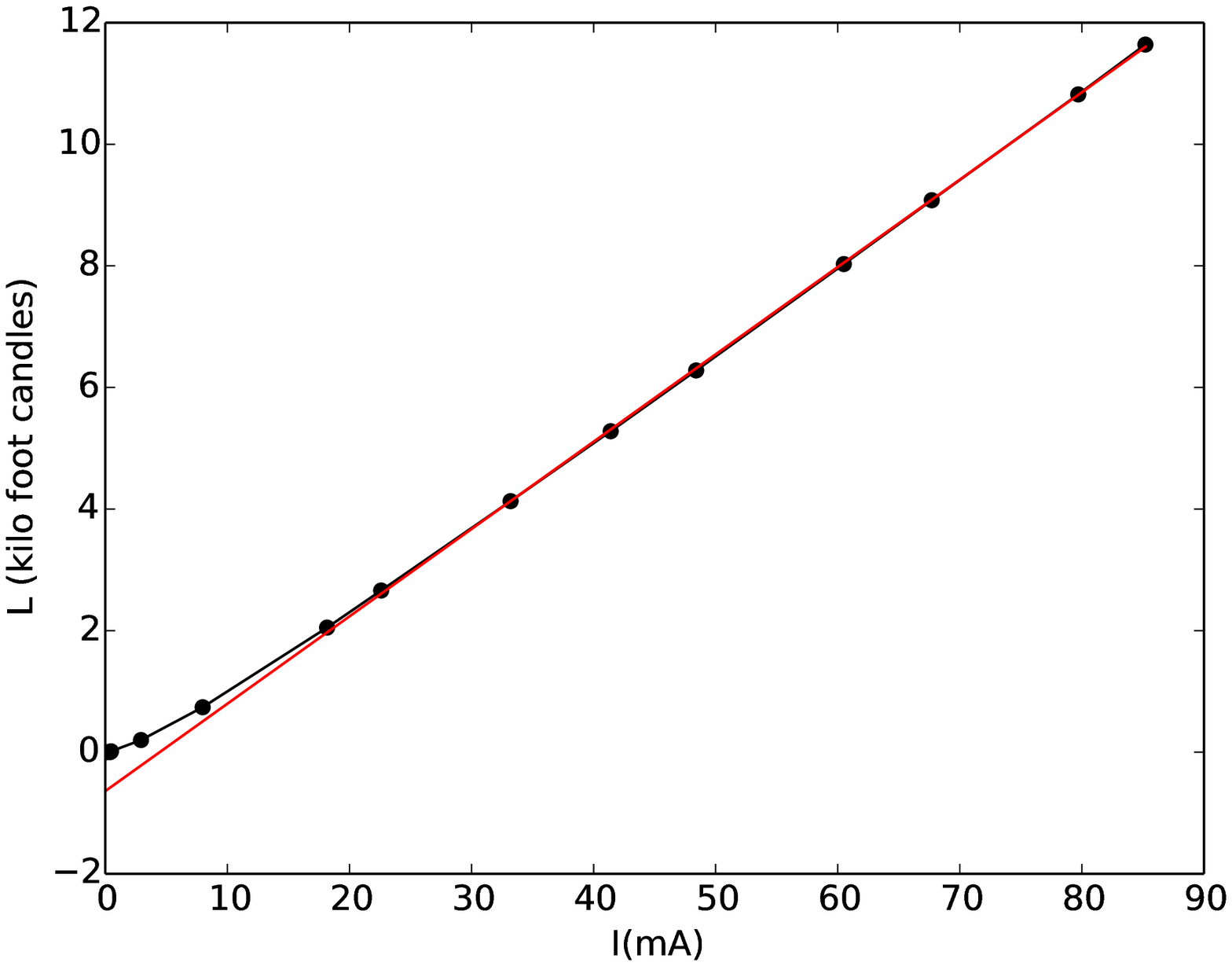}
    \label{LIdata}
  }
  \caption{
    \label{LIVcurves}
    Light-current-voltage behavior in stimulated emission. Voltage quickly rises after turn on and the transition to linear behavior at low current.
    }
\end{figure}

\begin{figure}[h]
  \subfigure[\ Light and current versus voltage at the transition (measured).] {
    \includegraphics[scale=0.5]{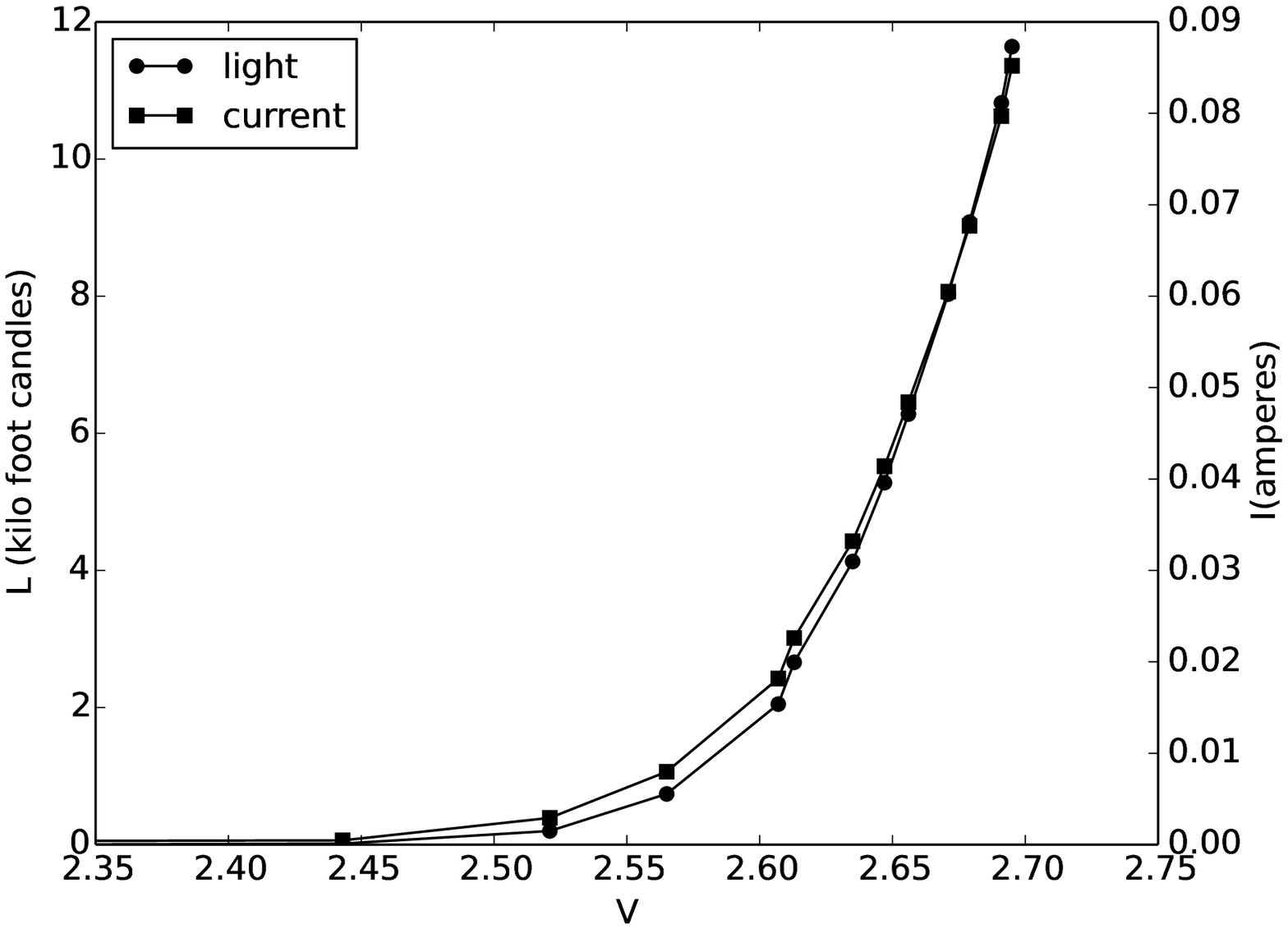}
    \label{LIVstep}
  }
  \subfigure[\ Mott-Gurney analysis at the transition (measured).] {
    \includegraphics[scale=0.5]{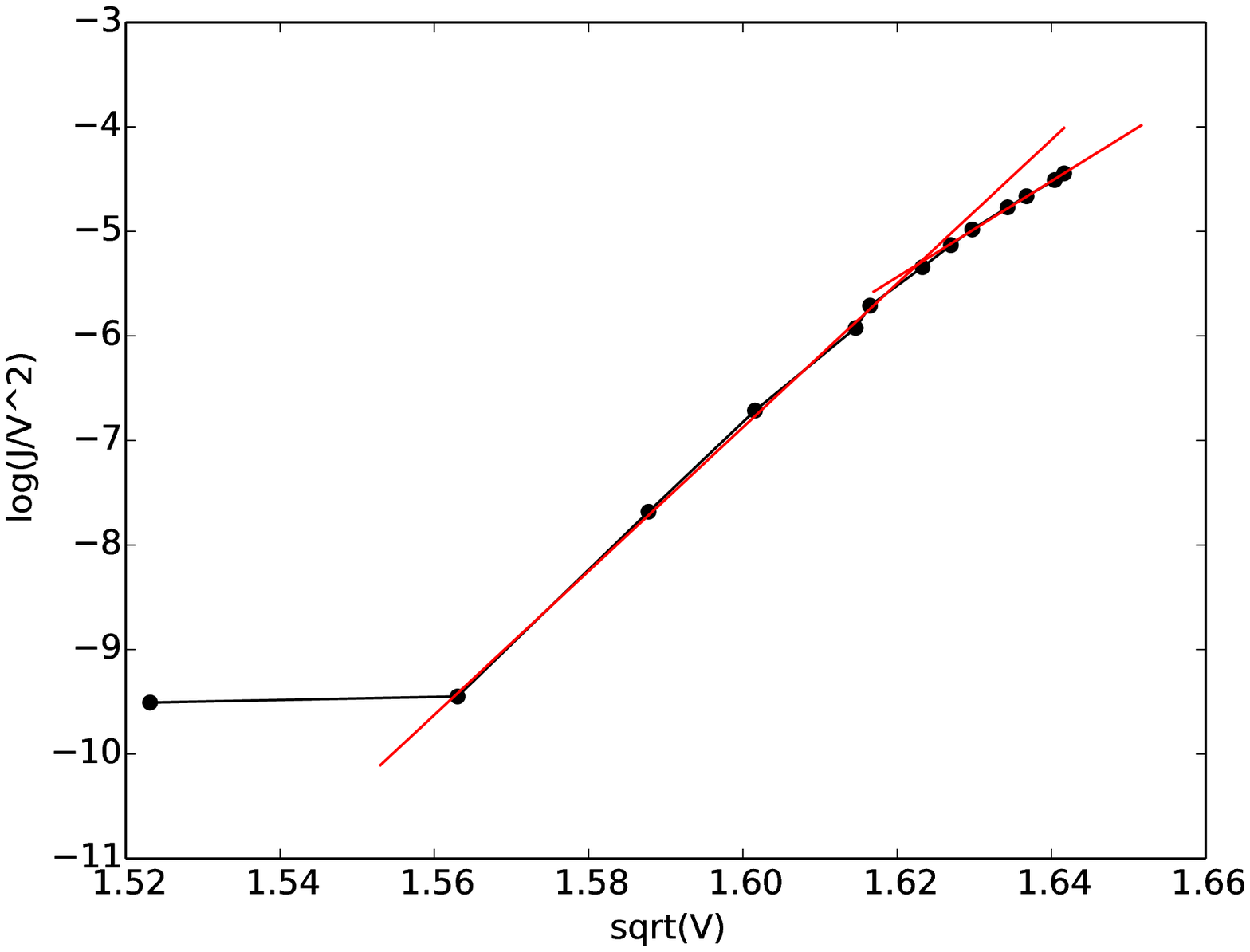}
    \label{FPMGstep}
  }
  \caption{
    \label{LIVtransition}
    Light-current behavior in stimulated emission at low current.  Measured data is plotted as $log(I/V^2)$ against $\sqrt{V}$. The abrupt change in the Poole-Frenkel coefficient occurs at the transition to linear output.}
\end{figure}

The treatment thus far is sufficient to allow us to address an electrical behavior observed in the region of the transition to linear output at low power where the excited state population and losses are still small.
Measured light, current and voltage data from a QW device are shown in FIG.~\ref{LIVdata}.
The device powers up over a swing of about 0.5V, and then voltage stays within 0.2V while light and current rise rapidly together.
The relationship between light and current is shown in FIG.~\ref{LIdata}. The device transitions to linear light output at low current and the linear fit exhibits a small negative offset as in equation \eqref{qwEQEbasic}.
The region of the transition is expanded in FIG.~\ref{LIVstep} where
a small but abrupt step is seen to accompany the transition to linear output.
At 2.62 eV, the step corresponds to the optical cavity length of the device at 475 nm.
A Mott-Gurney plot of the data in this region is shown in FIG.~\ref{FPMGstep}.
The Poole-Frenkel coefficient decreases by about a factor of two across the transition.
The following analysis suggests that this may be a signature of the transition to stimulated emission at low current.

The emitter layer can be viewed as two capacitors in series $1/C' = 1/C_e + 1/C_h$ where $C'$ is the geometric capacitance of the layer and, $C_{e,h} = (3/4) \epsilon/x_{e,h}$ and  $x_{e,h}/d = (\mu_{e,h} /\mu_h+\mu_e)$ are the capacitances and widths of two adjacent regions, $\epsilon$ is the dielectric constant, $d$ is the total thickness of the layer and $\mu_q$ are the charge carrier mobilities.\cite{Pitarch2006}
Then,
\begin{eqnarray}
  C' & = & {3\over 4} {\epsilon\over d} \\
  q_h & = & q_e = C' \mathbb{V} \label{nV}
\end{eqnarray}
where $\mathbb{V}$ is the voltage drop across the layer
and $q_h$ and $q_e$ are the charge in each region.
Charge carrier mobilities also enter into the kinetic rate constant for recombination,\cite{Lakhwani2014}
\begin{equation}
  \label{Kehmu}
  \Keh = {q\over\epsilon} ( \mu_e + \mu_h )
\end{equation}
Substituting this into equation \eqref{dnBasic}, and assuming low current with $N \sim 0$, gives us the Mott-Gurney current-voltage relationship
\begin{equation}
  J  =  {9\over 8} \epsilon (\mu_e + \mu_h) \mathbb{V}^2/d^3 \\
\end{equation}
where $J$ is current density and the carrier mobilities follow a Poole-Frenkel law, 
\begin{equation}
\mu_q = \mu_q(0) \exp\left[\alpha\sqrt{\mathbb{V}/d}\right] \label{PooleFrenkel}
\end{equation}
These are well studied behaviors in OLEDS.\cite{Torpey1984,Ioannidis98,Tsai2005,Bassler2012}
We are interested in the current-voltage behavior at small finite $N \neq 0$ where N changes with current density and then becomes fixed after the transition.

In spontaneous emission, from equation \eqref{dNBasic} we can write $N$ in terms of $n$
\begin{equation}
  N = {\Keh n^2 \over \Keh n^2 + 1/\tau'}
\end{equation}
and for small $n$ we can use the approximation $N \sim \Keh \tau' n^2$. Then, substituting this into equation \eqref{dnBasic} we have
\begin{equation}
  \geV I = \Keh n^2 - (\Keh)^2 \tau' n^4
\end{equation}
We write this in terms of the mobilities and Poole-Frenkel dependence
using equation \eqref{Kehmu} and equation \eqref{PooleFrenkel},
\begin{equation}
  {\gamma\over e} I = \mu^2_0\ e^{2\alpha\sqrt{\V_0}}\ \left({e^{-\alpha\sqrt{\V_0}}\over\mu_0} - \tau' \V^2_0\right)\ \V^2_0
\end{equation}
where $\mu_0 = \mu_e(0) + \mu_h(0)$.

The coefficient $\alpha$ is found from I-V data by graphing $\log I/V^2$ versus $\sqrt{V}$. For spontaneous emission, we expect
\begin{equation}
  \log\left(I\over \V^2_0\right) = 2 \alpha \sqrt{\V_0}
  + \log\left( \mu^2_0 (e V_a/\gamma) \left({e^{-\alpha\sqrt{\V_0}}\over\mu_0} - \tau' \V^2_0\right)\right)
\end{equation}
The second term is small and the graph of $\log(I/\V^2_0)$ versus $\sqrt{\V}$ will approximate a straight line with slope $2 \alpha$.

In stimulated emission $N = N_{SE}$ is constant so that
\begin{equation}
  \geV {I\over \V^2} = \mu_0 e^{\alpha \sqrt{V}} C^2 (1 - N_{SE})
\end{equation}
and for our M-G plot we have
\begin{equation}
  log\left( {I\over \V^2} \right) = \alpha\ \sqrt{\V} + log \left( {e V_a\over\gamma}\mu_0 C^2 (1 - N_{SE})\right)
\end{equation}
Graphing $\log(I/\V^2_0)$ versus $\sqrt{\V}$ we find a slope of $\alpha$ for stimulated emission. Therefore, for a transition from spontaneous emission to stimulation emission at low power we expect a factor of two change in the Poole-Frenkel coefficient.
The step seen in the L-I-V graph at the transition, might be explained by loss terms and the different forms of the charge current relationship in spontaneous emission (equation \eqref{nspcurrent}) compared to stimulated emission (equation \eqref{nsecurrent}).

\subsection{Annihilation, bleaching and roll-off} \label{ssec:quenching}
We now add higher order processes to our model.
Self quenching processes such as triplet-triplet annihilation (TTA), are described as second order in excited state species and can account for behaviors where efficiency decreases while output increases, both monotonically.\cite{Baldo2000}.
Behaviors where efficiency roll-off is accompanied by a plateau or roll-off in output, are also known.\cite{Yu2013,Wu2014,Gao2015} Asymptotic output can be produced in principle by polaron quenching, or by singlet-triplet annihilation, as will be discussed in \ref{ssec:threelevel}.  Output roll-off within the operating range of the device, seems to require a new loss mechanism which will be described here.
We continue with units chosen such that there is one molecule per unit volume so that $N_0 + N = 1$, and for numerical evaluations we set $\tsp = 1$ which gives us approximately $ns$ and $nm^3$ for typical materials.\cite{Brinkmann2000}

We begin with second order losses in the excited state, as in triplet-triplet (TTA) or singlet-singlet (SSA) annihilation.\cite{Baldo2000,Nakanotani2005,Reineke2007,Berger2010}
Steady state solutions for $n$ and $N$ give us
\begin{eqnarray}
  \geV I & = &\Keh n^2 (1 - N) \\
  K_{eh} n^2 (1 - N) & = & \KTT N^2 + \left({1 \over \tau'} + g P \right)\ N
\end{eqnarray}
and for spontaneous emission the relationship between light and current becomes
\begin{equation}
  {P_{sp(q)}\over\tcav} + \KTT {\tsp\tau'\over\chi} \left({P\over\tcav}\right)^2 = \chi\phi \geV I
\end{equation}
At high output $L \propto \sqrt{ I/\KTT }$ while efficiency falls off as $1/\sqrt{I}$,\cite{Baldo2000} and the charge limited output is $P/\tcav \lesssim\ (\chi/\tsp) (\sqrt{\Keh/\KTT})\ \nlim$.
For stimulated emission,
\begin{equation}
  {P_{sp(q)}\over\tcav} = \geV\ I\ -\ \alpha_Q
\end{equation}
where $\alpha_Q = \NSE/\tau' + \KTT \NSE^2$.
Output is linear in current, efficiency increases toward an asymptote at $\gamma$, and the onset current is incremented by the loss term evaluated at the transition.

\begin{figure}[h]
    \subfigure[\ Spontaneous emission ]{
      \includegraphics[scale=0.35]{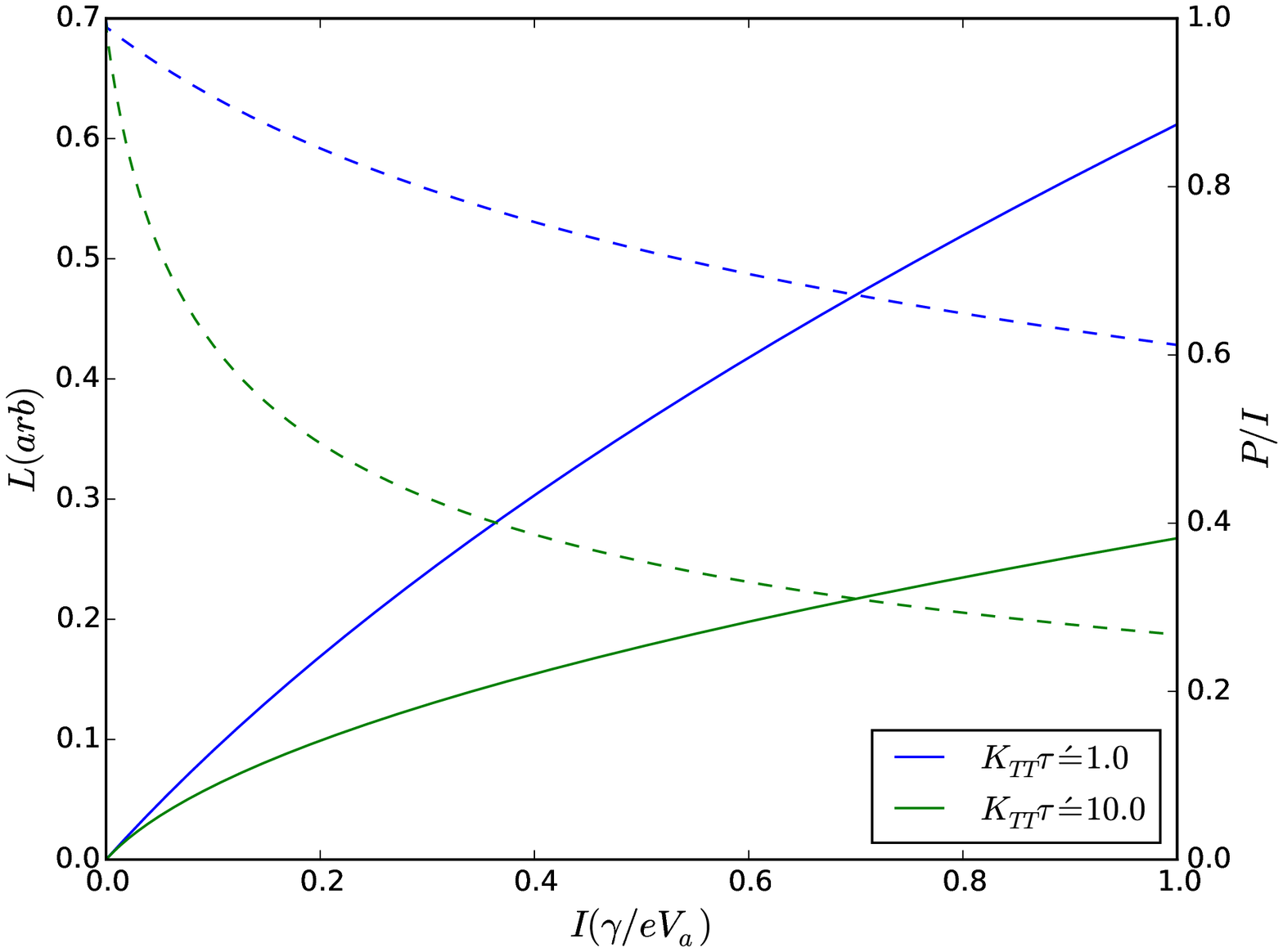}
      \label{fig:spontKTT}
    }
  \subfigure[\ Stimulated emission ]{
    \includegraphics[scale=0.35]{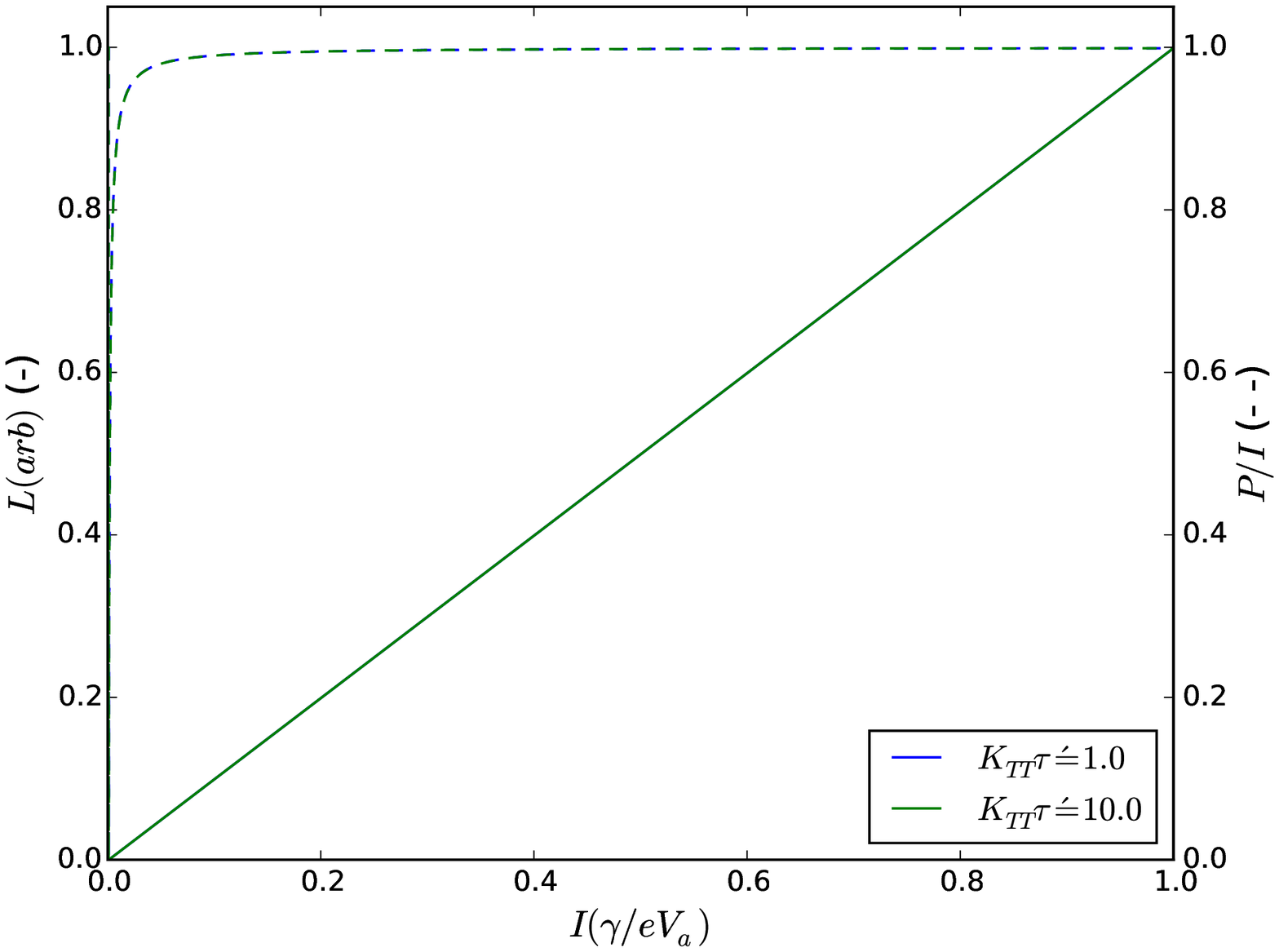}
    \label{fig:stimKTT}
  }
  \caption{
    \label{fig:KTT}
    Output and efficiency,  $\KTT \tau'$ as indicated.
    Efficiency rolls off while output rises monotonically.
  }
\end{figure}

Light output and external efficiency for spontaneous emission and stimulated emission with triplet-triplet annihilation are shown in FIG.~\ref{fig:KTT}. The $\KTT$ values used here are large compared to some reported values.\cite{Gartner2007}
Efficiency shows roll-off and output increases monotonically.
It is easily shown that $dN/dI > 0$ for $\KTT \geq 0$ and so $N^2$ losses in a 2 level system always produce monotonically increasing output.

\begin{figure}[h]
  \includegraphics[scale=0.35]{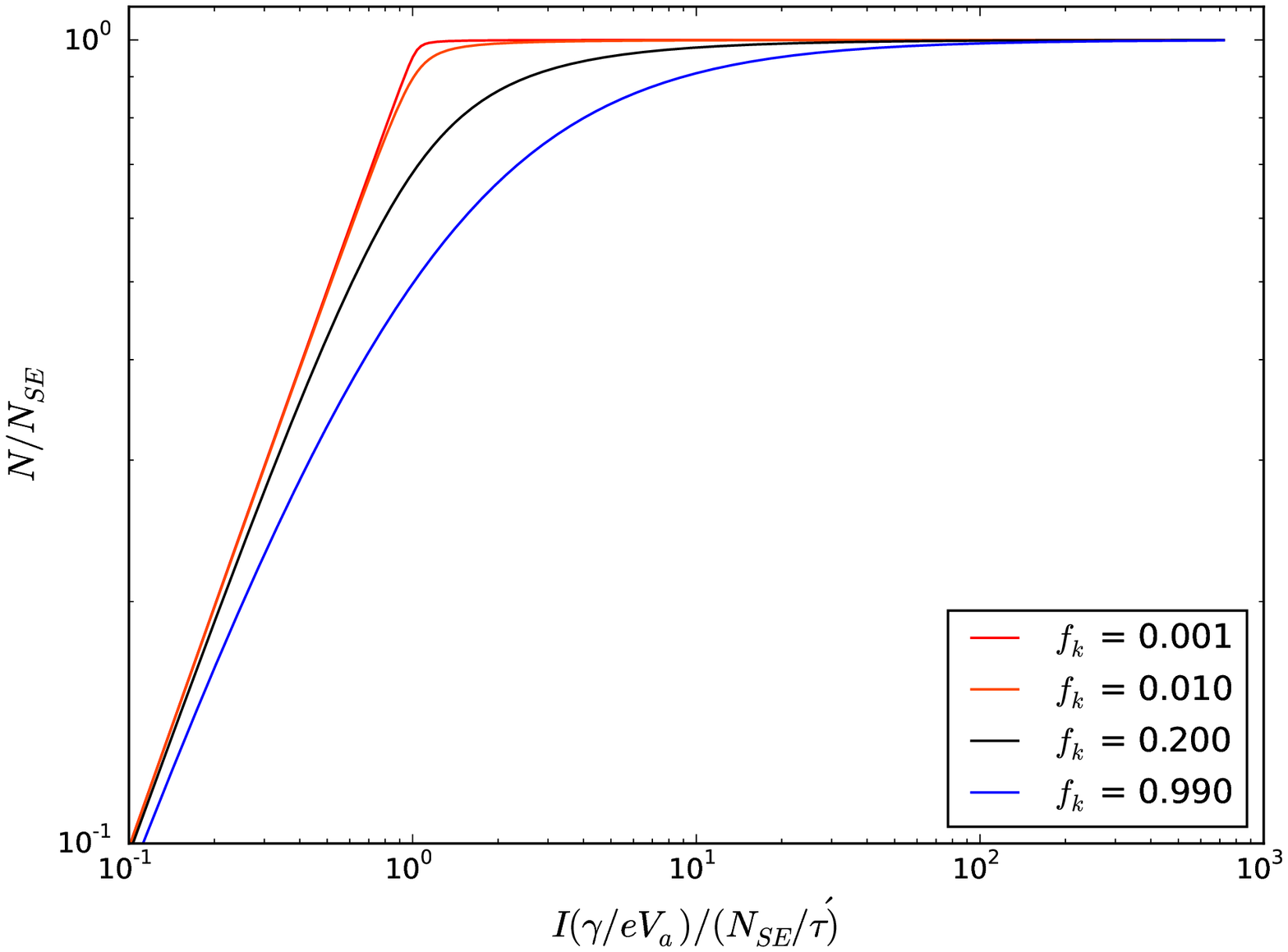}
  \caption{
    Onset of stimulated emission with a second order loss term.
  }
  \label{KTTonset}
\end{figure}

The transition to stimulated emission is found explicitly by starting with the general case solution where $\chi \neq 0$, $g\neq 0$ and $\NSE < 1$,
\begin{equation}
  \geV I = {N\over\tau'} \left( 1 + \KTT\ \tau'\ N + \chi\phi {N/\NSE\over 1 - N/\NSE} \right )
\end{equation}
As $\chi \rightarrow 0$, a sharp transition is obtained with an onset current that includes the $\KTT$ loss term evaluated at $\NSE$.
The transition region behavior is shown in FIG.~\ref{KTTonset} with $\KTT = 1/\tau'$, $\NSE = 10^{-3}$.

Losses in the form $n N$, described as charge or polaron quenching,\cite{Baldo2000,Reineke2007,Siboni2012} can be represented schematically as
\begin{eqnarray}
  n + N & \xrightarrow{K_{nN}} & N_0 + n^* \\
  n^* & \xrightarrow{1/\tau_n} & n
\end{eqnarray}
Rate equations for $n$, $N$ and $n^*$ are
\begin{eqnarray}
  {d n \over dt} & = & \geV I - \Keh n^2 (1 - N) - K_n n N + {1\over\tau_n} n^* \label{dnpolaron}\\ 
  {d N \over dt} & = & K_{eh} n^2 (1 - N) - K_n n N - \left({1 \over \tau'} + g P \right)\ N \label{dNpolaron}\\
  {d n^* \over dt} & = & K_n n N - {1\over\tau_n} n^* \label{dnstarpolaron}
\end{eqnarray}
For spontaneous emission with polaron quenching, the light current relationship is then
\begin{equation}
  \label{PnN}
  {P\over\tcav} = \chi\ \phi\ \geV I\ \left( {1\over K_n \tau' n + 1} \right)
\end{equation}
From equation \eqref{dnpolaron}, $n \sim \sqrt{\gev I / \Keh}$, and so attenuation builds as $1 + \Kn\tau'\sqrt{I}$.
We verify that there is no local maximum by first solving equation \eqref{dNpolaron} for $N$ as a function of $n$,
\begin{equation}
  \label{Npolaron}
  N = {K_{eh} n^2 \over (K_{eh} n^2 + K_n n + {1/\tau'})}
\end{equation}
and find that $d N / d n > 0$. Then substituting equation \eqref{Npolaron} into equation \eqref{dnpolaron} we obtain
\begin{equation}
  \label{Ipolaron}
  \geV I = { \Keh n^2 ( K_n n + 1/\tau')\over \Keh n^2 + K_n n + 1/\tau'}
\end{equation}
and find that $d I / d n > 0$.  Therefore, $ d N / dI > 0$ and output increases monotonically, while efficiency decreases monotonically.  The infinity in charge is avoided for $\NSE < 1$ after which the system transitions to stimulated emission.  Instead we consider a finite maximum charge density, and then equation \eqref{Npolaron} gives us a limit on the excited state population as,
\begin{equation}
  N \leq {\nlim \over \nlim + {K_n/\Keh} + {1/ \Keh \tau' \nlim} }  
\end{equation}
With $\Kn$ similar in order of magnitude with $\Keh$,\cite{Gartner2007} $N_{max}$ can be much larger than $\NSE$.

For stimulated emission with polaron quenching, the light-current relationship is found to be
\begin{equation}
  {P_{sp(nN)}\over\tcav} = {\geV} I - \alpha_{SE}/V_a - K_n n N_{SE}
\end{equation}
Output is approximately linear since the loss term is scaled by $\NSE$ and grows slowly.

\begin{figure}[h]
    \subfigure[\ Spontaneous emission ]{
      \includegraphics[scale=0.35]{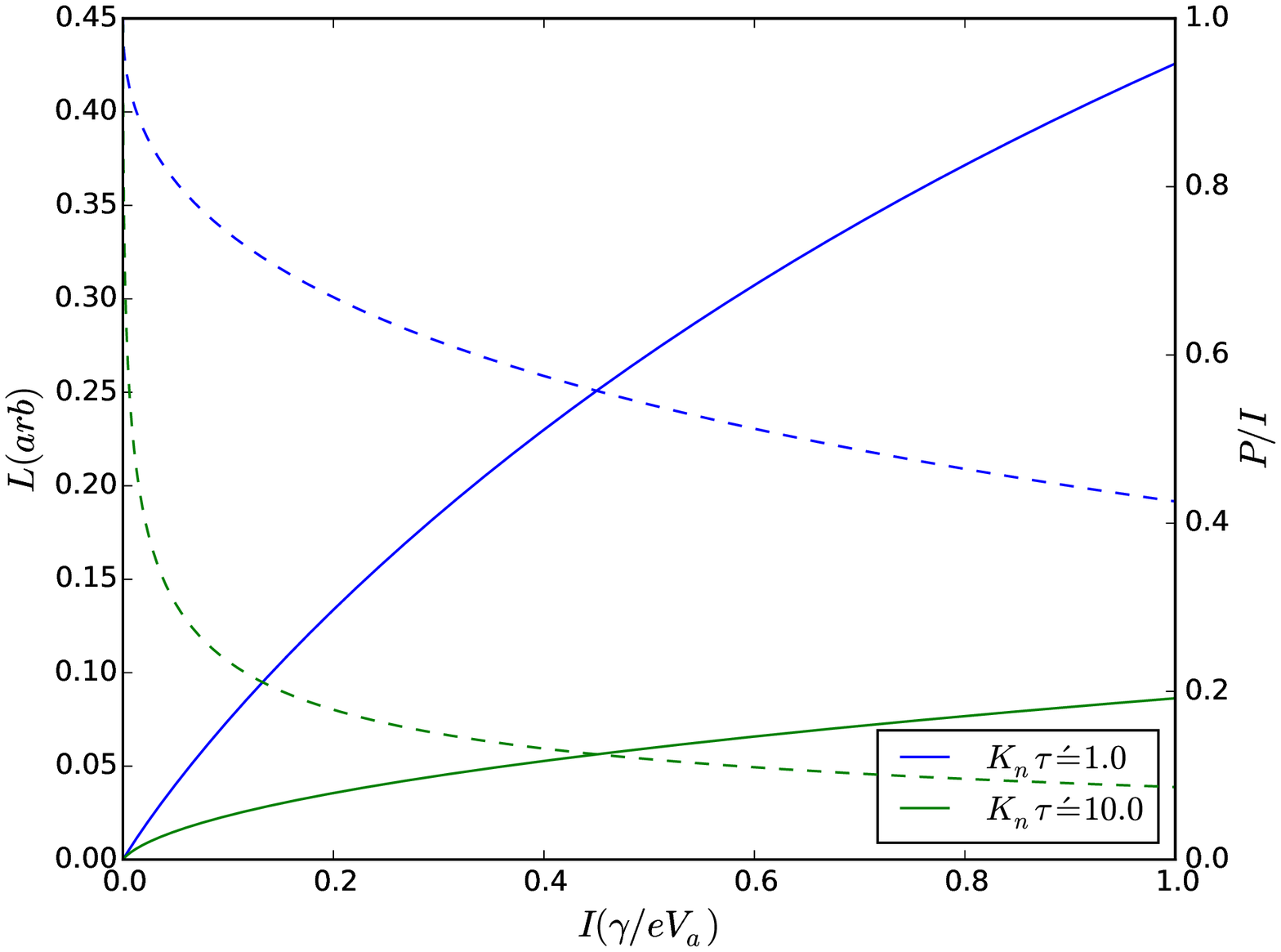}
      \label{fig:spontKn}
    }
  \subfigure[\ Stimulated emission ]{
    \includegraphics[scale=0.35]{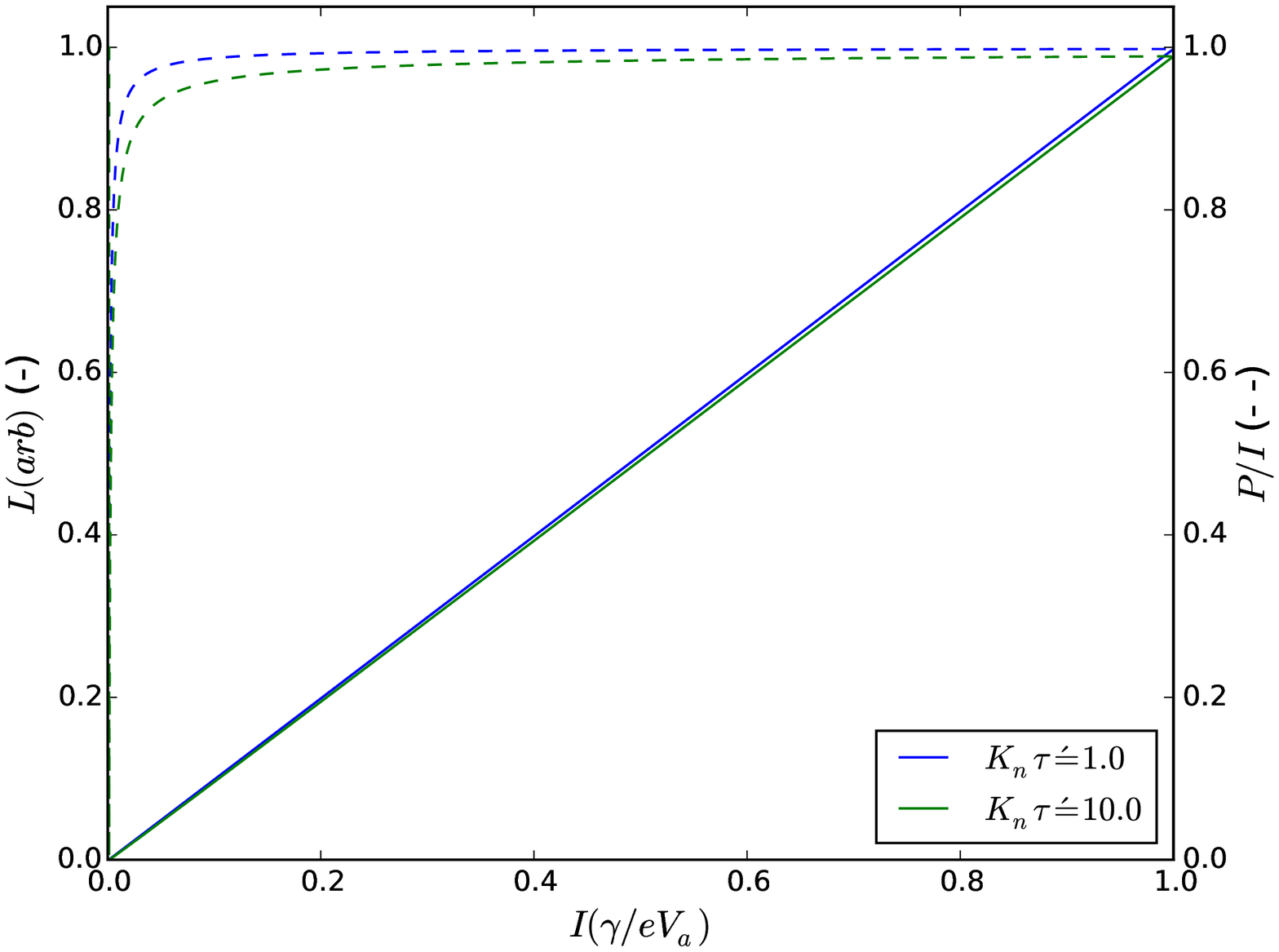}
    \label{fig:stimKn}
  }
  \caption{
    \label{fig:polaron}
    Polaron quenching losses, $\Kn \tau'$ as indicated.
  }
\end{figure}

Output and efficiency for spontaneous emission and stimulated emission with $n N$ quenching are shown in FIG.~\ref{fig:polaron}.
The curve for the spontaneous emission case is generated parametrically from equation \eqref{Npolaron} and equation \eqref{Ipolaron}.
$\Kn$ values in the range used here have been reported for a few materials.\cite{List2001,Baldo2002}
The losses are greater than those produced by TTA.
The stimulated emission device shows linear behavior with a slightly lower approach to the asymptote in efficiency.

\begin{figure}[h]
  \includegraphics[scale=0.35]{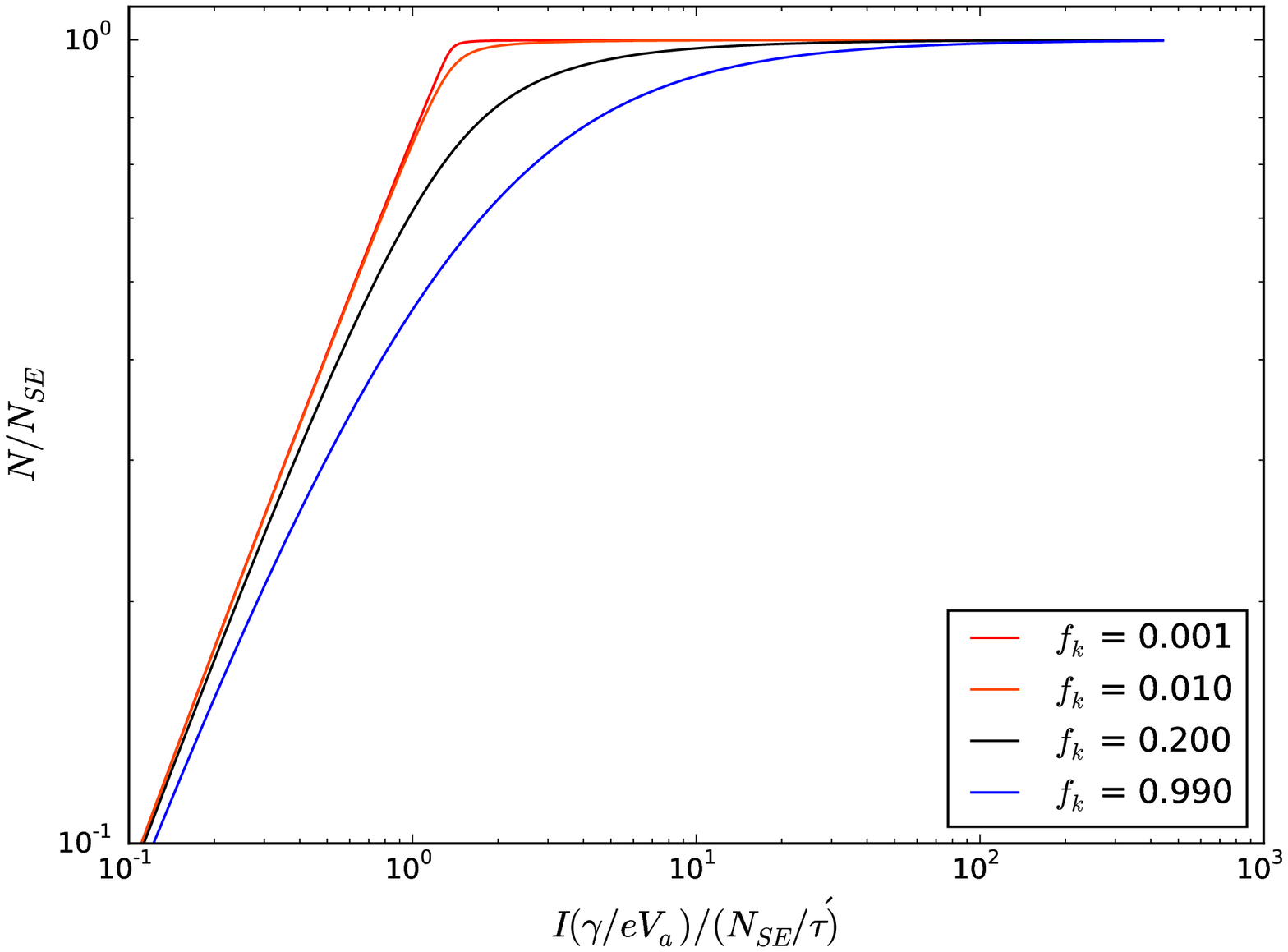}
  \caption{
    \label{fig:polarononset}
    Onset of stimulated emission with polaron quenching.
  }
\end{figure}

The transition to stimulated emission is again found in the general solution,
\begin{equation}
  \label{eq:polarontransition}
  \geV I = {N\over\tau'} \left( 1 + \Kn\tau' n + \chi\phi {N/\NSE\over 1 - N/\NSE} \right)
\end{equation}
The increment in the onset current is $\Kn \tau' \sqrt{\NSE/\Keh\tau'}$.  
FIG.~\ref{fig:polarononset} shows the result of evaluating equation \eqref{eq:polarontransition} at several values of $\chi$ with $\Kn = 10$.
As $\chi$ approaches zero, the transition again becomes sharp and moves to lower current.
Charge quenching increases with current and is thought to pose a challenge to reaching stimulated emission in conventional devices.\cite{Baldo2002,Gartner2007,GiebinkForrest2009}

We now consider a perhaps hypothetical loss mechanism that exhibits a maximum at finite current. The mechanism is represented schematically as
\begin{eqnarray}
  e + h + N & \xrightarrow{\KehN} & N_2 \\
  N_2 & \xrightarrow{ 1/\tau_{(2)} } & N_0
\end{eqnarray}
where $N_2$ is the population in a second excited state and the process might be described as recombination bleaching.
Charge build-up on emitter molecules has been reported\cite{Weichsel2012}
and higher excited states are accessible in molecules similar to some used in OLEDS.\cite{Itoh2012}
The rate constant could vary with applied voltage as $\exp[-\Delta E /\V]$, where $\Delta E$ is the energy increment to form the second excited state.  We will take this as constant for the present analysis.

Rate equations including the $n^2N$ loss term are written as,
\begin{eqnarray}
  {d n \over dt} & =
  & \geV I - K_{eh}\ n^2\ (1 - N - N_2) - \KehN\ n^2\ N\label{dnbleached} \\
  {d N \over dt} & =
  & K_{eh}\ n^2\ (1 - N - N_2) - \left( \KehN\ n^2 + {1 \over \tau'} + g P\right) N \label{dNbleached} \\
  {d N_2 \over dt} & =
  & \KehN\ n^2\ N - {1 \over \t2}\ N_2 \label{dN2bleached}
\end{eqnarray}
and for spontaneous emission the relationship between current and light is
\begin{equation}
  \label{Pspbleached}
       {P_{sp}\over\tcav} =
        \chi \phi \left( { 1 \over 1 + 2\KehN \tau' n^2 }\right) \geV I
\end{equation}
The bleaching interaction appears as an attenuation with second order dependence on charge. The excited state population $N$ as a function of $n$, is found from equations \eqref{dNbleached} and \eqref{dN2bleached},
\begin{equation}
  \label{Nrecomb2}
    N_{sp} = {\Keh n^2 \over \Keh\KehN\t2 n^4 + (\Keh + \KehN) n^2 + 1/\tau'}
\end{equation}
and the charge-current behavior is obtained from equation \eqref{dnbleached},
\begin{equation}
  \geV I = \Keh n^2\ { 2\KehN n^2 + 1/\tau' \over \Keh \KehN \t2 n^4 + (\Keh + \KehN) n^2 + 1/\tau'}
\end{equation}
From equation \eqref{Nrecomb2} it is found that $\Nsp$ has a maximum at
\begin{equation}
  \label{nmaxspontbleached}
  n^2_{sp(max)} = {1 \over \sqrt{ 2 \Keh \tau' \KehN \t2}}
\end{equation}
and so,
\begin{equation}
  \label{Nmaxspontbleached}
  N \leq\ { \Keh \tau'\over (\Keh + \KehN)\ \tau' + \sqrt{ 2 \Keh \tau' \KehN \t2} }
\end{equation}
which moves to lower charge density for longer lifetime emitters.
Maximum output is
\begin{equation}
  \label{Lmaxspontbleached}
  {\Psp\over\tcav}\ \leq\ \chi \phi
  \left( { \Keh\over (\Keh + \KehN)\ \tau' + \sqrt{ 2 \Keh \tau' \KehN \t2}
  }
  \right)
\end{equation}
which is reduced with longer lifetime emitters.

For stimulated emission, light output as a function of current is
\begin{equation}
  \label{Pstbleached}
  {P_{SE}\over\tcav} = \geV I - \left( {1\over\tau'} + \KehN n^2 \right) N_{SE}
\end{equation}
The loss terms are scaled by $\NSE$ and output is approximately linear in current.
The charge-current behavior for stimulated emission is
\begin{equation}
  \geV I = [ \Keh n^2 - (\Keh - \KehN)\NSE ] n^2 - \Keh \KehN \t2 \NSE n^4
\end{equation}
and the charge-light behavior is
\begin{equation}
  {\Pse\over\tcav} = \left[ \Keh (1 - N_{SE}) - \KehN N_{SE} \right] n^2
  - \Keh \KehN \t2 N_{SE}\ n^4
\end{equation}
which has a maximum at
\begin{equation}
n^2_{SE(max)} = {1\over 2} \left( {1 - N_{SE}\over N_{SE}} - {\KehN\over\Keh} \right) { 1\over\KehN\t2 }
\end{equation}
For small $N_{SE}$, the first term becomes large and the maximum in charge density is raised far above practical operating conditions. Output at the maximum is
\begin{equation}
  {\Pse\over\tcav} \lesssim {1\over 4}\ {\Keh\over\KehN\t2}\ { (1 - N_{SE})^2\over N_{SE} }
\end{equation}
But more likely, output will be charge density limited at $\Pse/\tcav \lesssim \Keh \nnlim$.

\begin{figure}[h]
    \subfigure[Spontaneous emission]{
      \includegraphics[scale=0.35]{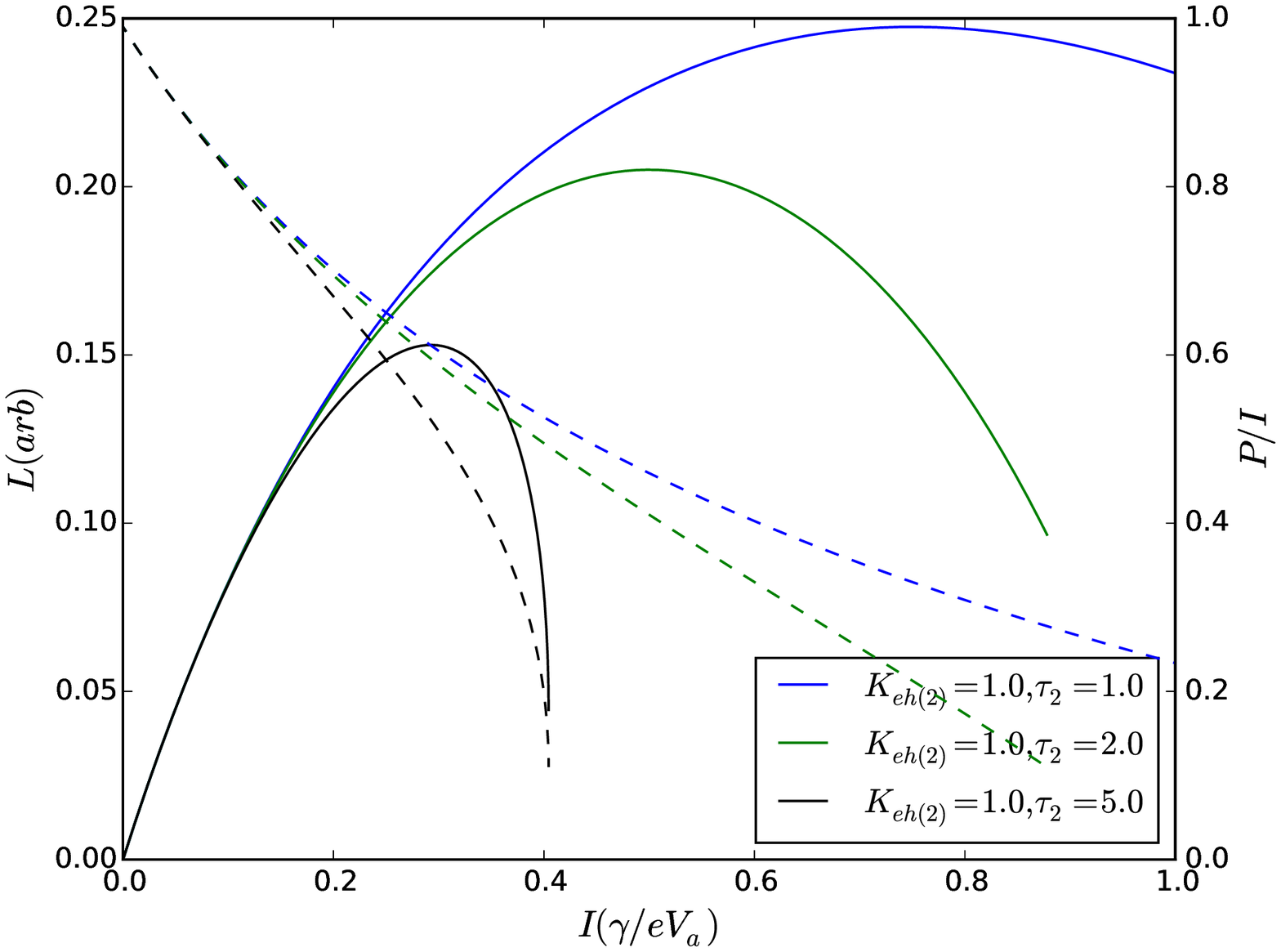}
      \label{SpontBleaching}
    }
  \subfigure[Stimulated emission]{
    \includegraphics[scale=0.35]{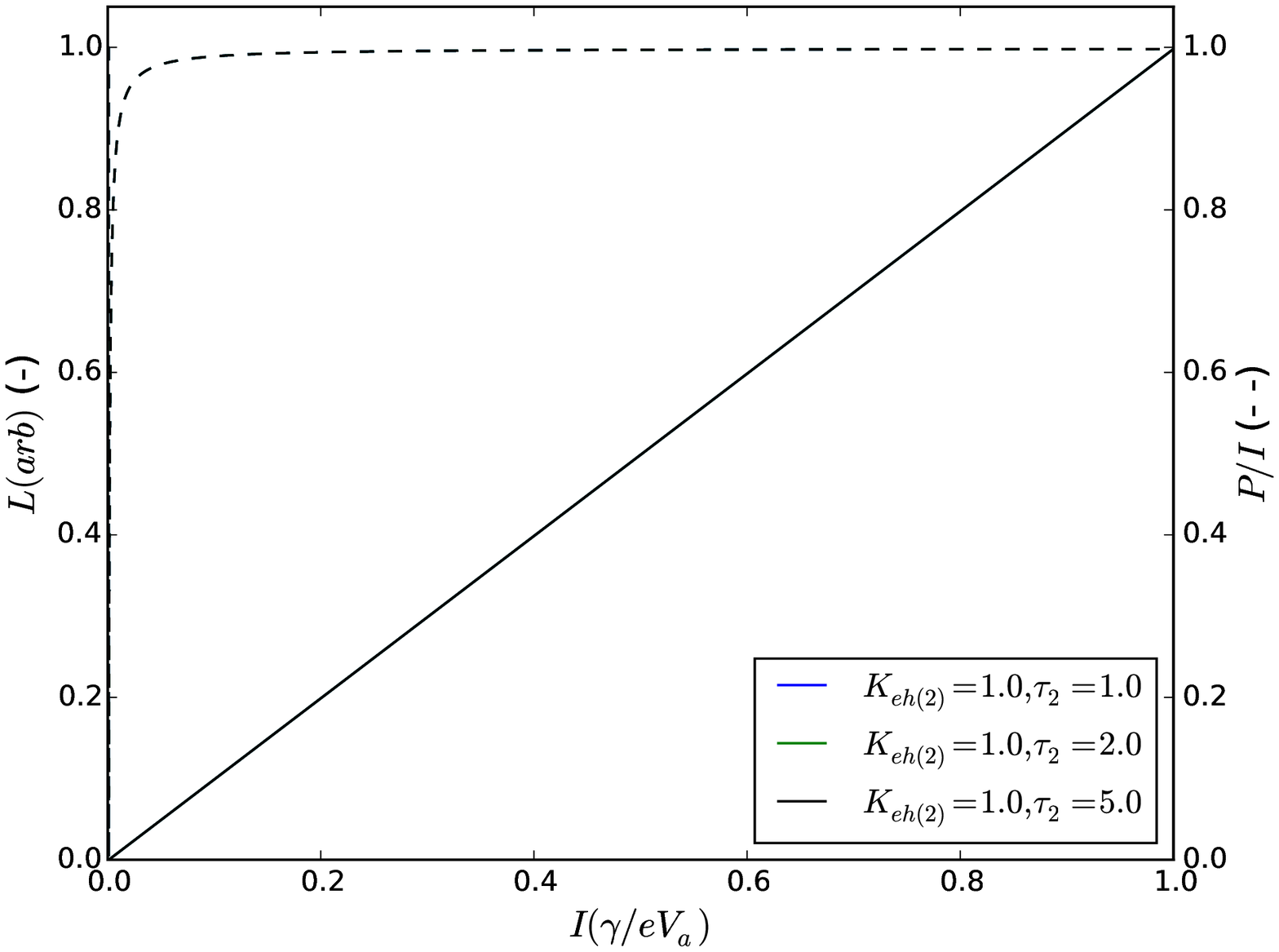}
    \label{StimBleaching}
  }
  \caption{
    \label{BleachingCalculation}
    Spontaneous and stimulated emission with $\KehN n^2 N$ losses.
    Output and efficiency calculated with $\KehN \t2 = 1, 2, 5$, and for the stimulated emission case $N_{SE} = 0.001$.
    Efficiency rolls off, output rises and then rolls off.
  }
\end{figure}

Output and current efficiency with $n^2 N$ losses is shown in FIG.~\ref{BleachingCalculation},
with $\KehN$ values chosen to illustrate the behavior qualitatively.
Stimulated emission with the same parameters immediately achieves ideal efficiency, output is 20 to 30 times greater at the same current density, and there is no roll-off.

\begin{figure}[h]
  \includegraphics[scale=0.35]{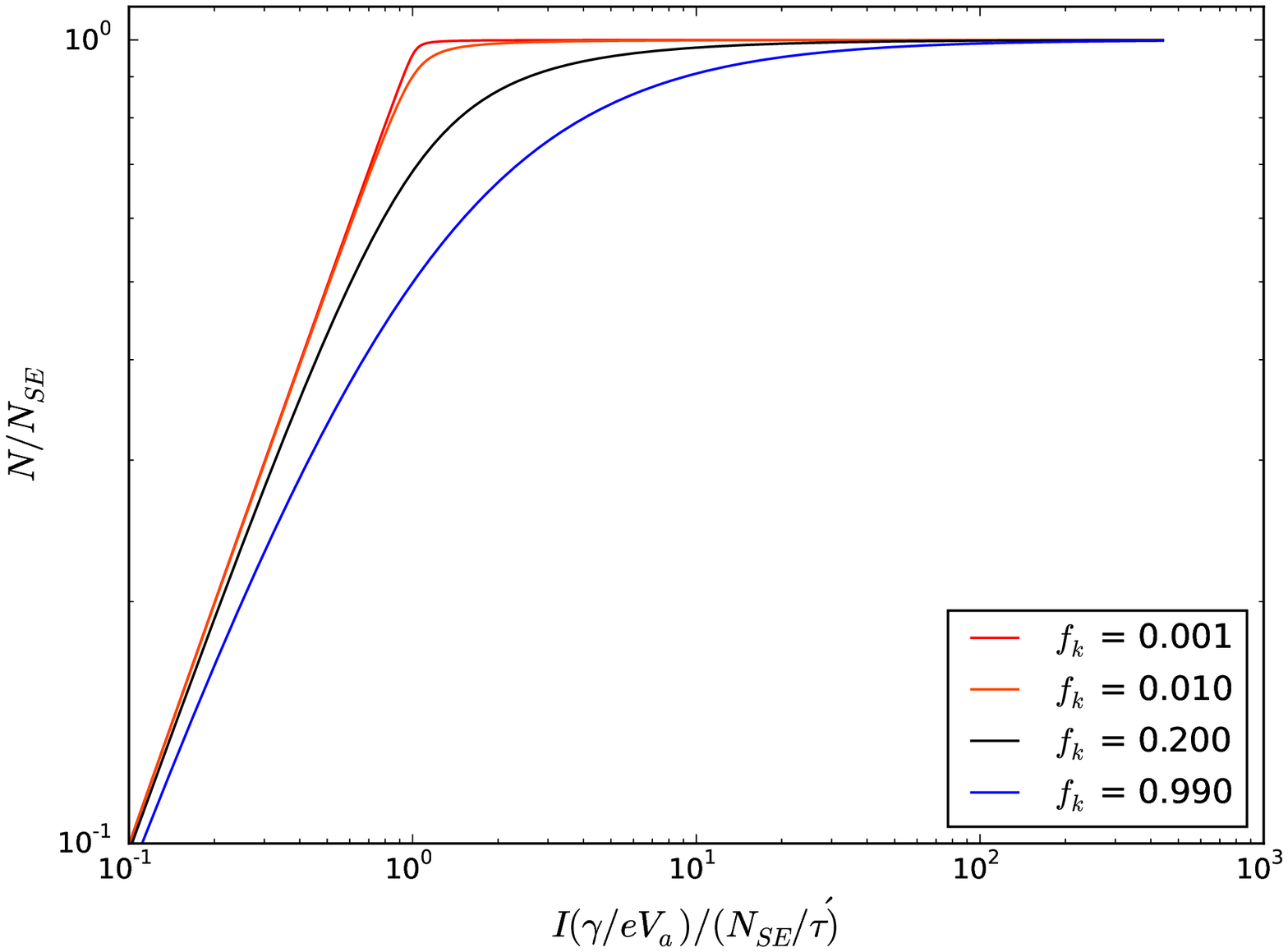}
  \caption{
    \label{Keh2onset}
    Onset of stimulated emission with charge bleaching.
  }
\end{figure}

The general solution in the transition region is
\begin{equation}
  \geV I = {N\over\tau'} \left( 1 + 2\KehN\tau' n^2 + \chi\phi {N/\NSE\over 1 - N/\NSE} \right)  
\end{equation}
and again, the loss term does not significantly alter the behavior as $\chi \rightarrow 0$ when the $n^2$ term is small in the region of the transition.  FIG.~\ref{Keh2onset} shows the onset behavior under conditions similar to those in the FIG.~\ref{BleachingCalculation}, the effect on the transition is minimal.

Behaviors of higher order loss mechanisms are summarized in Table~\ref{O2properties}.  The losses discussed so far do not preclude the transition to stimulated emission for $\chi \sim 0$ and the increments in the onset current are small for rate constants in the range of reported values.
However, in some materials the rate for intersystem crossing can be large compared to $1/\tau'$, as discussed in the next section.

\begin{table}[h]
  {
    \setlength{\tabcolsep}{1.0em}
    \renewcommand{\arraystretch}{2.0}%
    \begin{tabular}{c|c|c|c}
      \centering
      Loss
      & Spontaneous Em.
      & Stimulated Em.
      &
      $\displaystyle \Delta\ I \left({\gamma\over e V_a}\right)/\left({\NSE\over\tau'}\right) $\\
      \hline
      \multirow{2}{*}{
        $\displaystyle N^2$
      }
      &
      \multirow{2}{40mm}{
        $\displaystyle L \sim {\chi\over\tsp}\sqrt{ {V_a\over\KTT} {\gE} I}$
      }
      &
      \multirow{2}{20mm}{
        $\displaystyle L = \gE I - \alpha$
      }
      &
      \multirow{2}{*}{
        $\displaystyle \KTT\ \tau'\ \NSE$
      }\\
      & & & \\
      \hline
      \multirow{2}{*}{
        $\displaystyle n N$
      }
      &
      \multirow{2}{40mm}{
        $\displaystyle L = \chi\phi\ {(\gamma/e)\ I\over K_n\tau' n + 1}$
      }
      &
      \multirow{2}{20mm}{
        $\displaystyle L \sim \gE I - \alpha$
      }
      &
      \multirow{2}{*}{
        $\displaystyle \Kn\tau'\sqrt{\NSE\over \Keh\tau'}$
      }\\
      & & & \\
      \hline
      \multirow{2}{*}{
        $\displaystyle n^2 N$
      }
      &
      \multirow{2}{50mm}{
        $\displaystyle L = \chi\phi\ {(\gamma/e)\ I\over 2 \KehN\tau' n^2 + 1}$ \\[1em]
        $\displaystyle N \lesssim {\Keh \tau'\over 2[(\Keh+\KehN)\tau' + 1]}$
      }
      &
      \multirow{2}{25mm}{
        $\displaystyle L \sim \gE I - \alpha$
      }
      &
      \multirow{2}{*}{
        $\displaystyle {\KehN\over\Keh}\NSE$
      }
      \\
      & & & \\
    \end{tabular}
  }
  \caption{\label{O2properties} Comparison of light-current relationship and maximum output in spontaneous and stimulated emission with annihilation and bleaching losses, and contribution of each loss term to the onset current for stimulated emission with  $\chi \sim 0$, scaled to $\NSE/\tau'$.  At low power, $n^2 \sim \gev I/\Keh$.}
\end{table}

\subsection{Three level emitter and singlet/triplet fraction} \label{ssec:threelevel}
In this section we study efficiency and onset current in systems with singlet and triplet excited states.
The excited states are formed in a statistical ratio, but the populations are controlled by the rates of competing processes, such as depicted in FIG.~\ref{processes}.
The behavior of the base system with first order processes is studied first, which now include intersystem crossings, and then higher order losses are added,
which now include triplet-triplet annihilation and singlet-triplet annihilation.

In organic molecules, intersystem crossing (ISC) rates are typically $\lesssim 10^9/s$ (slow ISC) and in transition metal complexes ISC rates can be $10^{12}/s$ (fast ISC).\cite{Yersin2004,Forster2006}.
The reverse intersystem crossing (RISC) rates are temperature dependent,
$\Krisc = \Phi \exp[-\Delta E_{ST}/ k_B T]$ where $\Delta E_{ST} = E_S - E_T$ is the singlet-triplet energy gap and $k_B T$ is about 26 meV at room temperature.
Metal complexes and excimers with small singlet-triplet gaps are well known and
there are now organic molecules with
$\Delta E_{S_T} < 100\ {\rm meV}$ and
$\Krisc \sim 10^7/s$ at room temperature.\cite{Endo2011,Goushi2012,Uoyama2012,Dias2013}.
This creates a large space of interesting devices.
In our numerical evaluations we will consider three notional devices, one with a fast ISC emitter, one with a high gap (700 meV) slow ISC emitter, and one with a low gap (35 meV) slow ISC emitter.

Excited state production and loss by charge recombination and dissociation, is represented schematically as
\begin{equation}
  \begin{aligned}
  e + h + S_0 & \xrightarrow{\Keh} (\nST) S^* + (1-\nST) T^* \\
  S^* & \xrightarrow{\KdS} S_0 + e + h \\
  T^* & \xrightarrow{\KdT} S_0 + e + h
  \end{aligned}
\end{equation}
where $S_0$ represents the ground state.
The two excited state populations are connected by intersystem crossings,
\begin{equation}
  \begin{aligned}
    S^* & \xrightarrow{\Kisc} T^*\\
    T^* & \xrightarrow{\Krisc} S^*
  \end{aligned}
\end{equation}
and each excited state can relax non-radiatively, or by spontaneous or stimulated emission into any of the allowed modes,
\begin{equation}
  \begin{aligned}
      S^* & \xrightarrow{1/\tnrS} S_0 + {\rm phonon}\\
      S^* & \xrightarrow{\fkS/\tspS} S_0 + P_{k_S}\\
      P_{k_S} + S^* & \xrightarrow{\gkS} S_0 + 2\ P_{k_S}\\
  \end{aligned}
\end{equation}
\begin{equation}
  \begin{aligned}
      T^* & \xrightarrow{1/\tnrT} S_0 + {\rm phonon}\\
      T^* & \xrightarrow{\fkT/\tspT} S_0 + P_{k_T}\\
      P_{k_T} + T^* & \xrightarrow{\gkT} S_0 + 2  P_{k_T}   
  \end{aligned}
\end{equation}
where $P_{k_x}$ represents a photon produced into mode $k$ from state $x \in S,T$.
The rate equation for charge density is
\begin{equation}
  \label{st:dnbase}
  { d \n \over dt } = { \gamma \over e V_a } I  - \Keh \n^2 \Sa + \KdS \Sb + \KdT \Tb
\end{equation}
where $d[n] = d[ e ] = d[ e ]$ is the change in recombination charge density,
$K_{d_x}$ is the rate constant for dissociation of charge from $x$, and in our units, $\Sa + \Sb + \Tb = 1$.
Rate equations for the excited state populations are written as
  \begin{equation}
    \label{st:dNbase}
    \begin{aligned}
      { d \Sb \over dt } =\ &\nST\ \Keh \Sa \n^2 + \Krisc \Tb\\
      &- \left[ \Kisc + \KdS + {1\over\tnrS}
        + \sum_{k_S}\left( {\fkS\over\tspS} + \gkS \PkS \right) \right] \Sb\\[0.5em]
      { d \Tb \over dt } =\ &(1 - \nST)\ \Keh \Sa \n^2  + \Kisc \Sb\\
      &- \left[ \Krisc + \KdT + {1\over\tnrT}
        + \sum_{k_T}\left( {\fkT\over\tspT} + \gkT \PkT \right) \right] \Tb
    \end{aligned}
  \end{equation}
where $\nST$ is the fraction of recombination events leading to singlets,
$\tnrS$ is the lifetime for non-radiative relaxation,
$\tspS$ is the lifetime for spontaneous emission,
$\fkS$ is the cavity factor for mode $k$ with $\sum_{k_S}\fkS = 1$,
and $\gkS$ is the stimulated emission rate constant.
Rate equations for photon production and loss in each of the optical modes are written as,
\begin{equation}
  \label{st:dPbase}
  \begin{aligned}
  { d \PkS \over dt } & = \left( {\fkS\over\tspS} + \gkS \PkS \right)\ \Sb - {\PkS\over\tkS} \\[0.5em]
  { d \PkT \over dt } & = \left( {\fkT\over\tspT} + \gkT \PkT \right)\ \Tb - {\PkT\over\tkT}
  \end{aligned}
\end{equation}
where $\tau_{k_x}$ is the cavity lifetime for mode $k_x$ and the equations are repeated for all of the modes in the device. Light is coupled to the outside through a vertical mode $k'_x$.
Applying steady state conditions and adding the rate equations, we obtain
\begin{equation}
  \label{st:Ibase}
  \geV  I =
        {1\over\tnrS}\ \Sb
        + {1\over\tnrT}\ \Tb
        + \sum_{k_S} {1\over\tkS}\ \PkS
        + \sum_{k_T} {1\over\tkT}\ \PkT
\end{equation}
The recombination current is balanced by non-radiative relaxation and production of photons in all of the allowed modes.

For spontaneous emission in both the singlet and triplet, the steady state solutions are
\begin{equation}
  \label{st:NbaseSp}
  \begin{aligned}
    \nST\ \geV I = \left( \Kisc + (1-\nST)\KdS
    + {1\over\tS} \right) \Sb - \left(\Krisc + \nST\KdS\right)\Tb\\[0.5em]
    (1 - \nST)\ \geV I = \left( \Krisc + \nST \KdT
    + {1\over\tT}\right) \Tb - \left(\Kisc + (1-\nST) \KdS\right)\Sb
  \end{aligned}
\end{equation}
and the light-current relationship becomes,
\begin{equation}
  \label{st:IbaseSp}
  \gE I = {1\over\fkSout\phi_S}\ \Lk[S] + {1\over\fkTout\phi_T}\ \Lk[T]
\end{equation}
with $L_{k'_x} = V_a [P_{k'_x}]/\tau_{k'_x}$, where $[P_{k'_x}]/\tau_{k'_x} = [x] f_{k'_x}/\tau_{sp_x}$
and $1/\tau'_{x} = 1/\tau_{nr_x} + 1/\tau_{sp_x}$.
Equations \eqref{st:NbaseSp} can be solved to obtain the ratio of singlet to triplet excited state populations,
\begin{equation}
  \label{st:RbaseSp}
        {\Sb\over\Tb} =
        {
          \Krisc  + \nST \left(\KdT + 1/\tT\right)
          \over
          \Kisc + \left(1 - \nST\right)\left(\KdS + 1/\tS\right)
        }
\end{equation}
and light-current relationships for the singlet and triplet can then be obtained as
\begin{equation}
  \label{st:LbaseSp}
  \begin{aligned}
    \Lk[S]\ & =\ \left({R_{S/T}\ \tT/\tS\over R_{S/T} + \tT/\tS}\right) \FkPrime[S]\ \phi_S \gE I \\[1em]
    \Lk[T]\ & =\ \left({\tS/\tT \over R_{S/T}\ + \tS/\tT}\right) \FkPrime[T] \phi_T\ \gE I
  \end{aligned}
\end{equation}
where $R_{S/T} = \Sb/\Tb$ is the singlet/triplet ratio from the preceding equation.
If the intersystem crossing and dissociation rates could be set to zero ($K_{[r]isc} = K_{d_x} = 0$), then equations \eqref{st:LbaseSp} would reduce to the traditional\cite{Tsutsui13} efficiency relations
\begin{equation}
  \label{tsutsuieqs}
  \begin{aligned}
    \LkSout & = \nST\ \fkSout\ \phi_S\ \ge I \\[0.5em]
    \LkTout & = (1 - \nST)\ \fkTout\ \phi_T\ \ge I
  \end{aligned}
\end{equation}
With the linear processes present, the fluorescence/phosphorescence yield ratio is generally not equal to the singlet/triplet ratio, and the singlet/triplet ratio is generally very different from $\nST/(1-\nST)$.
When higher order loss terms are included, the relationship between singlet and triplet populations, or outputs, will be a function of current, or charge.

For stimulated emission in the singlet state, the rate equations become
\begin{equation}
  \label{st:NbaseStS}
  \begin{aligned}
    &\nST \geV I = \left( \Lambda_S + \Rprime[S] \right) \Sb_{(SE)} - \Lambda_T \Tb + \LP[S] \\[0.5em]
    &\nTS \geV I = \left( \Lambda_T + \Rprime[T] \right) \Tb - \Lambda_S\ \Sb_{(SE)}
  \end{aligned}
\end{equation}
where we have introduced the quantities $\Lambda_S = \Kisc + \nTS\KdS$, $\Lambda_T = \Krisc + \nST\KdT$
and $\nTS = 1 - \nST$, and $\Sb_{(SE)} = (\gkS \tkS)^{-1}$ is the singlet excited state in stimulated emission and is constant.
The light-current relationship for the singlet in stimulated emission is then
\begin{equation}
  \begin{aligned}
    \LP[S]\ &= \geV I \left( 1- \beta_S\right) - \alpha_S \\[1em]
    &\alpha_S = \left({\Rprime[S]} + {\Lambda_S\over\Lambda_T\tau'_T + 1}\right)\Sb_{(SE)} \\[0.5em]
    &\beta_S = {\nTS\over\Lambda_T\tau'_T + 1}
  \end{aligned}
\end{equation} 
while output from the triplet becomes
\begin{equation}
  \LP[T] =  \chi_T \phi_T \beta_S\ \geV I 
  + \left({ \Lambda_S \over\Lambda_T\tau'_T + 1}\right) \Sb_{(SE)}
\end{equation}
where $\phi_T = \tT/\tspT$.
The equation for the complimentary case with the triplet in stimulated emission is
\begin{equation}
  \begin{aligned}
    \LP[T]\ &= \geV I \left( 1- \beta_T\right) - \alpha_T\\[1em]
    &\alpha_T = \left( {\Rprime[T]} + {\Lambda_T \over \Lambda_S\tau'_S + 1}\right) \Tb_{(SE)} \\[0.5em]
    &\beta_T = {\nST\over\Lambda_S\tau'_S + 1}
    \end{aligned}
\end{equation}
while output from the singlet becomes
\begin{equation}
  \LP[S] =  \chi_S \phi_S \beta_T\ \geV I 
  + \left({ \Lambda_T \over\Lambda_S\tau'_S + 1}\right) \Tb_{(SE)}
\end{equation}
In each of the singlet and triplet, crossing from the other contributes to efficiency and reduces the offset current in stimulated emission.
Fast intersystem crossing in a high gap emitter for example, can be expected to achieve near 100\% efficiency in the triplet in stimulated emission.

The transition to stimulated emission is again found in the general solution,
\begin{equation}
  \begin{aligned}
    \nST \geV I = &
    \left[ \Lambda_S
      + {1\over\tS}\left(1 + \chi_S\phi_S {\Sb/\Sb_{SE}\over 1- \Sb/\Sb_{SE}} \right)\right]\Sb
    - \Lambda_T \Tb \\[1em]
    \nTS \geV I = &
    \left[ \Lambda_T + {1\over\tT}\left(1 + \chi_T\phi_T {\Tb/\Tb_{SE}\over 1- \Tb/\Tb_{SE}} \right)\right]\Tb
    - \Lambda_S \Sb
  \end{aligned}
\end{equation}
The onset current at $\chi_S \sim 0$, for the singlet to reach stimulated emission with the triplet still in spontaneous emission, is
\begin{equation}
      \geV I_{onset_S} =
      \left( 1 + {\nST \Lambda_S \tS - \nTS\Lambda_T \tT \over \Lambda_T\tT + \nST}\right) {\Sb_{SE}\over\tS}
\end{equation}
Similarly, the onset current for the triplet to reach stimulated emission at $\chi_T \sim 0$ with the singlet still in spontaneous emission, is
\begin{equation}
  \geV I_{onset_T} =
  \left( 1 + {\nTS \Lambda_T \tT - \nST\Lambda_S \tS \over \Lambda_S\tS + \nST}\right) {\Tb_{SE}\over\tT}
\end{equation}
For the singlet, a fast intersystem crossing rate can produce a large onset current, but in a low gap device this can be offset by the reverse rate scaled by the longer lifetime of the triplet.
Onset current for the triplet will generally be small.

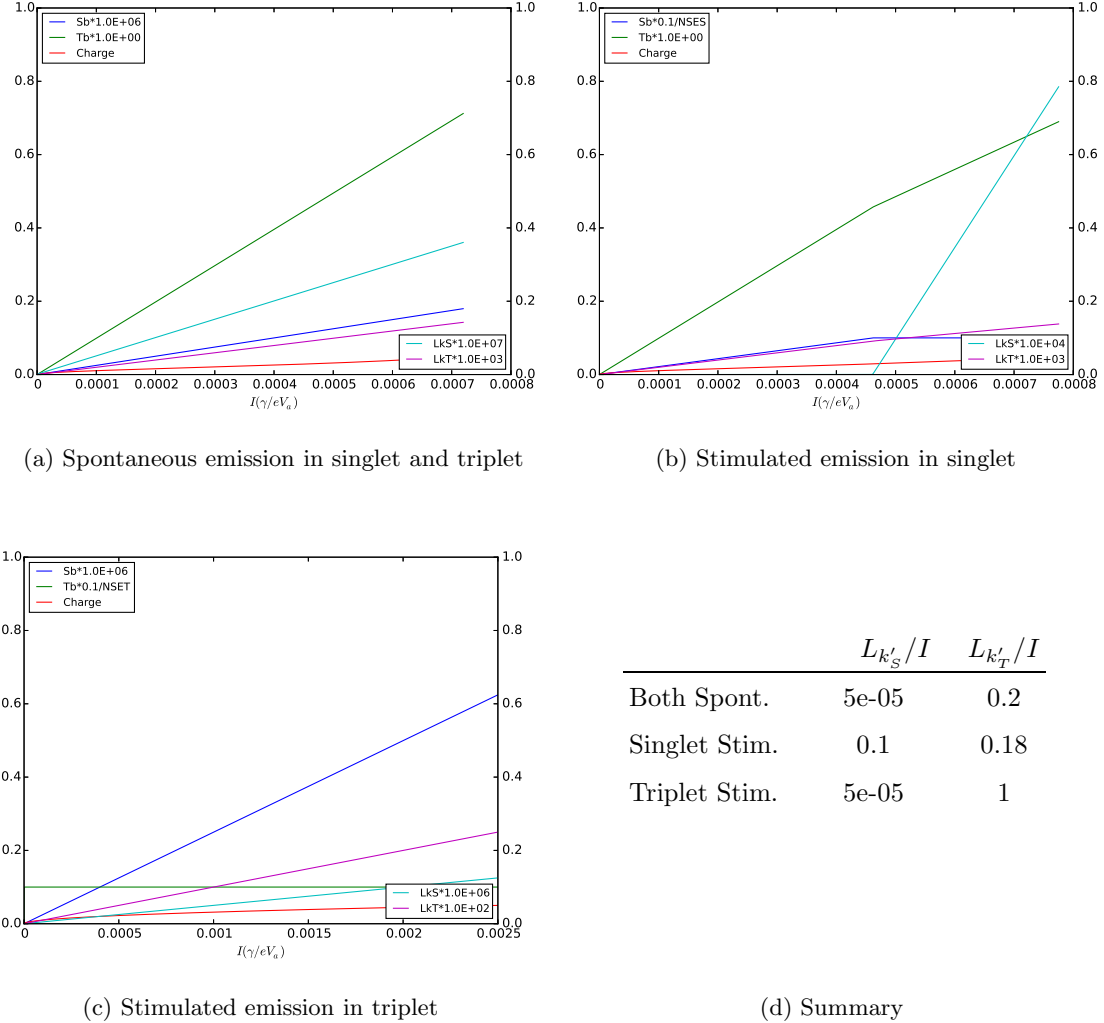
\begin{figure}[h]
  \input{FIG.FastISC.tex}

  \caption{
    \label{fig:FastISC}
    Fast Intersystem Crossing ($S^*\rightarrow T^*$), no losses:
    Excited state populations, charge and light output from singlet and triplet states.
    {(Pars: {\it $\Keh=1\ \tspS=1\ \tspT=0.001$, \pars}, units: $ ns, nm, nm^3, N_0 + N = 1$)}
    }
\end{figure}

FIG.~\ref{fig:FastISC} shows results of numerical evaluation with fast ISC.  The singlet is depopulated and efficiency in the triplet is at $\chi$. Stimulated emission in the singlet shows the expected increase in onset current, and stimulated emission in the triplet reaches near ideal efficiency.

The high gap slow ISC emitter is evaluated and shown in FIG.~\ref{fig:HighGap}. The singlet is again depopulated and the triplet efficiency is again close to $\chi$. In stimulated emission, efficiency from the singlet is at $\nST$. The triplet in stimulated emission is above $(1-\nST)$.

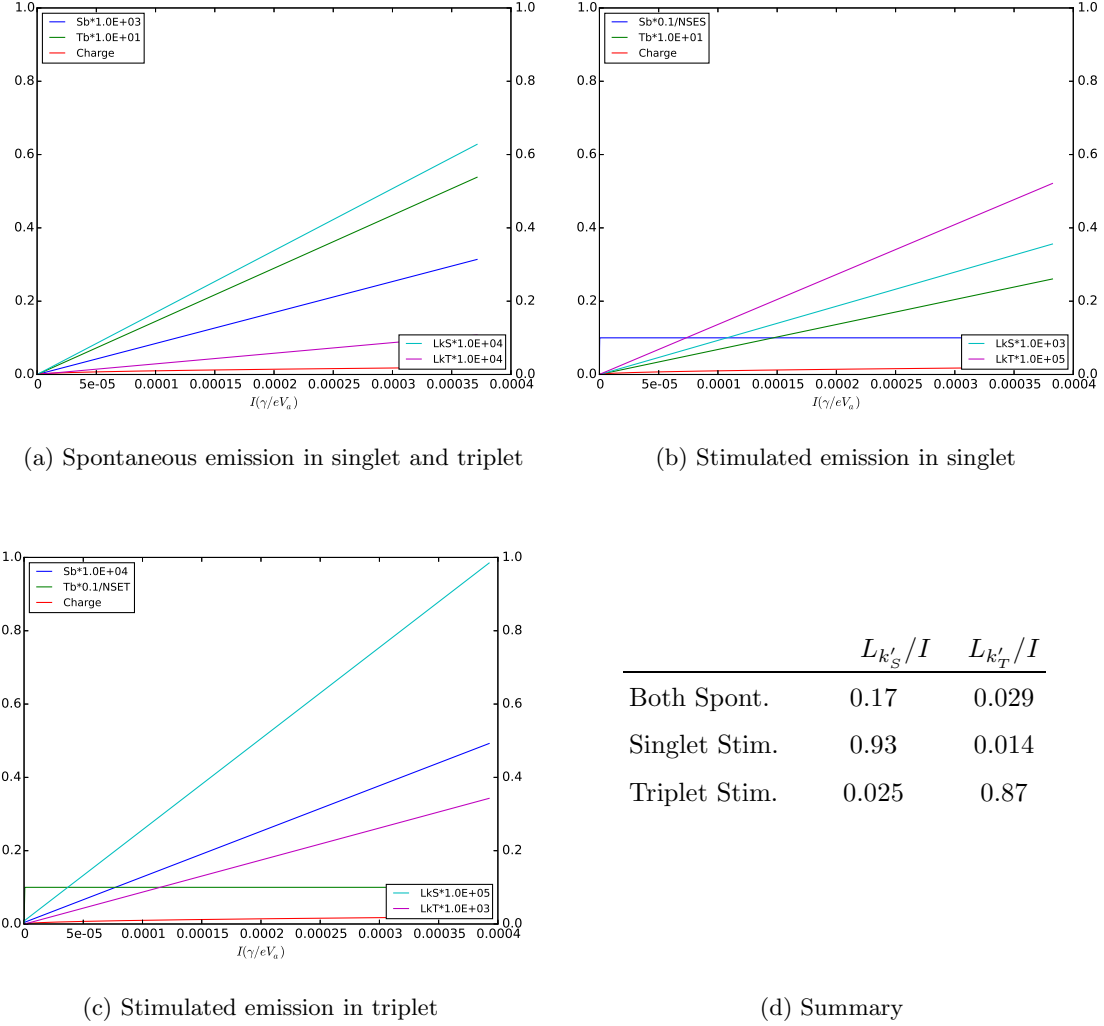
\begin{figure}[h]
  \input{FIG.LowGap.tex}

  \caption{
    \label{fig:LowGap}
    Slow Intersystem Crossing, Low Gap ($S^*\leftrightarrow T^*$, $\Delta E_{S-T} = 35\ meV$), no losses:  Excited state populations, charge and light output from singlet and triplet states.
    {(Pars: {\it $\Keh=1\ \tspS=1\ \tspT=0.001$, \pars}, units: $ ns, nm, nm^3, N_0 + N = 1$)}
  }
\end{figure}

Results for the low gap slow ISC emitter are shown in FIG.~\ref{fig:LowGap}. In spontaneous emission, efficiency in the singlet is close to $\chi$ while the triplet with its slow emission rate, is depleted by reverse ISC. In stimulated emission the singlet reaches near ideal efficiency since the forward ISC is small at $\Sb_{SE}$ and the triplet in stimulated emission is again above $(1-\nST)$.

\begin{figure}[h]
  \input{FIG.HighGap.tex}

  \caption{
    \label{fig:HighGap}
    Slow Intersystem Crossing ($S^*\rightarrow T^*$, $\Delta E_{S-T} = 700\ meV$), no losses:  Excited state populations, charge and light output from singlet and triplet states.
    {(Pars: {\it $\Keh=1\ \tspS=1\ \tspT=0.001$, \pars}, units: $ ns, nm, nm^3, N_0 + N = 1$)}
  }
\end{figure}

Triplet-triplet annihilation (TTA) is represented schematically in two channels where one channel contributes to the singlet excited state population,
\begin{equation}
  \begin{aligned}
      T^* + T^* & \xrightarrow{\KTTT} T^{**} + S_0\\[0.5em]
      T^* + T^* & \xrightarrow{\KTTS} S^{*} + S^*\\  
  \end{aligned}
\end{equation}
When both the singlet and triplet states are in spontaneous emission, the steady state solutions are
\begin{equation}
  \label{st:NspQ}
  \begin{aligned}
    &\nST\geV I = \left(\Lambda_S + \Rprime[S] \right) \Sb
    - \Lambda_T \Tb - \KTTS \Tb^2 \\[0.5em]
    &\nTS\geV I = \left(\Lambda_T + \Rprime[T] \right) \Tb
    - \Lambda_S \Sb + \KTT' \Tb^2
  \end{aligned}
\end{equation}
where $\KTT = \KTTT + \KTTS$ and $\KTT' = \KTT + \KTTS$.
The $\Tb^2$ term appears as a loss in the triplet and as a source in the singlet.
Eliminating $\Tb^2$ from equations \eqref{st:NspQ} gives us
\begin{equation}
  \label{st:IST}
  \left( {\nTS\over\KTT'} + {\nST\over\KTTS}\right)\geV I =
  \left( {\Lambda_T+1/\tT\over\KTT'} - {\Lambda_T\over\KTTS} \right)\Tb
  +\left( {\Lambda_S+1/\tS\over\KTTS} - {\Lambda_S\over\KTT'} \right)\Sb
\end{equation}
We then obtain relationship between current and the excited state populations in spontaneous emission as,
\small
\begin{equation}
  \label{st:TTASinglet}  
  \begin{aligned}
    \left[\nST + {\KTTS\over\KTT} + \left(\Lambda_T \tT - {\KTTS\over\KTT}\right) {\beta_I\over\beta_T\tT}\right] &\geV I = \\
    &\left[1+\Lambda_s\tS+{\KTTS\over\KTT}+\left(\Lambda_T \tT - {\KTTS\over\KTT}\right){\beta_I\tS\over\beta_T\tT} \right] {\Sb\over\tS}
  \end{aligned}
\end{equation}
\normalsize
and
\begin{equation}
  \label{st:TTATriplet}  
  \begin{aligned}
    \left(\nTS + \Lambda_S {\beta_I\over\beta_S}\right)\geV I = \left(1+\Lambda_T\tT + \Lambda_S {\beta_T\tT\over\beta_S}\right){\Tb\over\tT} + \KTT'\Tb^2
  \end{aligned}
\end{equation}
where $\beta_I$, $\beta_T$ and $\beta_S$ are the expressions inside the parentheses in equation \eqref{st:IST}. The light-current relationships are
\begin{equation}
  \begin{aligned}
   & L_{k'_S} = \Gamma_{I/S}\ \chi\phi \gE I\\[0.5em]
   & L_{k'_T} \sim {\fkT\over\tspT} \sqrt{ \left(\nTS + \Lambda_S {\beta_I\over\beta_S}\right) {V_a\over\KTT} {\gamma\over e} I}
  \end{aligned}\\[0.5em]
\end{equation}
where $\Gamma_{S/I}$ is the ratio of the two square brackets in equation \eqref{st:TTASinglet}.  Examining the coefficient expressions in detail, we see the expected contributions from intersystem crossing in each direction and the contribution from the singlet producing channel of the annihilation mechanism.  The leading term in each is of course $\nST$ or $\nTS$,  but the overall coefficient is very different from these quantities.
We note that the TTA mechanism produces only monotonically increasing output and no asymptote in the spontaneous emission limit.

With the singlet in stimulated emission, the steady state solutions are
\begin{equation}
  \begin{aligned}
    \label{tripletstim}
    &\nST\geV I = \left(\Lambda_S + \Rprime[S]\right) \Sb_{SE} - \Lambda_T \Tb - \KTTS \Tb^2_{SE}+\LP[S]\\[0.5em]
    &\nTS\geV I = \left(\Lambda_T + \Rprime[T]\right) \Tb - \Lambda_S \Sb_{SE} + \KTT' \Tb^2_{SE}
  \end{aligned}
\end{equation}
The relation for current and light output from the singlet are,
\small
\begin{equation}
  \begin{aligned}
    \Biggl[
    \nST + \nTS{\KTTS\over\KTT}
    + & \left(\Lambda_T-\Gamma_T{\KTTS\over\KTT}\right){\beta_I\over\beta_T}
    \Biggr] \geV I =\\[1em]
    &\left[\Gamma_S-\Lambda_S{\KTTS\over\KTT} -
      \left(\Lambda_T-\Gamma_T{\KTTS\over\KTT}\right){\beta_S\over\beta_T}\right]\Sb_{SE} + \LP[S]
  \end{aligned}
\end{equation}
\normalsize
where $\Gamma_S = \Lambda_S + 1/\tS$ and $\Gamma_T = \Lambda_T + 1/\tT$.  The relation for current and the triplet population is
\begin{equation}
  \nTS \geV I = \Gamma_T \Tb - \Lambda_S \Sb_{SE} + \KTT' \Tb^2
\end{equation}
Efficiency in the singlet is $\nST$ plus contributions from the singlet channel in TTA and from reverse crossing.
In the triplet we see the expected square root dependence as power is increased,  $L_T \sim \sqrt{\nTS (V_a/\KTT') (\gamma/e) I}$.

For stimulated emission in the triplet, the steady state solutions are,
\small
\begin{equation}
  \begin{aligned}
    \label{tripletstim}
    &\nST\geV I = \left(\Lambda_S + \Rprime[S]\right) \Sb - \Lambda_T \Tb_{SE} - \KTTS \Tb^2_{SE}\\[0.5em]
    &\nTS\geV I = \left(\Lambda_T + \Rprime[T]\right) \Tb_{SE} - \Lambda_S \Sb + \KTT' \Tb^2_{SE}+\LP[T]
  \end{aligned}
\end{equation}
\normalsize
and the excited state and output relations are
\begin{equation}
  \begin{aligned}
    \left(1-{\nST\over\Gamma_S\tS}\right) \geV I = \left(1+{\Lambda_T\tT\over\Lambda_S\tS}\right){\Tb_{SE}\over\tT}
    +\left(\KTT+{1\over\Gamma_S\tS}\right)\Tb^2_{SE} + \LP[T]
  \end{aligned}
\end{equation}
and
\begin{equation}
  \nST\geV I = \Gamma_S \Sb - \Lambda_T \Tb_{SE} - \KTTS \Tb^2_{SE}
\end{equation}
With the triplet in stimulated emission both outputs are linear, and with singlet production from TTA or reverse crossing in a low gap emitter, efficiency in the singlet can exceed $\nST \chi \phi$.

Results of numerical evaluation for TTA in the fast ISC system are shown in FIG.~\ref{fig:KTTfastisc}, where we see losses in the triplet, and gains in the singlet compared to FIG.~\ref{fig:FastISC}.
In FIG.~\ref{fig:KTThighgap} with slow ISC, stimulated emission in the singlet is now above $\nST$.
In FIG.~\ref{fig:KTTlowgap} with the low gap emitter, in spontaneous emission we see loss in the triplet and gain in the singlet, and in stimulated emission the losses are mitigated and the results are as in FIG.~\ref{fig:LowGap}.

\begin{figure}[h]
  \input{FIG.KTT.FastISC.tex}

  \caption{
    \label{fig:KTTfastisc}
    Triplet-Triplet Annihilation with fast ISC.
    {(Pars: {\it $\Keh=1\ \tspS=1\ \tspT=0.001$, \pars}, units: $ ns, nm, nm^3, N_0 + N = 1$)}
    }
\end{figure}

\begin{figure}[h]
  \input{FIG.KTT.HighGap.tex}

  \caption{
    \label{fig:KTThighgap}
    Triplet-Triplet Annihilation with slow ISC.
    {(Pars: {\it $\Keh=1\ \tspS=1\ \tspT=0.001$, \pars}, units: $ ns, nm, nm^3, N_0 + N = 1$)}
    }
\end{figure}

\begin{figure}[h]
  \input{FIG.KTT.LowGap.tex}

  \caption{
    \label{fig:KTTlowgap}
    Triplet-Triplet Annihilation with low gap emitter.
    {(Pars: {\it $\Keh=1\ \tspS=1\ \tspT=0.001$, \pars}, units: $ ns, nm, nm^3, N_0 + N = 1$)}
    }
\end{figure}

In triplet-singlet annihilation (TSA) both the singlet and the triplet are consumed. The mechanism is represented schematically as
\begin{equation}
  \begin{aligned}
      T^* + S^* & \xrightarrow{\KTS} T^{**} + S_0\\
  \end{aligned}
\end{equation}
In spontaneous emission, the steady state solutions are
\begin{equation}
  \label{eq:TSAspont}
  \begin{aligned}
    &\nST\geV I = \left(\Lambda_S + \Rprime[S] \right) \Sb - \Lambda_T \Tb + \KTS \Tb \Sb \\
    &\nTS\geV I = \left(\Lambda_T + \Rprime[T] \right) \Tb - \Lambda_S \Sb + \KTS \Tb \Sb
  \end{aligned}
\end{equation}
and we find relationships between current and excited state populations as,
\begin{equation}
  \def\LT{{\Lambda_{T}}}
  \def\LS{{\Lambda_{S}}}
  \begin{aligned}
    \geV I = {
      { 1 + \LS\tS + \LT\tT + \left(2\LT\tT+1\right)\KTS\tS\Sb }
      \over
          { \LT\tT + \nST - \left(\nTS-\nST\right)\KTS\tT\Sb }
          }\ {\Sb\over\tS}\\[1em]
    \geV I = {
      { 1 + \LS\tS + \LT\tT + \left(2\LS\tS+1\right)\KTS\tT\Tb }
      \over
          { \LS\tS + \nTS + \left(\nTS-\nST\right)\KTS\tS\Tb }
          }\ {\Tb\over\tT}
  \end{aligned}  
\end{equation}
For small excited state populations, the behaviors are quadratic.
We see that the singlet has an asymptote,
\begin{equation}
  \Sb_{TSA} < {\Lambda_T\tT + \nST \over \KTS\tT \left(\nTS-\nST\right) }
\end{equation}
which is generally order of $(1/\tT)/\KTS$ and larger than $\Sb_{SE}$.
So, we expect that TSA will typically not prevent a QW device from reaching stimulated emission.

For stimulated emission in the triplet, the steady state equations are
\begin{equation}
  \begin{aligned}
    &\nST\geV I = (\Lambda_S + \Rprime[S]) \Sb - \Lambda_T \Tb_{SE} + \KTS \Sb\Tb_{SE} \\
    &\nTS\geV I = (\Lambda_T + \Rprime[T]) \Tb_{SE} - \Lambda_S \Sb + \KTS\Sb\Tb_{SE} +
    {P_{k'_T}\over\tau_{k'_ST}}
  \end{aligned}  
\end{equation}
and solving for the light current relationship we obtain,
\begin{equation}
  \begin{aligned}
    {P_{k'_T}\over\tkT} = &\left(\nTS + \nST \rho_T\right) \geV I - \alpha_T\\[1em]
    &\rho_T = {\Lambda_S - \KTS\Tb_{SE} \over \Lambda_S + 1/\tS + \KTS\Tb_{SE}}\\[0.5em]
    &\alpha_T = \left(1/\tT + \Lambda_T { 1/\tS + 2 \KTS \Tb_{SE}\over \Lambda_S + 1/\tS + \KTS \Tb_{SE}}\right)\Tb_{SE}\\[1em]
    {P_{k'_S}\over\tkS} = &{\chi_S\over\tspS} {\nST \gev I + \Lambda_T \Tb_{SE}\over \Lambda_S + 1/\tS + \KTS\Tb_{SE}}\\[1em]
    \end{aligned}
\end{equation}
We see that the singlet becomes linear when the triplet has reached stimulated emission and that efficiency increases and the offset current is reduced by crossing from the singlet.
For the case where the singlet is in stimulated emission we obtain,
\begin{equation}
  \begin{aligned}
    {P_{k'_S}\over\tkS} = &\left(\nST + \nTS \rho_S\right) \geV I - \alpha_S\\[1em]
    &\rho_S = {\Lambda_T - \KTS\Sb_{SE} \over \Lambda_T + 1/\tT + \KTS\Sb_{SE}}\\[0.5em]
    &\alpha_S = \left(1/\tS + \Lambda_S { 1/\tT + 2 \KTS \Sb_{SE}\over \Lambda_T + 1/\tT + \KTS \Sb_{SE}}\right)\Sb_{SE}\\[1em]
  {P_{k'_T}\over\tkT} = &{\chi_T\over\tspT} {\nTS \gev I + \Lambda_S \Sb_{SE}\over \Lambda_T + 1/\tT + \KTS\Sb_{SE}}[1em]
  \end{aligned}
\end{equation}
and we see that the $\KTS$ loss mechanism reduces efficiency in the singlet.
In a low gap emitter the efficiency can be larger than $\nST$ if $\Lambda_T > \KTS \Sb_{SE}$.

FIG.~\ref{fig:KTSfastisc} shows numerical evaluation with TSA in the fast ISC model. In spontaneous emission and stimulated emission the results are similar to the linear fast ISC system.
In FIG.~\ref{fig:KTShighgap} with slow ISC and in FIG.~\ref{fig:KTSlowgap} with the low gap emitter, efficiency in spontaneous emission is reduced in both the singlet and triplet, and in stimulated emission the results are similar to the linear system.

Summarizing this section, we see that intersystem crossing is important in determining which of the singlet and triplet can reach stimulated emission with a small onset current, and in stimulated emission with a low excited state population the losses become insignificant.  TSA introduces an asymptote, but it will typically be greater than the excited state population required for stimulated emission.

\begin{figure}[h]
  \input{FIG.KTS.FastISC.tex}

  \caption{
    \label{fig:KTSfastisc}
    Singlet-Triplet Annihilation with fast ISC.
    {(Pars: {\it $\Keh=1\ \tspS=1\ \tspT=0.001$, \pars}, units: $ ns, nm, nm^3, N_0 + N = 1$)}
    }
\end{figure}

\begin{figure}[h]
  \input{FIG.KTS.HighGap.tex}

  \caption{
    \label{fig:KTShighgap}
    Singlet-Triplet Annihilation with slow ISC.
    {(Pars: {\it $\Keh=1\ \tspS=1\ \tspT=0.001$, \pars}, units: $ ns, nm, nm^3, N_0 + N = 1$)}
    }
\end{figure}

\begin{figure}[h]
  \input{FIG.KTS.LowGap.tex}

  \caption{
    \label{fig:KTSlowgap}
    Singlet-Triplet Annihilation in a low gap emitter.
    {(Pars: {\it $\Keh=1\ \tspS=1\ \tspT=0.001$, \pars}, units: $ ns, nm, nm^3, N_0 + N = 1$)}
    }
\end{figure}

\subsection{Numerical Evaluations} \label{ssec:numerical}
In the preceding analyses we have considered processes in OLED emitters individually. We now combine the known processes and write a more comprehensive set of rate equations\cite{Gartner2007} which we solve numerically. Results are shown for our limiting cases and notional devices and then for a known emitter in conventional and QW configurations using realistic coefficients for spontaneous and stimulated emission, crossings, dissociations, and losses. Rate equations for charge and excited state populations are
\small
\begin{equation}
  \label{eq:allterms}
  \begin{aligned}
    { d \n \over dt } = &{ \gamma \over e V_a } I  - \Keh \n^2 \Sa + \KdS \Sb + \KdT \Tb
    - \KehS \n^2 \Sb - \KehT \n^2 \Tb \\[2em]
    { d \Sb \over dt } = &
    \ \nST K_{eh} \n^2 \Sa - \KehS \n^2 \Sb - \KSn \n \Sb - \KdS \Sb - (\Kisc \Sb - \Krisc \Tb)\\
    &- \KTS \Tb \Sb + \KTTS \Tb^2 - (2-\xi_S)\KSS \Sb^2 - \left(1/\tnrS + 1/\tspS + \gkS \PkS\right)\Sb\\[1em]
    { d \Tb \over dt } = &
    \ \nTS K_{eh} \Sa \n^2  - \KehT \n^2 \Tb - \KTn \n \Tb - \KdT \Tb + (\Kisc \Sb - \Krisc \Tb) \\
    &- \KTS\Tb\Sb - (\KTTT+2\KTTS)\Tb^2 + (1-\xi_S) \KSS \Sb^2 \\
    &-\left(1/\tnrT+1/\tspT + \gkT\PkT\right)\Tb\\[1em]
    {d \Sbb \over dt} = &\KehS\ n^2\ \Sb - {1 \over \tSbb}\ \Sbb\\[1em]
    {d \Tbb \over dt} = &\KehT\ n^2\ \Tb - {1 \over \tTbb}\ \Tbb\\
  \end{aligned}
\end{equation}
A singlet-singlet annihilation mechanism $\KSS\Sb^2$ is included that is functionally similar to the triplet-triplet annihilation mechanism.   The rate equations for photon production are
\begin{equation}
  \label{eq:alltermsphotons}
  \begin{aligned}
    { d \PkS \over dt } = &( \fkS/\tspS + \gkS \PkS ) \Sb - \PkS/\tkS\\[1em]
    { d \PkT \over dt } = &( \fkT/\tspT + \gkT \PkT ) \Tb - \PkT/\tkT\\
  \end{aligned}
\end{equation}
\normalsize
Equations \eqref{eq:allterms} and equations \eqref{eq:alltermsphotons} were transcribed into a python program\cite{available} for numerical evaluation using Powell's method\cite{Powell1964}.
Results shown in our analyses in the preceding section were generated from this program with loss coefficients enabled as indicated in the figures.
Results for combined losses from TTA, TSA and charge quenching in the fast ISC emitter are shown in FIG.~\ref{fig:Allfastisc}, for the high gap slow ISC emitter in FIG.~\ref{fig:Allhighgap}, and for the low gap emitter in FIG.~\ref{fig:Alllowgap}.

\begin{figure}[h]
  \input{FIG.KTTKTSKnT.FastISC.tex}

  \caption{
    \label{fig:Allfastisc}
    TTA, TSA, CTA with fast ISC.
    {(Pars: {\it $\Keh=1\ \tspS=1\ \tspT=0.001$, \pars}, units: $ ns, nm, nm^3, N_0 + N = 1$)}
    }
\end{figure}

\begin{figure}[h]
  \input{FIG.KTTKTSKnT.HighGap.tex}

  \caption{
    \label{fig:Allhighgap}
    TTA, TSA, CTA with slow ISC emitter.
    {(Pars: {\it $\Keh=1\ \tspS=1\ \tspT=0.001$, \pars}, units: $ ns, nm, nm^3, N_0 + N = 1$)}
    }
\end{figure}

\begin{figure}[h]
  \input{FIG.KTTKTSKnT.LowGap.tex}

  \caption{
    \label{fig:Alllowgap}
    TTA, TSA, CTA in low gap emitter.
    {(Pars: {\it $\Keh=1\ \tspS=1\ \tspT=0.001$, \pars}, units: $ ns, nm, nm^3, N_0 + N = 1$)}
    }
\end{figure}

\begin{figure}[h]
  \subfigure[Cavity L($\lambda$) = 460 nm, x($\lambda$) = 240 nm] {
    \includegraphics[scale=0.4]{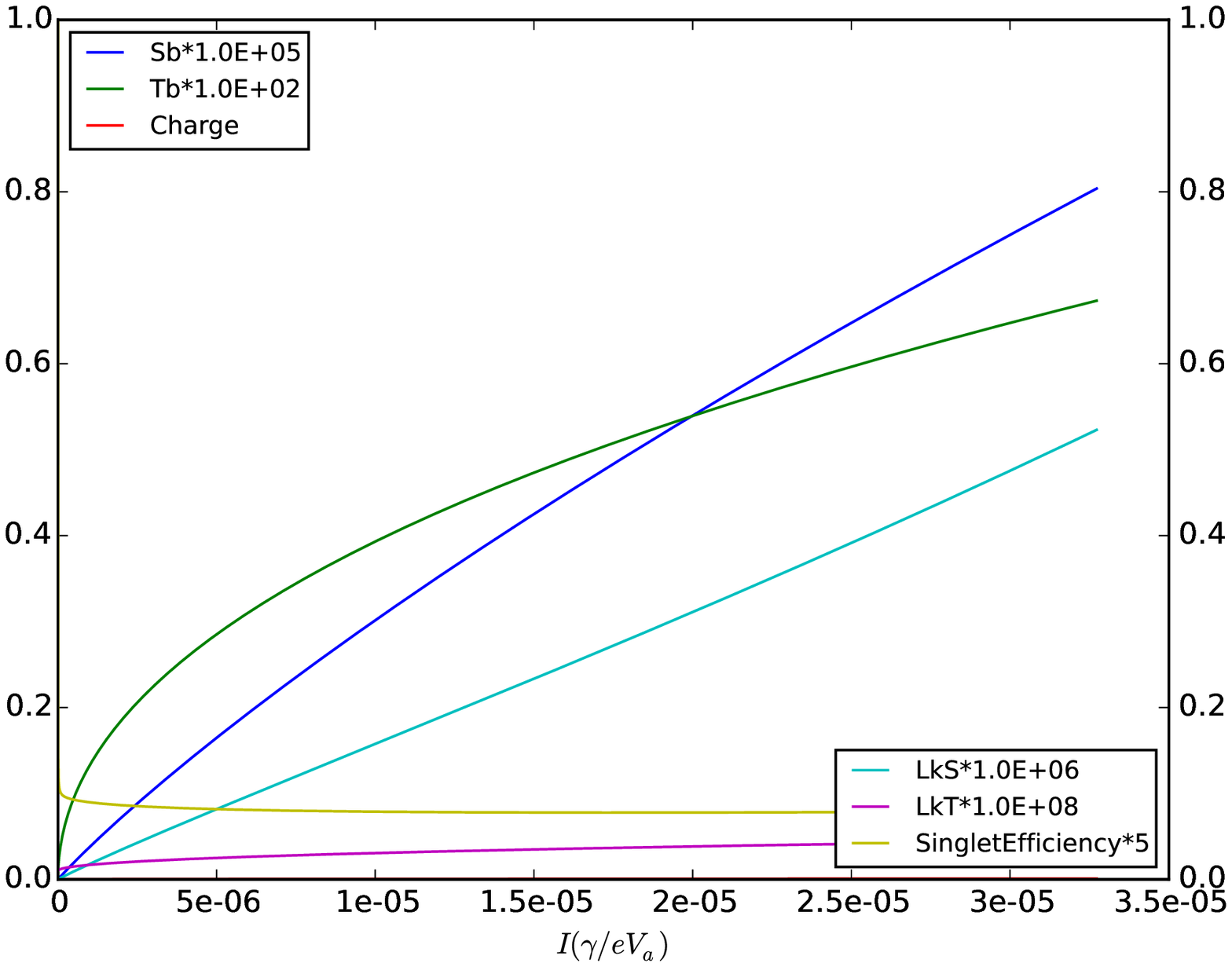}
    \label{fig:alq3lit}
  }
  \subfigure[Cavity L($\lambda$) = 520 nm, x($\lambda$) = 130 nm] {
    \includegraphics[scale=0.4]{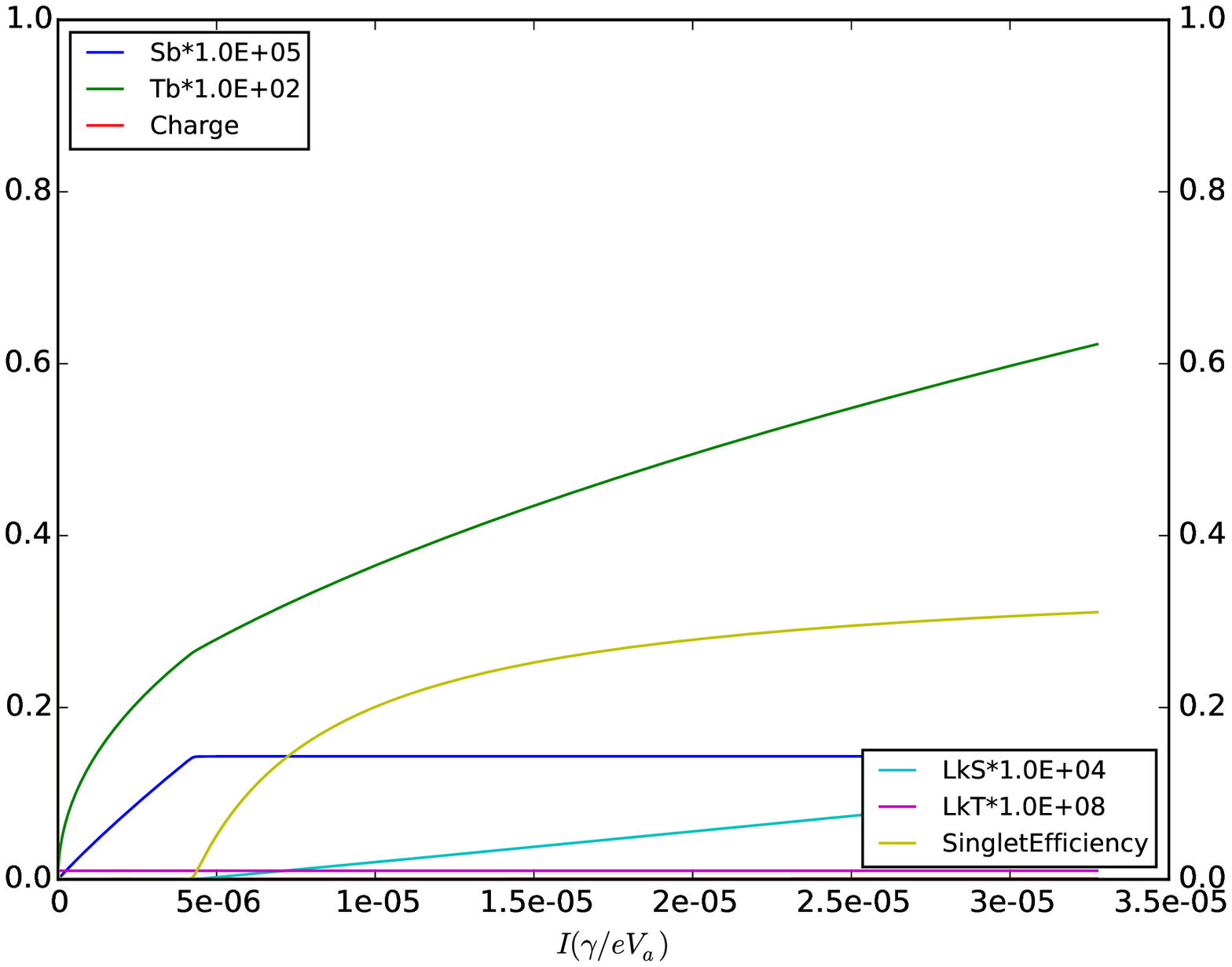}
    \label{fig:alq3qwS}
    }
  \subfigure[Cavity L($\lambda$) = 700 nm, x($\lambda$) = 175 nm, $\tspT = 15 ms$] {
    \includegraphics[scale=0.4]{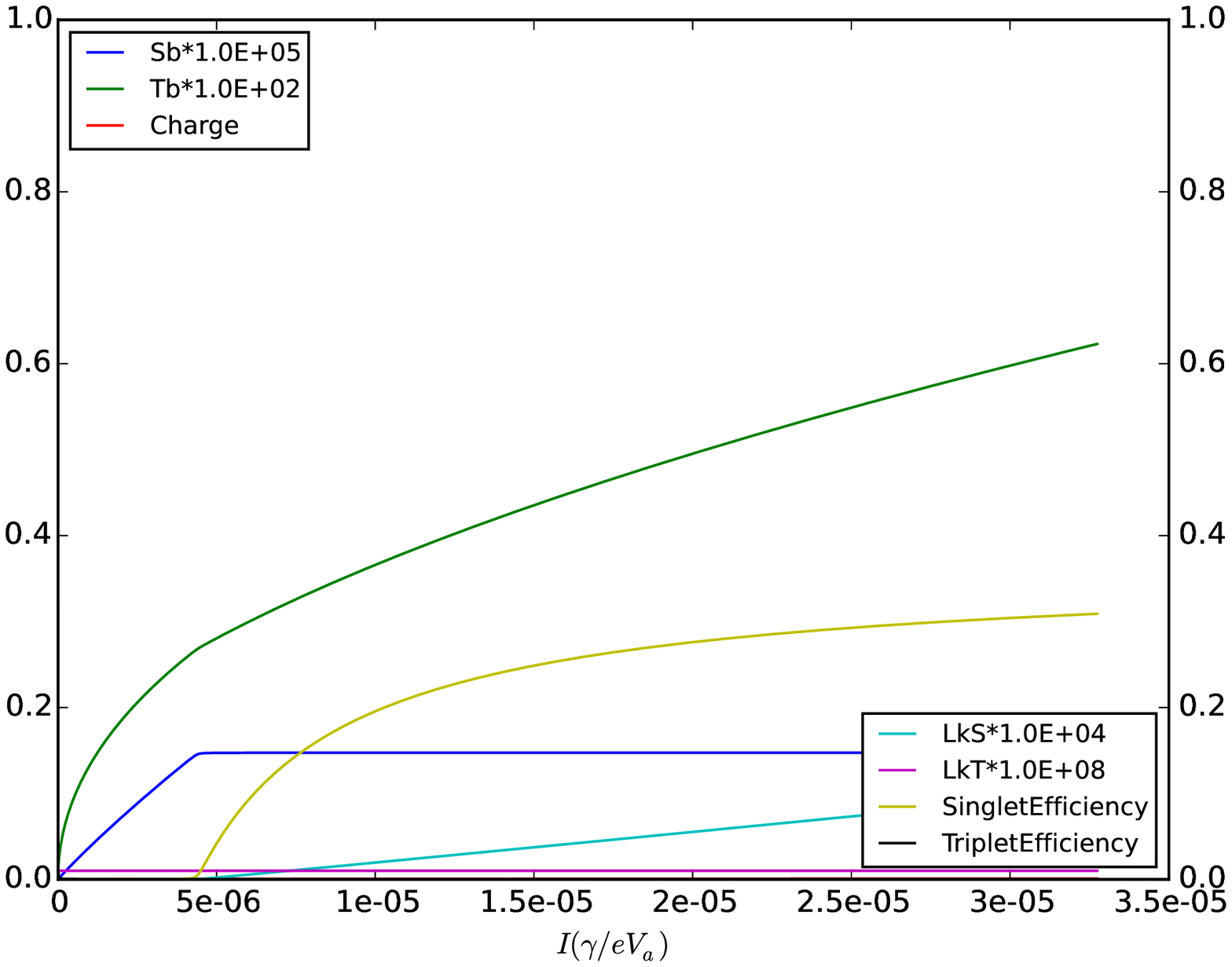}
    \label{fig:alq3qwT}
    }
  \subfigure[Cavity L($\lambda$) = 700 nm, x($\lambda$) = 175 nm, $\tspT = 15 \mu s$] {
    \includegraphics[scale=0.4]{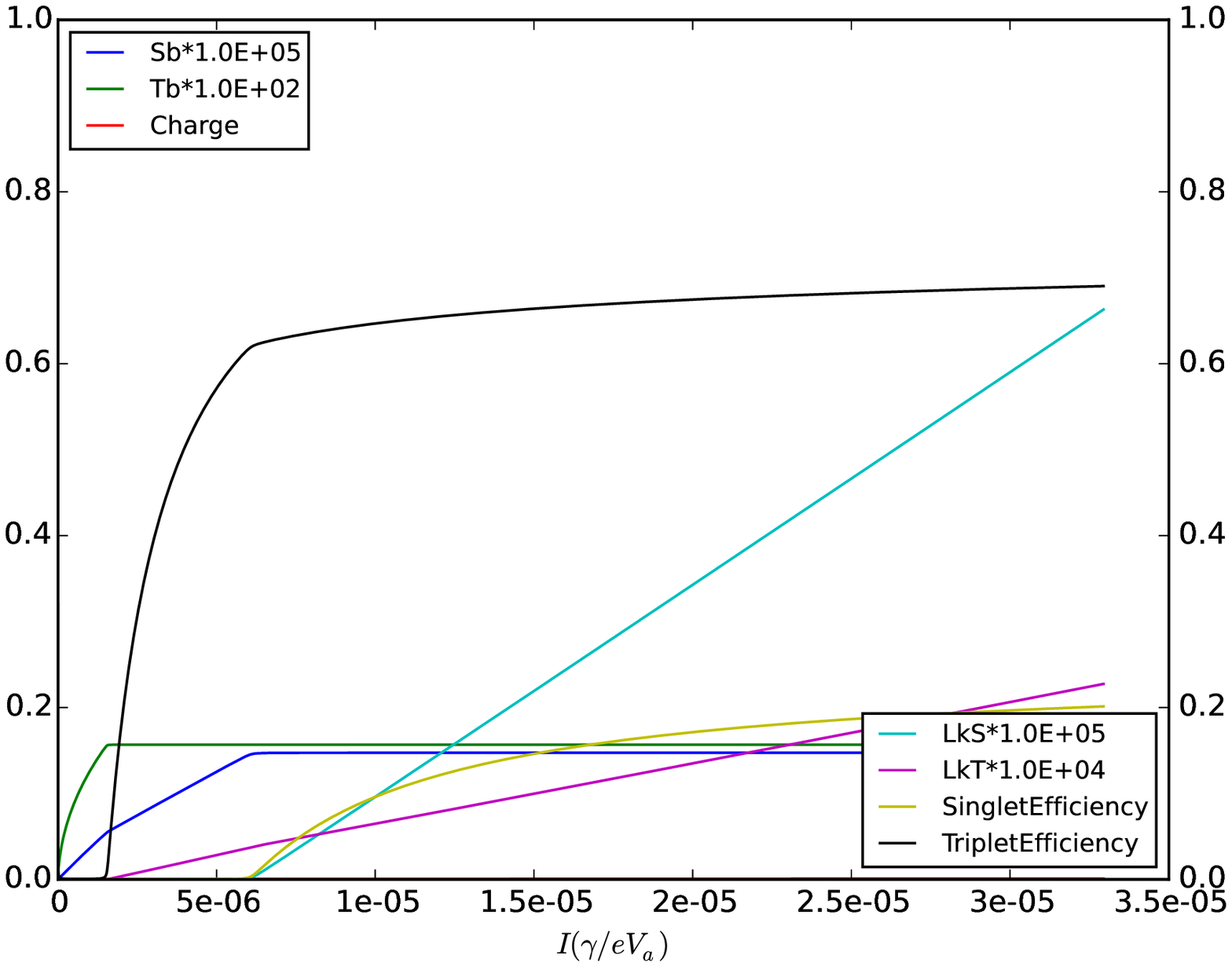}
    \label{fig:alq3qwTfast}
    }
  \caption{\label{fig:alq3device}
    Notional device using rate constants similar to Alq3 in conventional and QW configurations and with different phosphorescent lifetimes.
    }
\end{figure}

In FIG.~\ref{fig:alq3device} we show results using rate constants similar to those reported for Alq3\cite{Shim2006,Brutting2005,Peng2011}.
Where a rate constant for a process has not been found in the literature, a typical value typical is used.\cite{Gartner2007}
This is a challenging emitter because its intersystem crossing rate is $\sim 20$ times faster than its fluorescence ($17 ns$),
which contributes to low yield from the singlet and increases its onset current, and because its slow phosphorescence ($15 ms$) attenuates output from the triplet and increases its excited state population for stimulated emission.
In FIG.~\ref{fig:alq3lit} results are shown with the device configured with a non-resonant cavity.  Efficiency in the singlet is order of 1\%.
In FIG.~\ref{fig:alq3qwS} the configuration is altered to make the cavity resonant with the singlet and the emitter is located at the quarter wave position. The singlet efficiency in the resonant QW device reaches 30\%.
In FIG.~\ref{fig:alq3qwT} the device is resonant with the triplet at 700 nm and the phosphorescent lifetime is 15 ms.
The device does not reach stimulated emission in the triplet at any current because $\Tb_{SE} \sim 1.5$.
Choosing an emitter with a shorter lifetime decreases the population at the transition linearly ($\NSE \propto \tsp$) and for losses faster than the emission rate ($\KTT \NSE >> 1/\tsp$) the onset current decreases as $\tspT^2$.
The transition is accessible for lifetimes shorter than $1 ms$.
In FIG.~\ref{fig:alq3qwTfast} the phosphorescent lifetime is reduced to $15 \mu s$ and the total efficiency in both the triplet and singlet together is approximately 90\%. The onset current in the singlet is increased slightly because it looses the benefit of singlet production from triplet annihilation when the triplet transitions to stimulated emission.

\section{Summary and Conclusions}
In this work we have studied OLEDS in the limiting behaviors of spontaneous emission and stimulated emission, and in the transition to stimulated emission in a device where optical interference suppresses spontaneous emission into the vertical mode, which we call a quarter wave (QW) OLED.
Requirements for realizing a QW device include
(a) a back mirror with reflectivity close to 1,
(b) a thin emission region located at the quarter wave point and
(c) $\NSE << 1$, where $\NSE = (g \tcav)^{-1}$.
Low reflectivity at the exit may be of advantage because $\chi$ is small over a larger region around the quarter-wave point.
Example devices have used phosphorescent emitters and exit reflectivity order of 3\%.
The QW configuration produces a sharp transition to stimulated emission at a minimum current density.
When losses at the transition are not large compared to the spontaneous emission rate, the onset current is determined by optical parameters.
This is in contrast to architectures where emission into the cavity is enhanced and stimulated emission is approached asymptotically with no clear threshold.
In our analysis of electrical properties we find a previously observed behavior that may be a signature of the transition to stimulated emission at low current density ($\sim 1 mA/mm^3$).
In studying loss mechanisms in OLEDS we have proposed a charge bleaching mechanism that may explain  output roll-off as distinct from efficiency roll-off, both of which are known in OLEDS.
Apart from fast ISC, we have not found any loss mechanisms that prevent the transition to stimulated emission in typical materials in the QW architecture.
We plan further experimental work and we are extending the model to treat multi-component emitters.
We are optimistic that the QW architecture will provide a practical route to useful high efficiency devices.

\end{document}

%% file: FIG.FastISC.tex
  \subfigure[\ Spontaneous emission in singlet and triplet]{
    \includegraphics[scale=0.4]{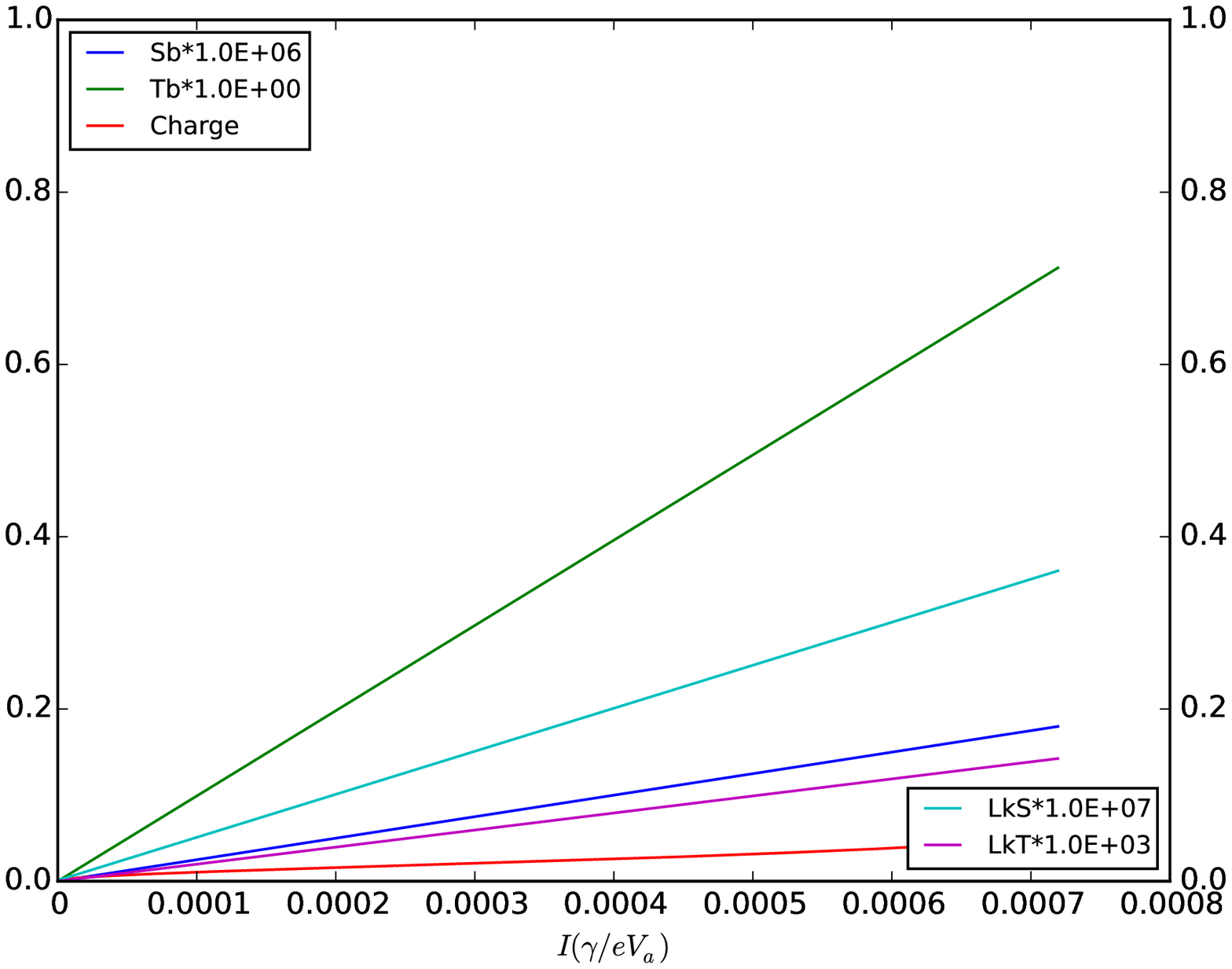}
    \label{fig:FIG.FastISCSpont}
  }
  \subfigure[\ Stimulated emission in singlet]{
    \includegraphics[scale=0.4]{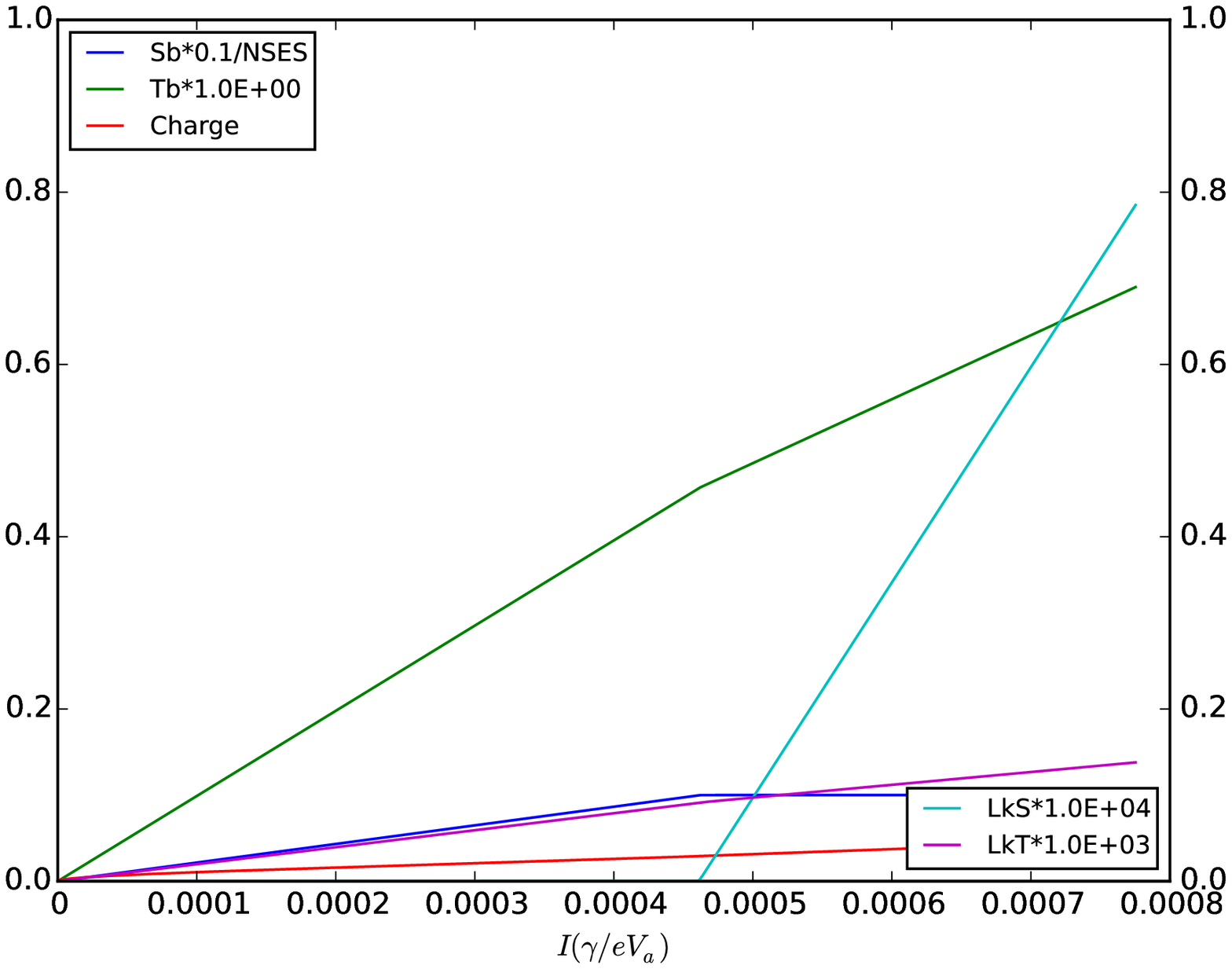}
    \label{fig:FIG.FastISCSinglet}
  } \\[5ex]
  \subfigure[\ Stimulated emission in triplet]{
    \includegraphics[scale=0.4]{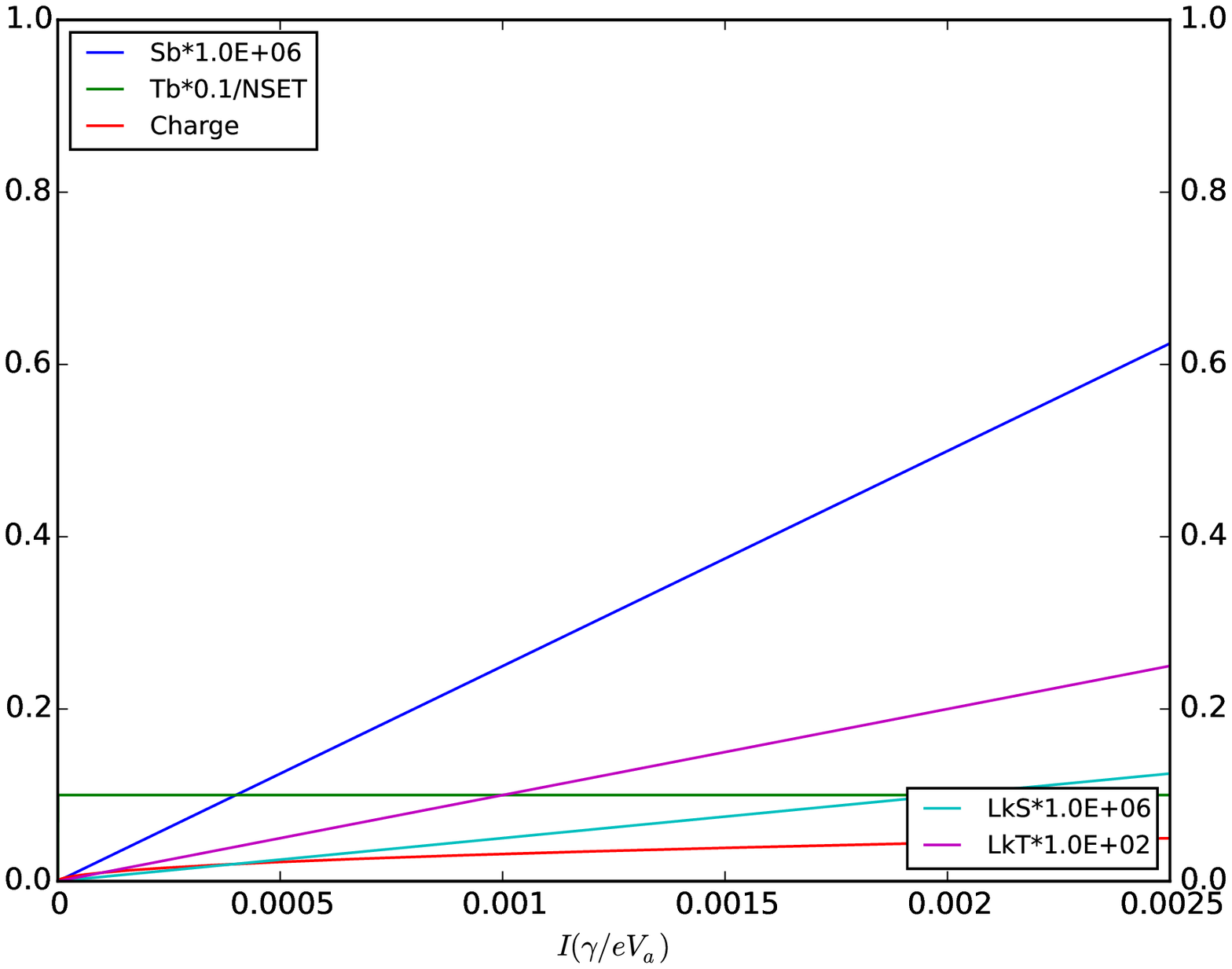}
    \label{fig:FIG.FastISCTriplet}
  }
  \subtable[b][\ Summary] {
   {\vbox to 1.8in{\hbox to 2.85in{\hspace{2mm}
    \begin{tabular}{lcc}
       & \hspace{20pt} $\ \Lk[S] / I$\hspace{10pt} & $ \Lk[T] / I$ \\
       \hline
      Both Spont.    & 5e-05 & 0.2 \\
      Singlet Stim.  & 0.1 & 0.18 \\
      Triplet Stim.  & 5e-05 & 1 \\
    \end{tabular}
    }
    \vfill
    } }
   }
  \def\pars{Kisc=1000.0 Krisc=0.0}

%% file: FIG.HighGap.tex
  \subfigure[\ Spontaneous emission in singlet and triplet]{
    \includegraphics[scale=0.4]{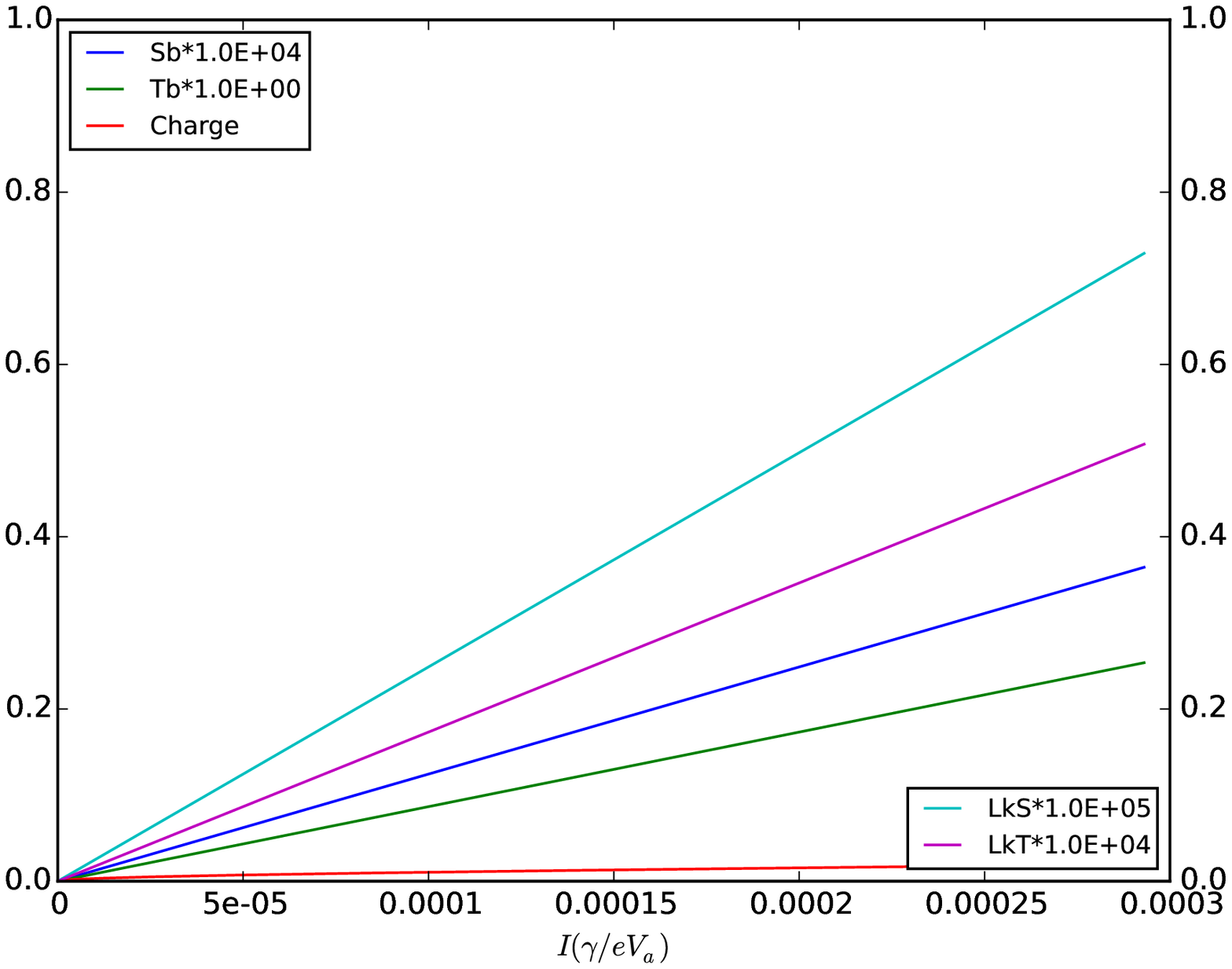}
    \label{fig:FIG.HighGapSpont}
  }
  \subfigure[\ Stimulated emission in singlet]{
    \includegraphics[scale=0.4]{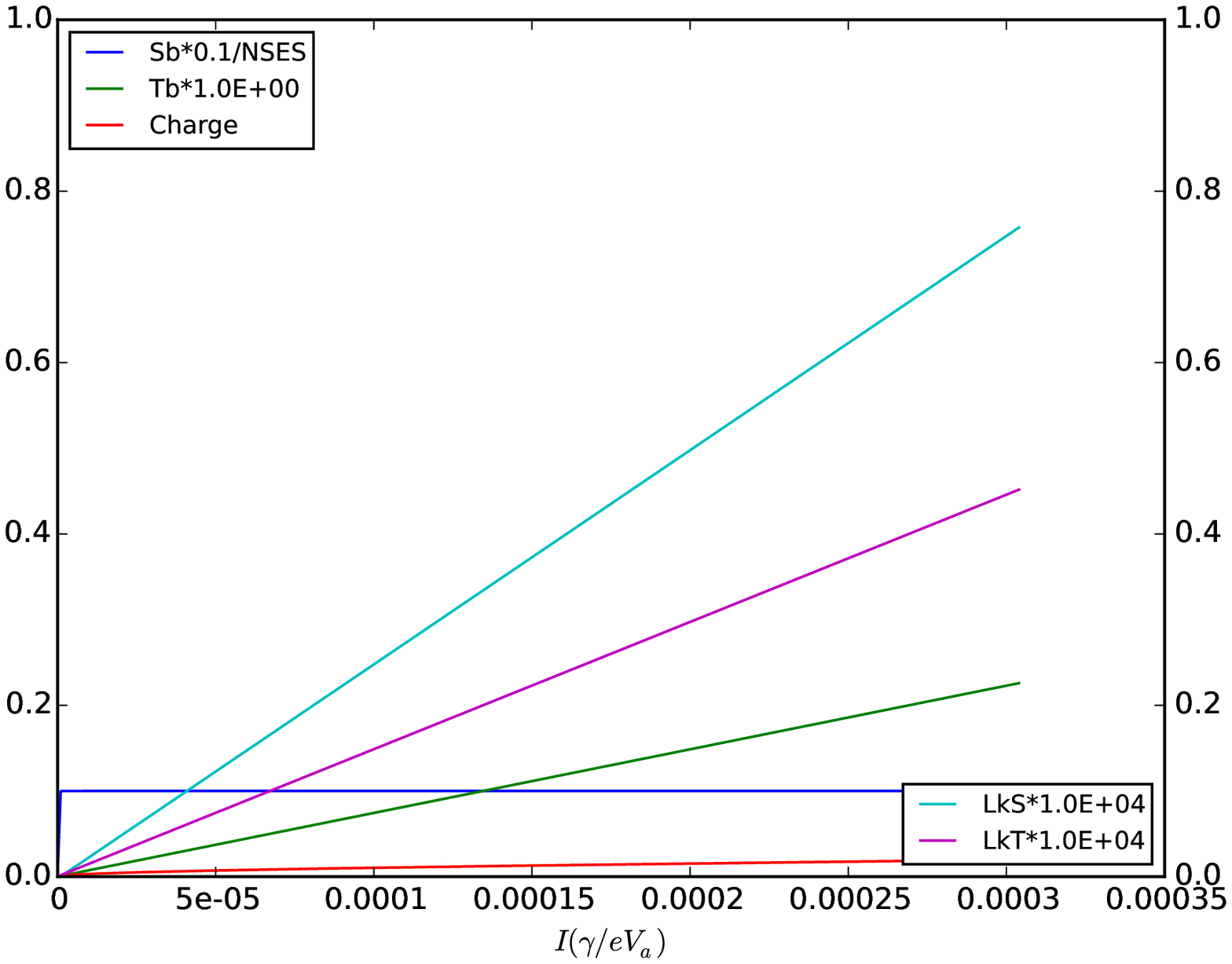}
    \label{fig:FIG.HighGapSinglet}
  } \\[5ex]
  \subfigure[\ Stimulated emission in triplet]{
    \includegraphics[scale=0.4]{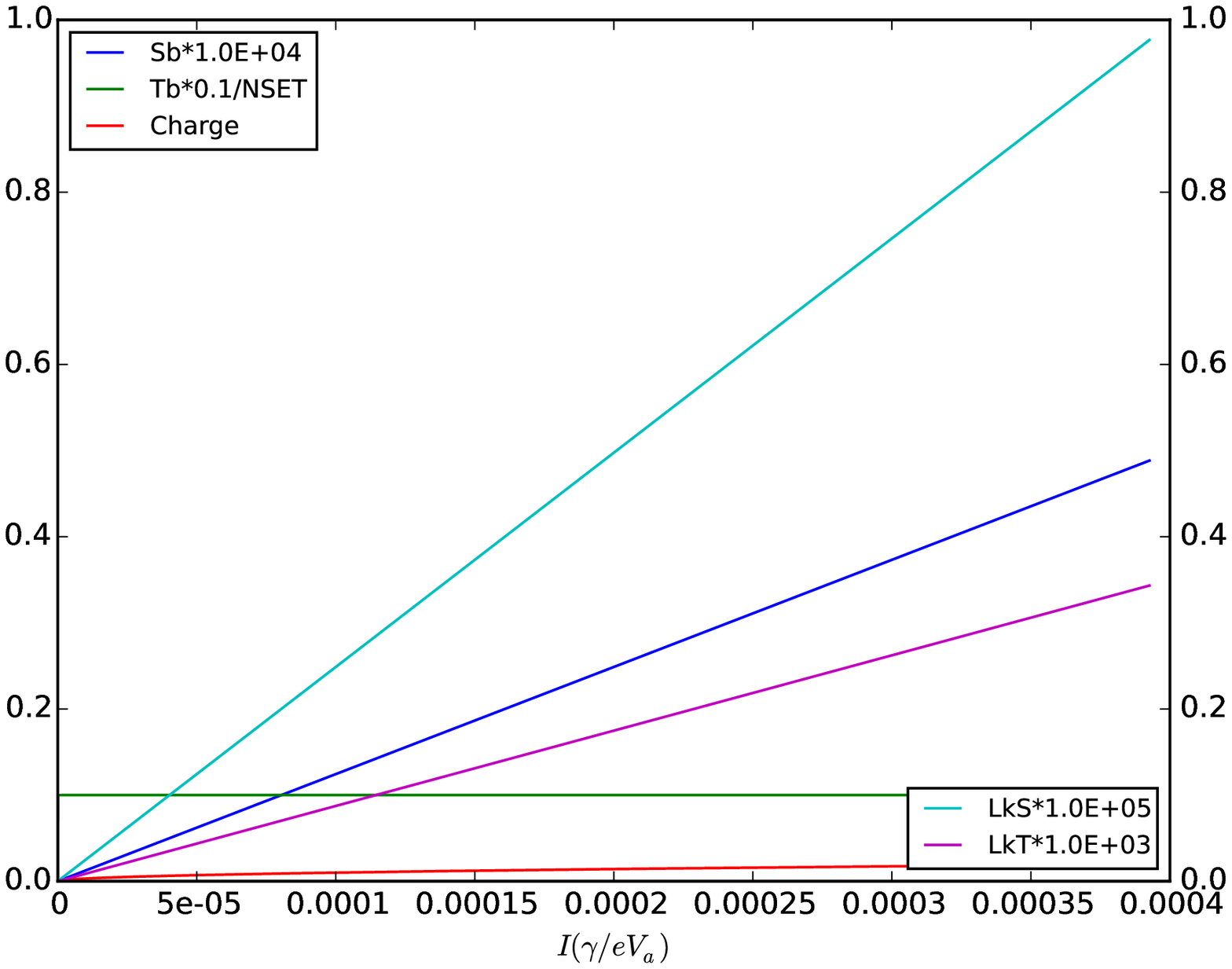}
    \label{fig:FIG.HighGapTriplet}
  }
  \subtable[b][\ Summary] {
   {\vbox to 1.8in{\hbox to 2.85in{\hspace{2mm}
    \begin{tabular}{lcc}
       & \hspace{20pt} $\ \Lk[S] / I$\hspace{10pt} & $ \Lk[T] / I$ \\
       \hline
      Both Spont.    & 0.025 & 0.17 \\
      Singlet Stim.  & 0.25 & 0.15 \\
      Triplet Stim.  & 0.025 & 0.87 \\
    \end{tabular}
    }
    \vfill
    } }
   }
  \def\pars{Kisc=1.0 Krisc=7.14e-12}

%% file: FIG.LowGap.tex
  \subfigure[\ Spontaneous emission in singlet and triplet]{
    \includegraphics[scale=0.4]{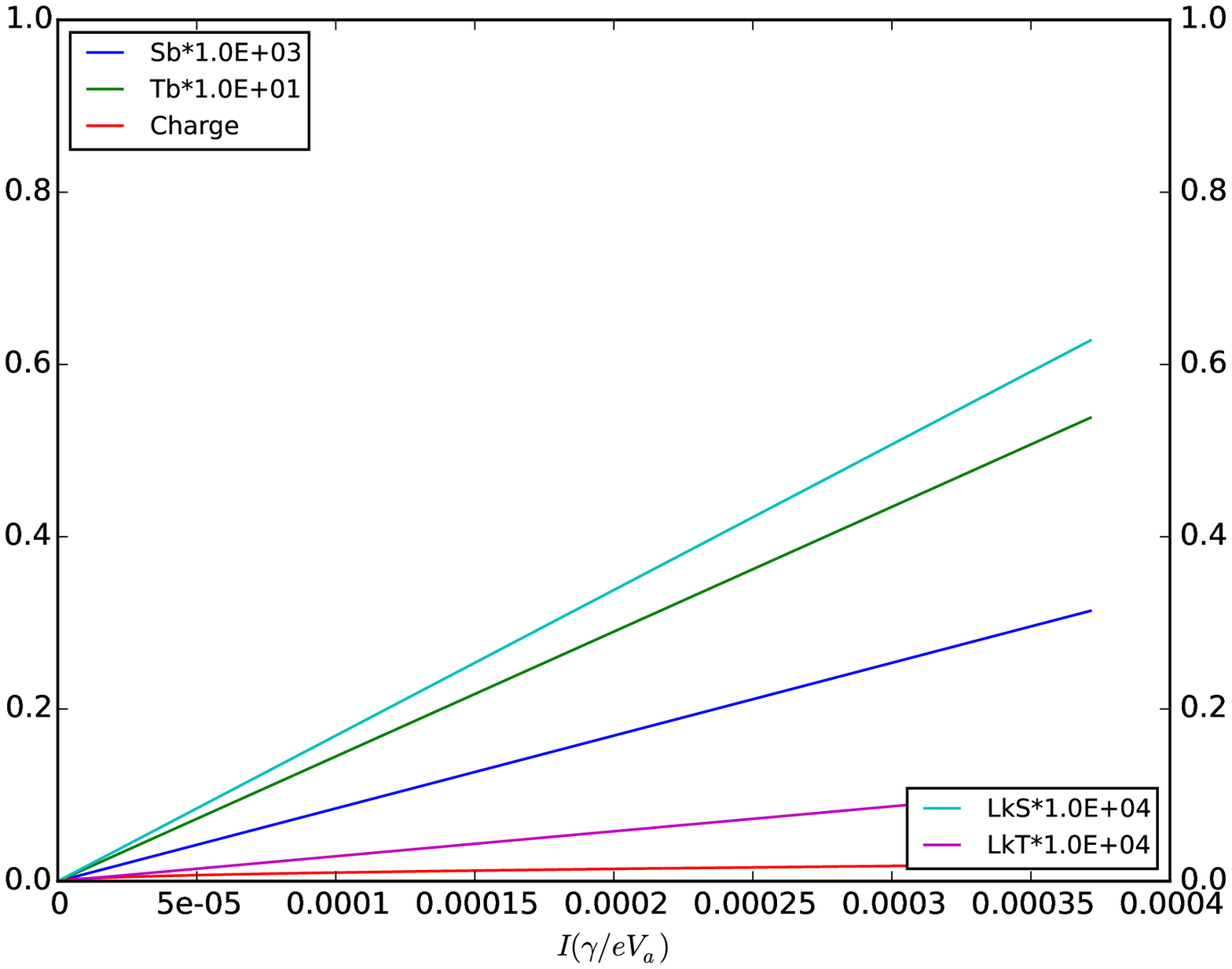}
    \label{fig:FIG.LowGapSpont}
  }
  \subfigure[\ Stimulated emission in singlet]{
    \includegraphics[scale=0.4]{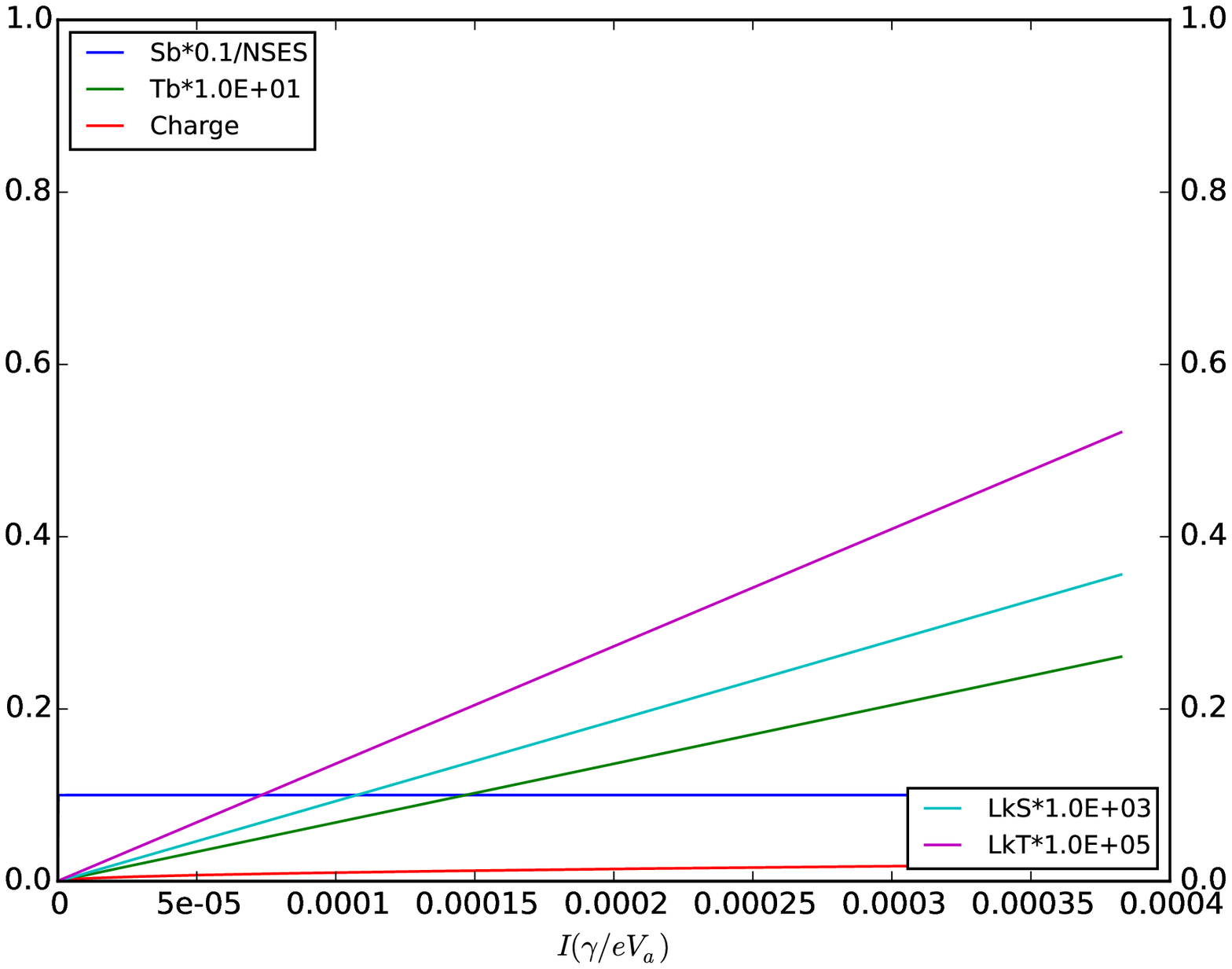}
    \label{fig:FIG.LowGapSinglet}
  } \\[5ex]
  \subfigure[\ Stimulated emission in triplet]{
    \includegraphics[scale=0.4]{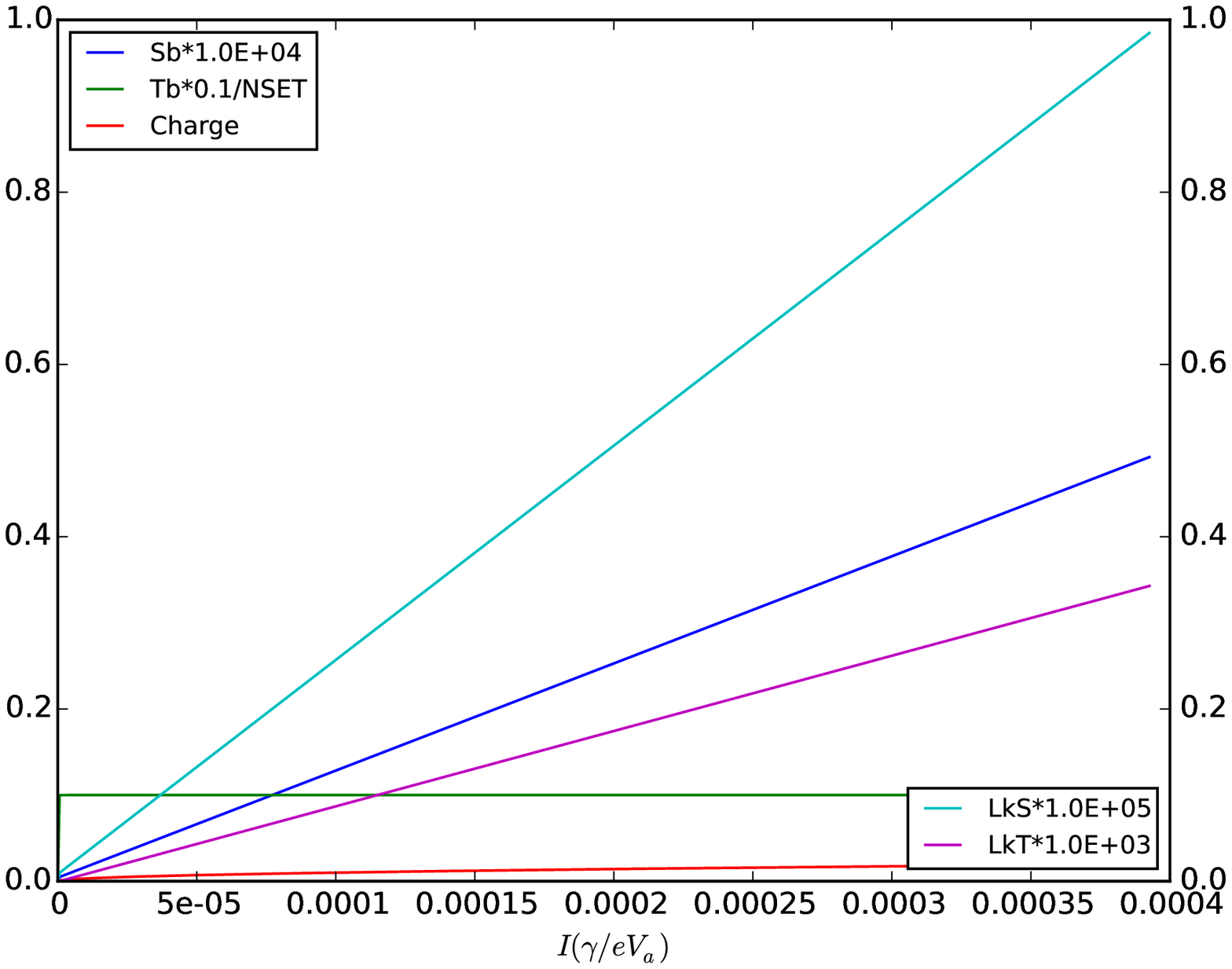}
    \label{fig:FIG.LowGapTriplet}
  }
  \subtable[b][\ Summary] {
   {\vbox to 1.8in{\hbox to 2.85in{\hspace{2mm}
    \begin{tabular}{lcc}
       & \hspace{20pt} $\ \Lk[S] / I$\hspace{10pt} & $ \Lk[T] / I$ \\
       \hline
      Both Spont.    & 0.17 & 0.029 \\
      Singlet Stim.  & 0.93 & 0.014 \\
      Triplet Stim.  & 0.025 & 0.87 \\
    \end{tabular}
    }
    \vfill
    } }
   }
  \def\pars{Kisc=1.0 Krisc=0.01}

%% file: FIG.KTT.FastISC.tex
  \subfigure[\ Spontaneous emission in singlet and triplet]{
    \includegraphics[scale=0.4]{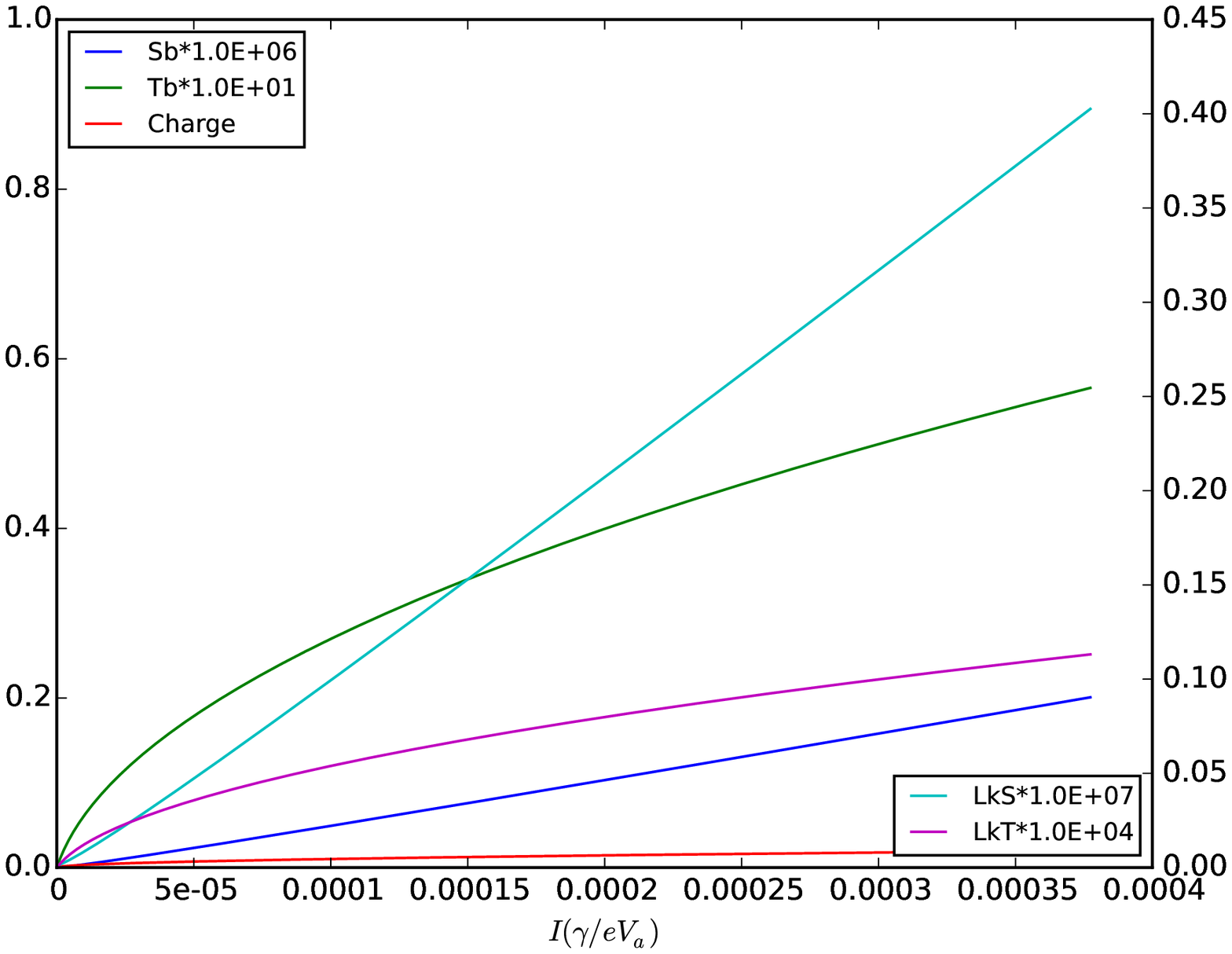}
    \label{fig:FIG.KTT.FastISCSpont}
  }
  \subfigure[\ Stimulated emission in singlet]{
    \includegraphics[scale=0.4]{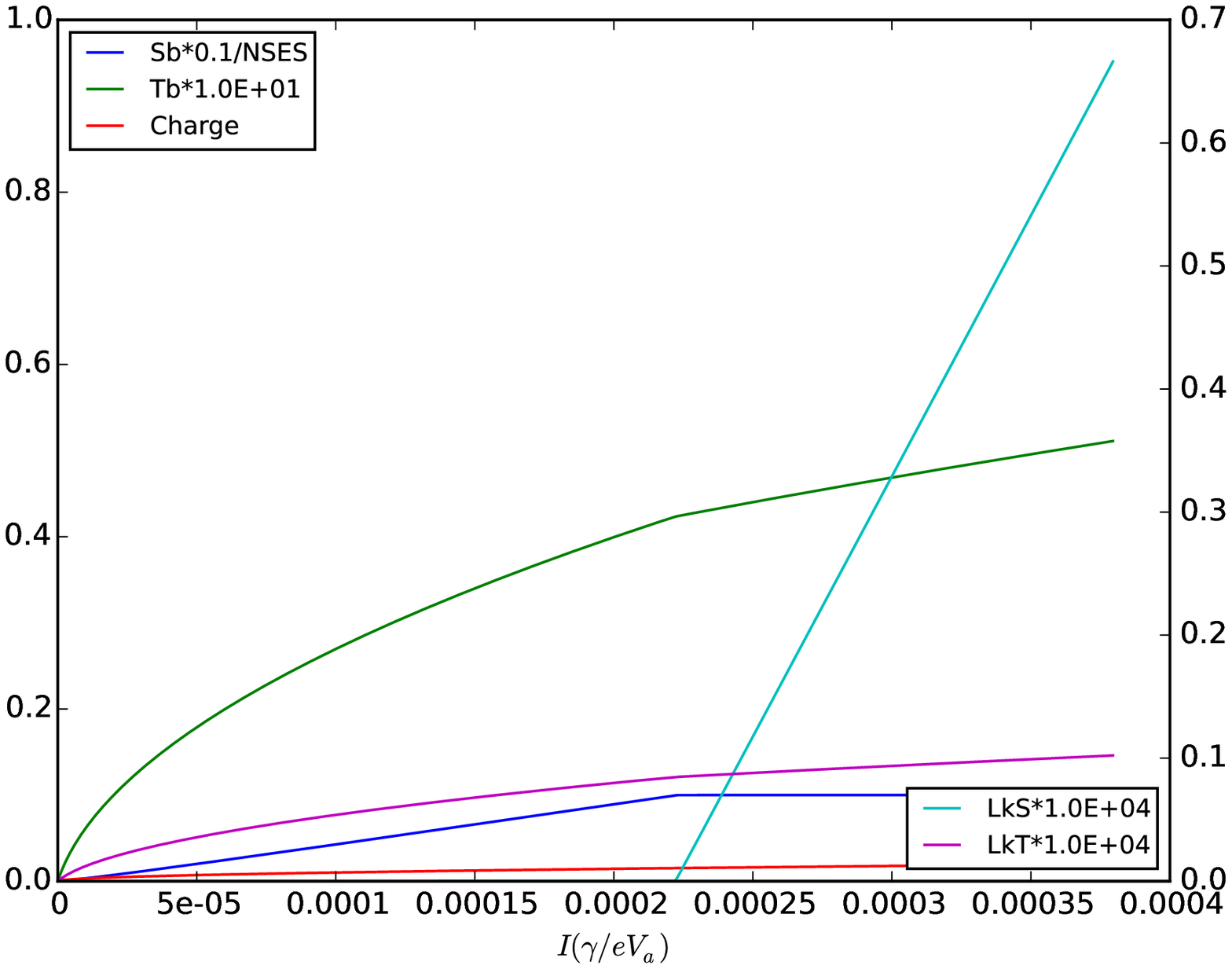}
    \label{fig:FIG.KTT.FastISCSinglet}
  } \\[5ex]
  \subfigure[\ Stimulated emission in triplet]{
    \includegraphics[scale=0.4]{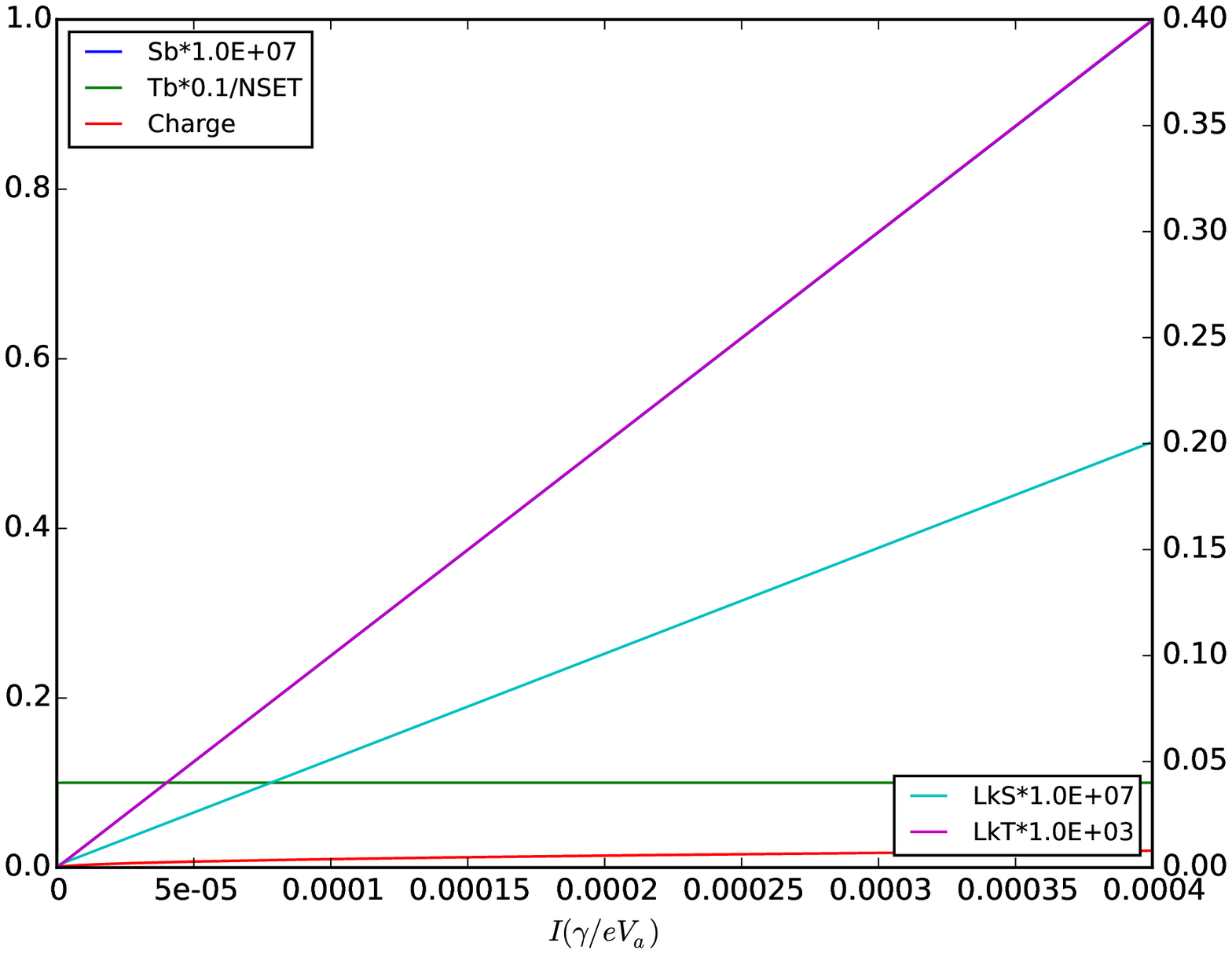}
    \label{fig:FIG.KTT.FastISCTriplet}
  }
  \subtable[b][\ Summary] {
   {\vbox to 1.8in{\hbox to 2.85in{\hspace{2mm}
    \begin{tabular}{lcc}
       & \hspace{20pt} $\ \Lk[S] / I$\hspace{10pt} & $ \Lk[T] / I$ \\
       \hline
      Both Spont.    & 0.00011 & 0.03 \\
      Singlet Stim.  & 0.18 & 0.027 \\
      Triplet Stim.  & 5e-05 & 1 \\
    \end{tabular}
    }
    \vfill
    } }
   }
  \def\pars{KTT=0.1 Kisc=1000.0 Krisc=0.0}

%% file: FIG.KTT.HighGap.tex
  \subfigure[\ Spontaneous emission in singlet and triplet]{
    \includegraphics[scale=0.4]{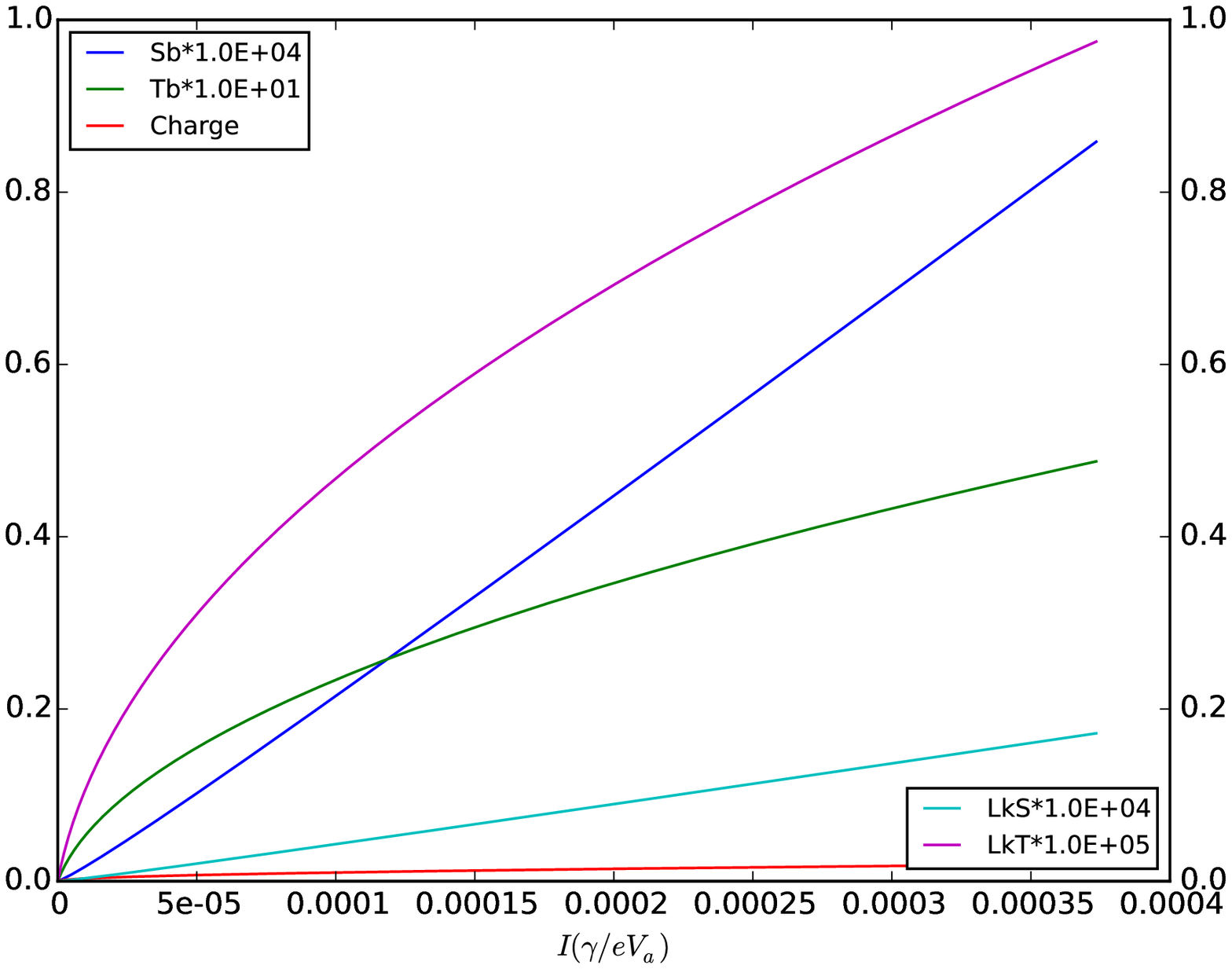}
    \label{fig:FIG.KTT.HighGapSpont}
  }
  \subfigure[\ Stimulated emission in singlet]{
    \includegraphics[scale=0.4]{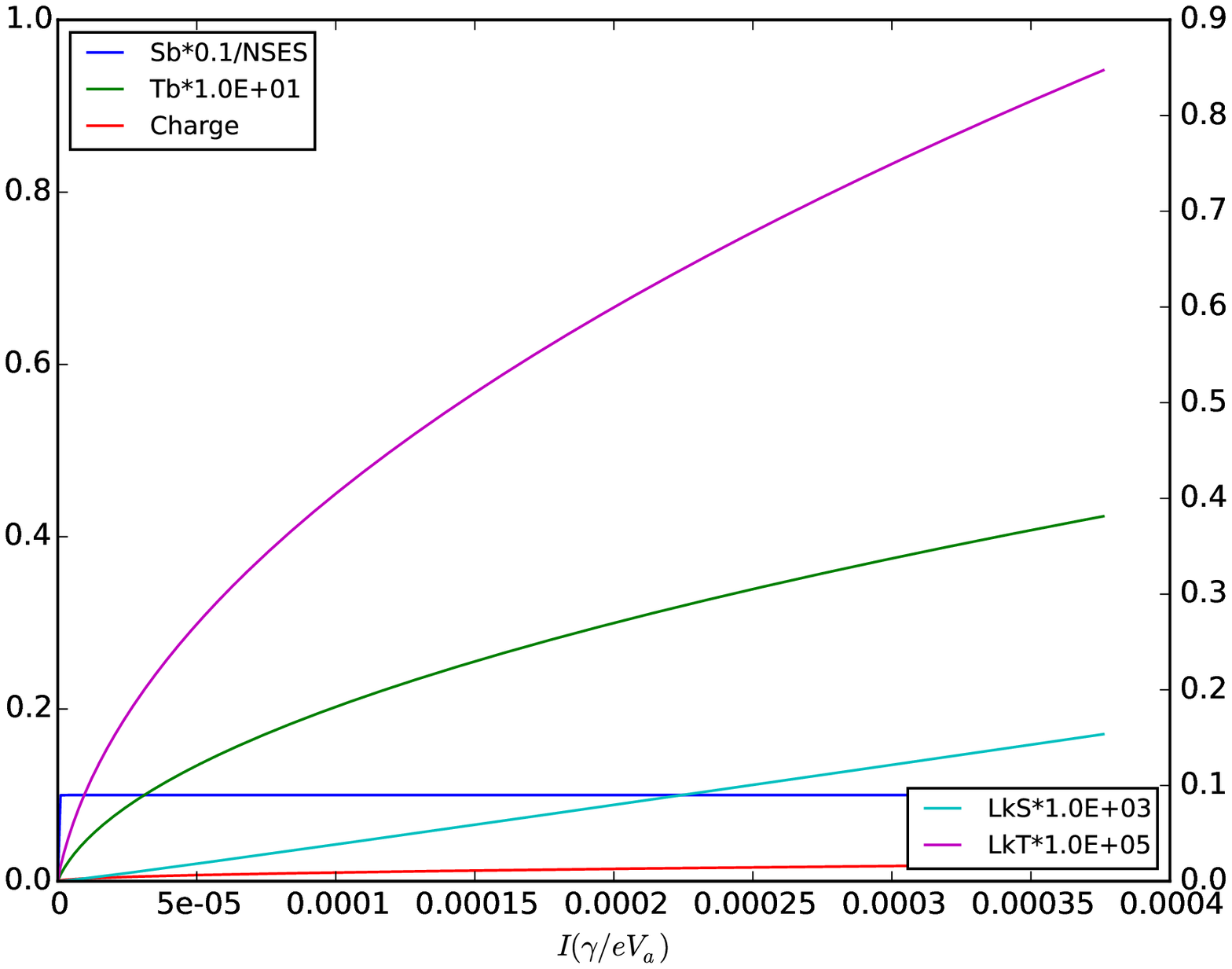}
    \label{fig:FIG.KTT.HighGapSinglet}
  } \\[5ex]
  \subfigure[\ Stimulated emission in triplet]{
    \includegraphics[scale=0.4]{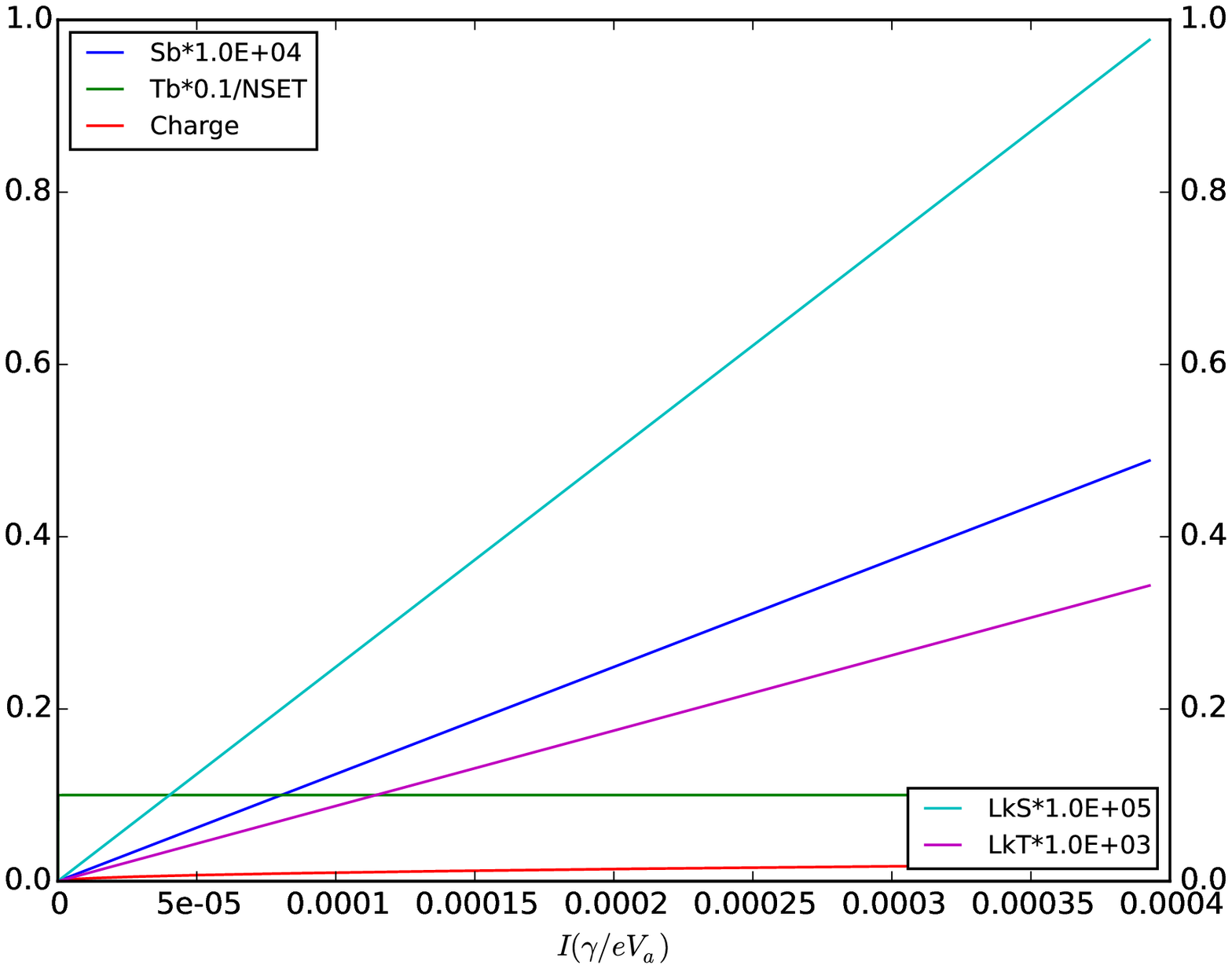}
    \label{fig:FIG.KTT.HighGapTriplet}
  }
  \subtable[b][\ Summary] {
   {\vbox to 1.8in{\hbox to 2.85in{\hspace{2mm}
    \begin{tabular}{lcc}
       & \hspace{20pt} $\ \Lk[S] / I$\hspace{10pt} & $ \Lk[T] / I$ \\
       \hline
      Both Spont.    & 0.046 & 0.026 \\
      Singlet Stim.  & 0.41 & 0.022 \\
      Triplet Stim.  & 0.025 & 0.87 \\
    \end{tabular}
    }
    \vfill
    } }
   }
  \def\pars{KTT=0.1 Kisc=1.0 Krisc=7.14e-12}

%% file: FIG.KTT.LowGap.tex
  \subfigure[\ Spontaneous emission in singlet and triplet]{
    \includegraphics[scale=0.4]{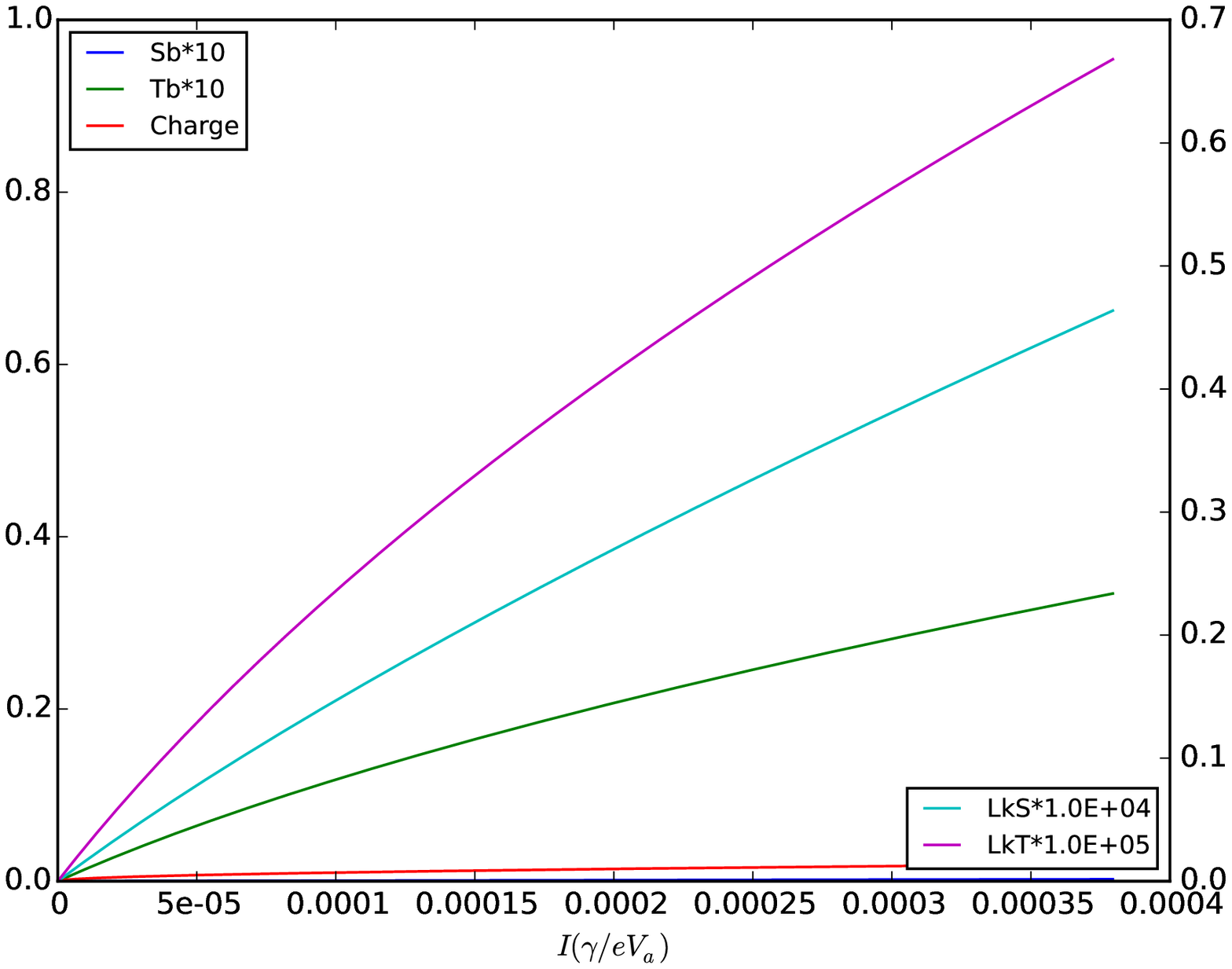}
    \label{fig:FIG.KTT.LowGapSpont}
  }
  \subfigure[\ Stimulated emission in singlet]{
    \includegraphics[scale=0.4]{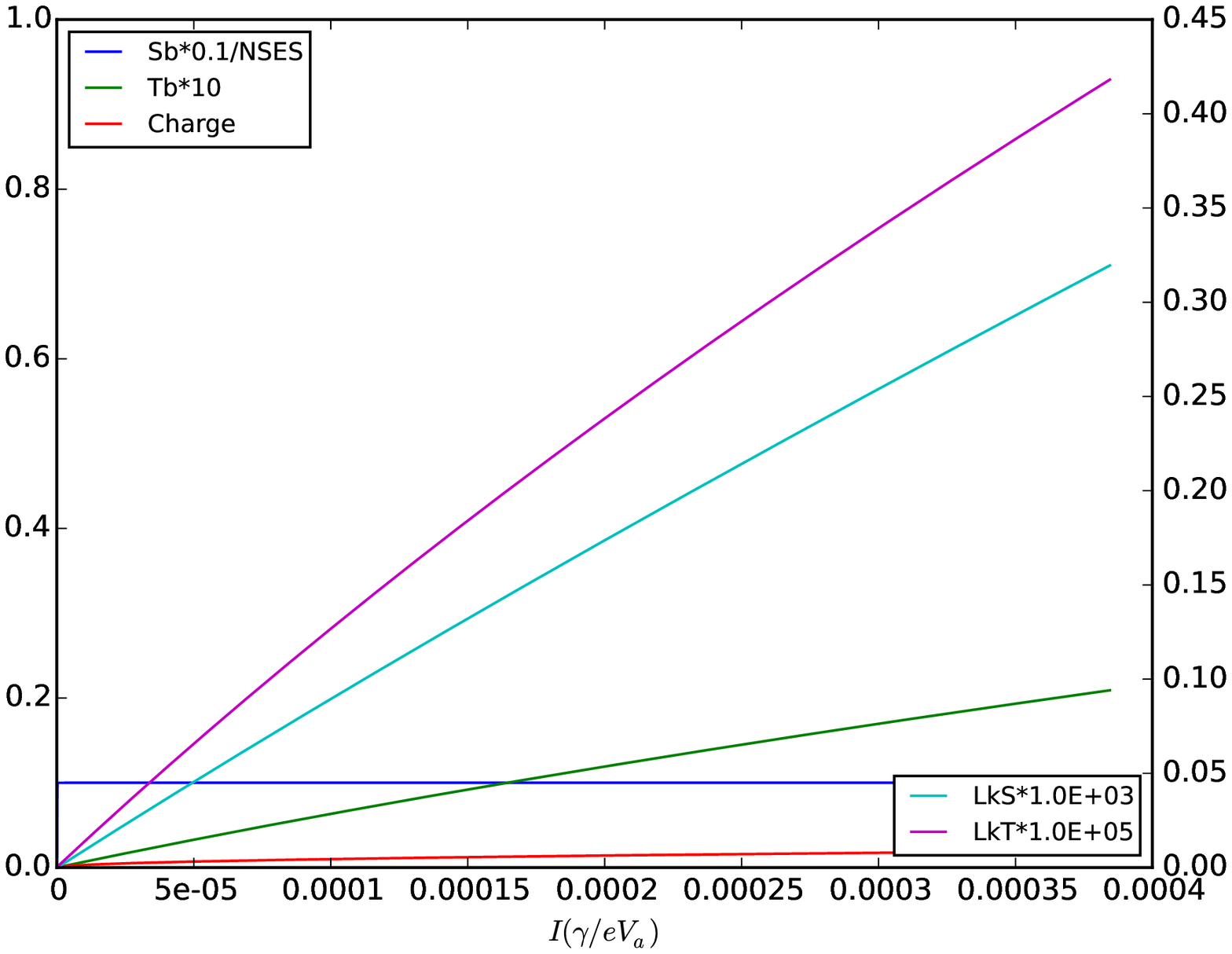}
    \label{fig:FIG.KTT.LowGapSinglet}
  } \\[5ex]
  \subfigure[\ Stimulated emission in triplet]{
    \includegraphics[scale=0.4]{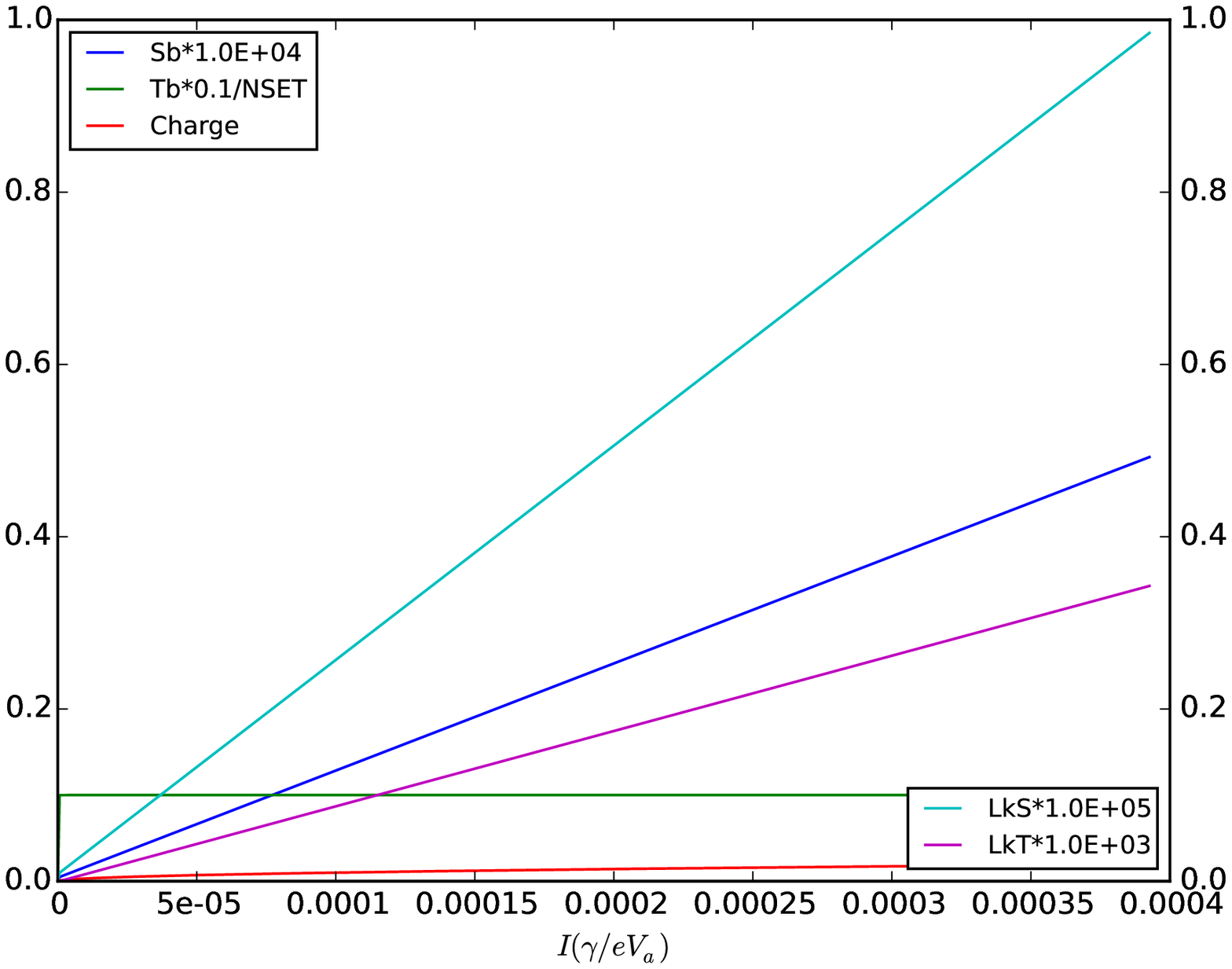}
    \label{fig:FIG.KTT.LowGapTriplet}
  }
  \subtable[b][\ Summary] {
   {\vbox to 1.8in{\hbox to 2.85in{\hspace{2mm}
    \begin{tabular}{lcc}
       & \hspace{20pt} $\ \Lk[S] / I$\hspace{10pt} & $ \Lk[T] / I$ \\
       \hline
      Both Spont.    & 0.12 & 0.018 \\
      Singlet Stim.  & 0.92 & 0.011 \\
      Triplet Stim.  & 0.025 & 0.87 \\
    \end{tabular}
    }
    \vfill
    } }
   }
  \def\pars{KTT=0.1 Kisc=1.0 Krisc=0.01}

%% file: FIG.KTS.FastISC.tex
  \subfigure[\ Spontaneous emission in singlet and triplet]{
    \includegraphics[scale=0.4]{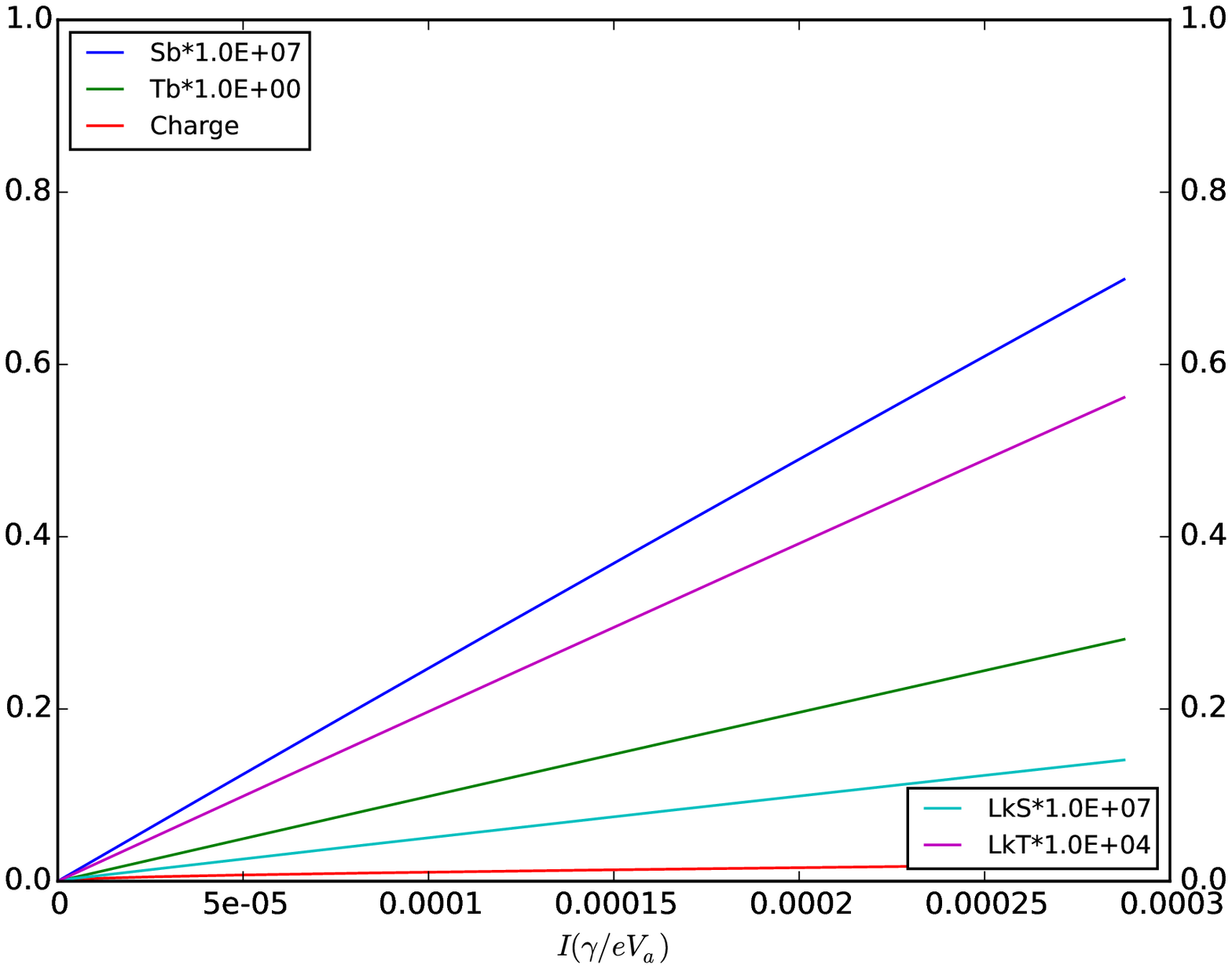}
    \label{fig:FIG.KTS.FastISCSpont}
  }
  \subfigure[\ Stimulated emission in singlet]{
    \includegraphics[scale=0.4]{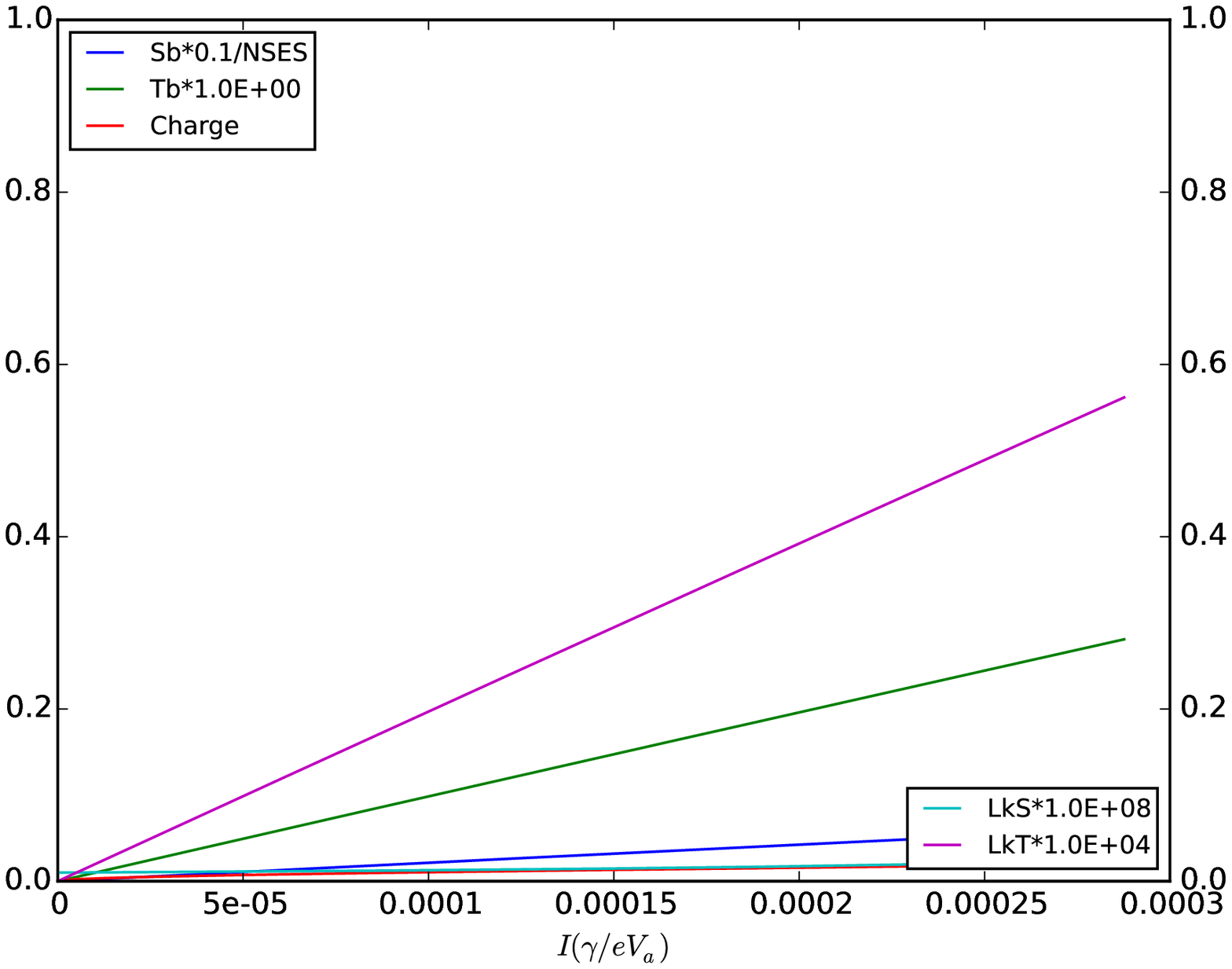}
    \label{fig:FIG.KTS.FastISCSinglet}
  } \\[5ex]
  \subfigure[\ Stimulated emission in triplet]{
    \includegraphics[scale=0.4]{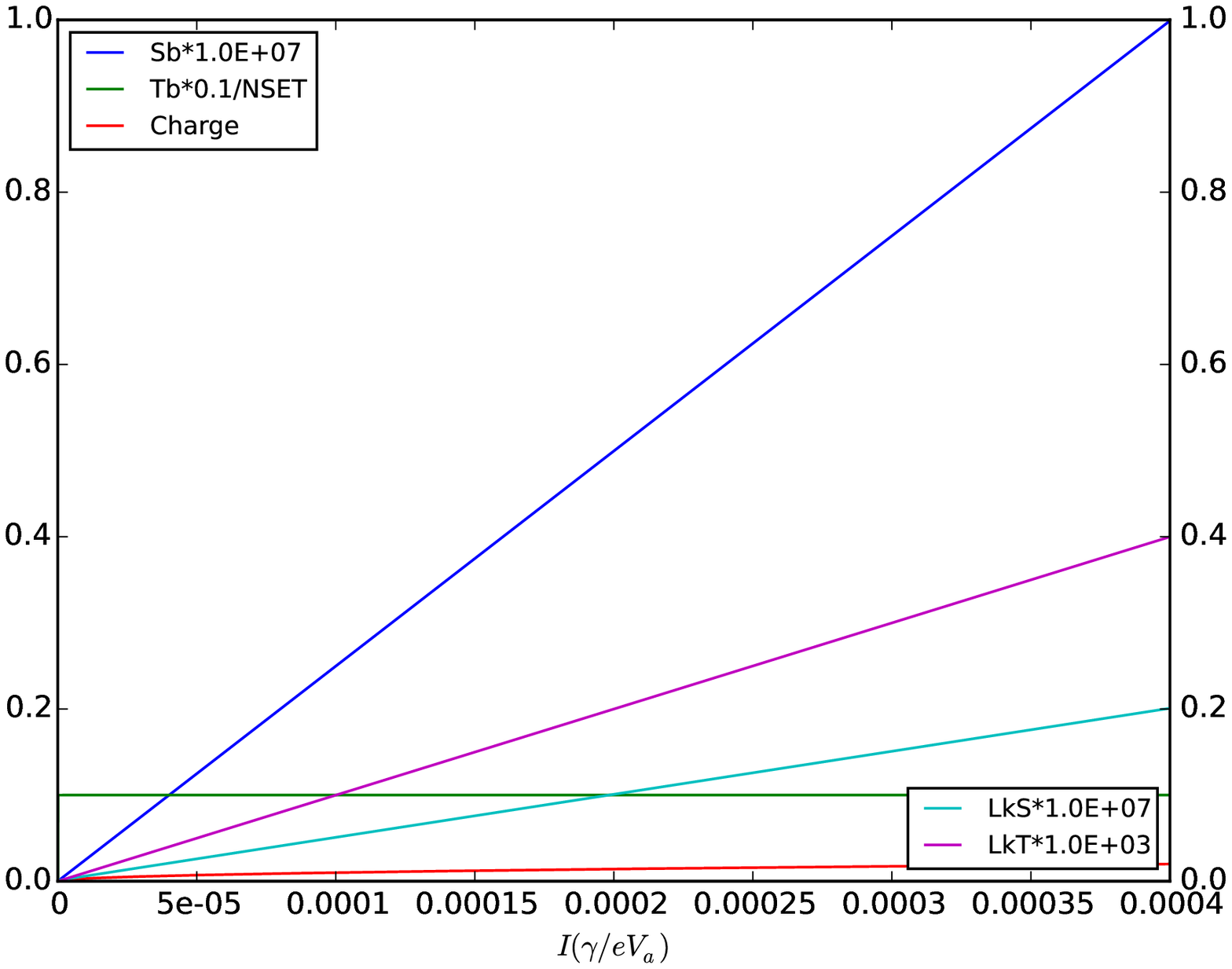}
    \label{fig:FIG.KTS.FastISCTriplet}
  }
  \subtable[b][\ Summary] {
   {\vbox to 1.8in{\hbox to 2.85in{\hspace{2mm}
    \begin{tabular}{lcc}
       & \hspace{20pt} $\ \Lk[S] / I$\hspace{10pt} & $ \Lk[T] / I$ \\
       \hline
      Both Spont.    & 4.9e-05 & 0.2 \\
      Singlet Stim.  & 8.8e-07 & 0.2 \\
      Triplet Stim.  & 5e-05 & 1 \\
    \end{tabular}
    }
    \vfill
    } }
   }
  \def\pars{KTS=100.0 Kisc=1000.0 Krisc=0.0}

%% file: FIG.KTS.HighGap.tex
  \subfigure[\ Spontaneous emission in singlet and triplet]{
    \includegraphics[scale=0.4]{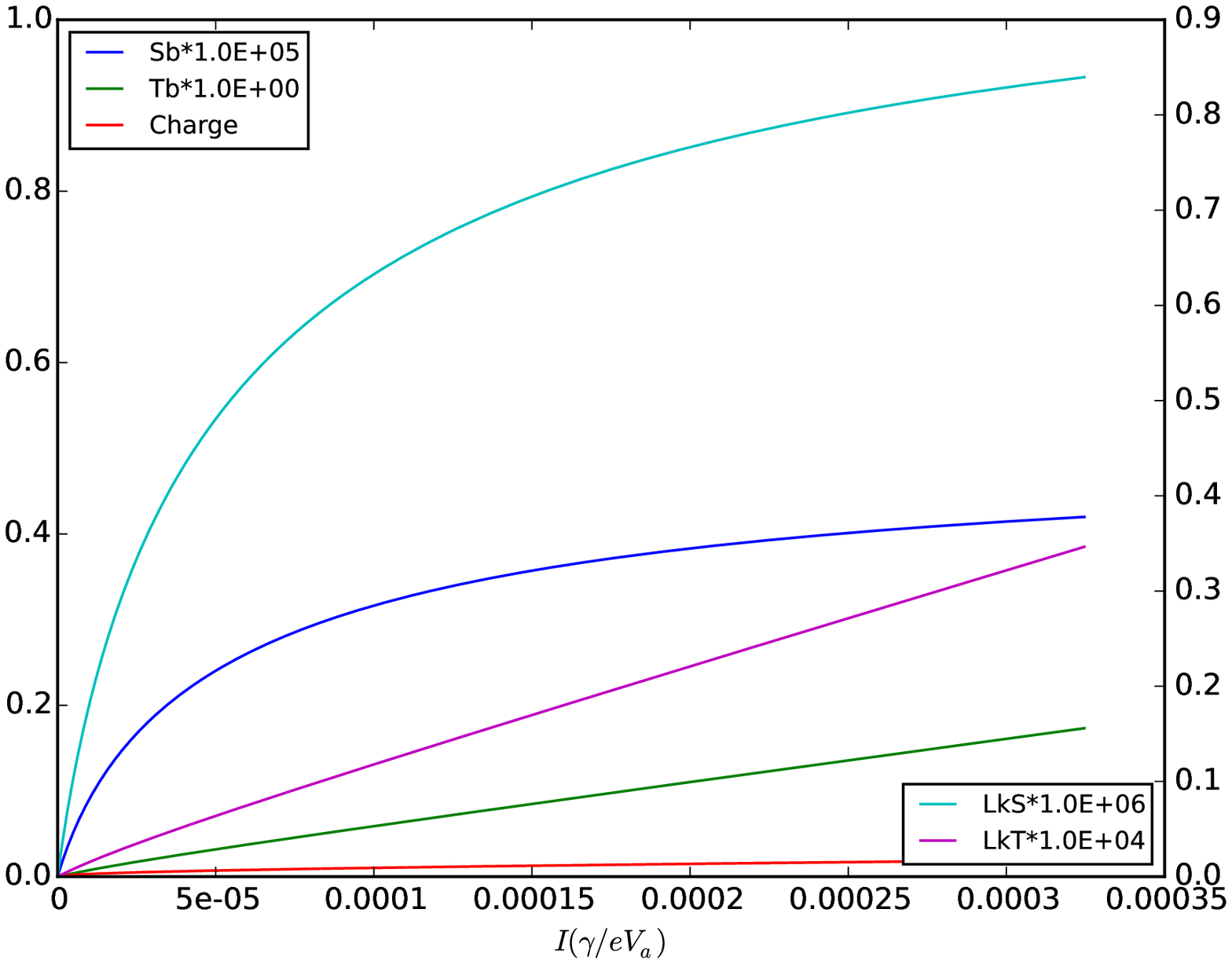}
    \label{fig:FIG.KTS.HighGapSpont}
  }
  \subfigure[\ Stimulated emission in singlet]{
    \includegraphics[scale=0.4]{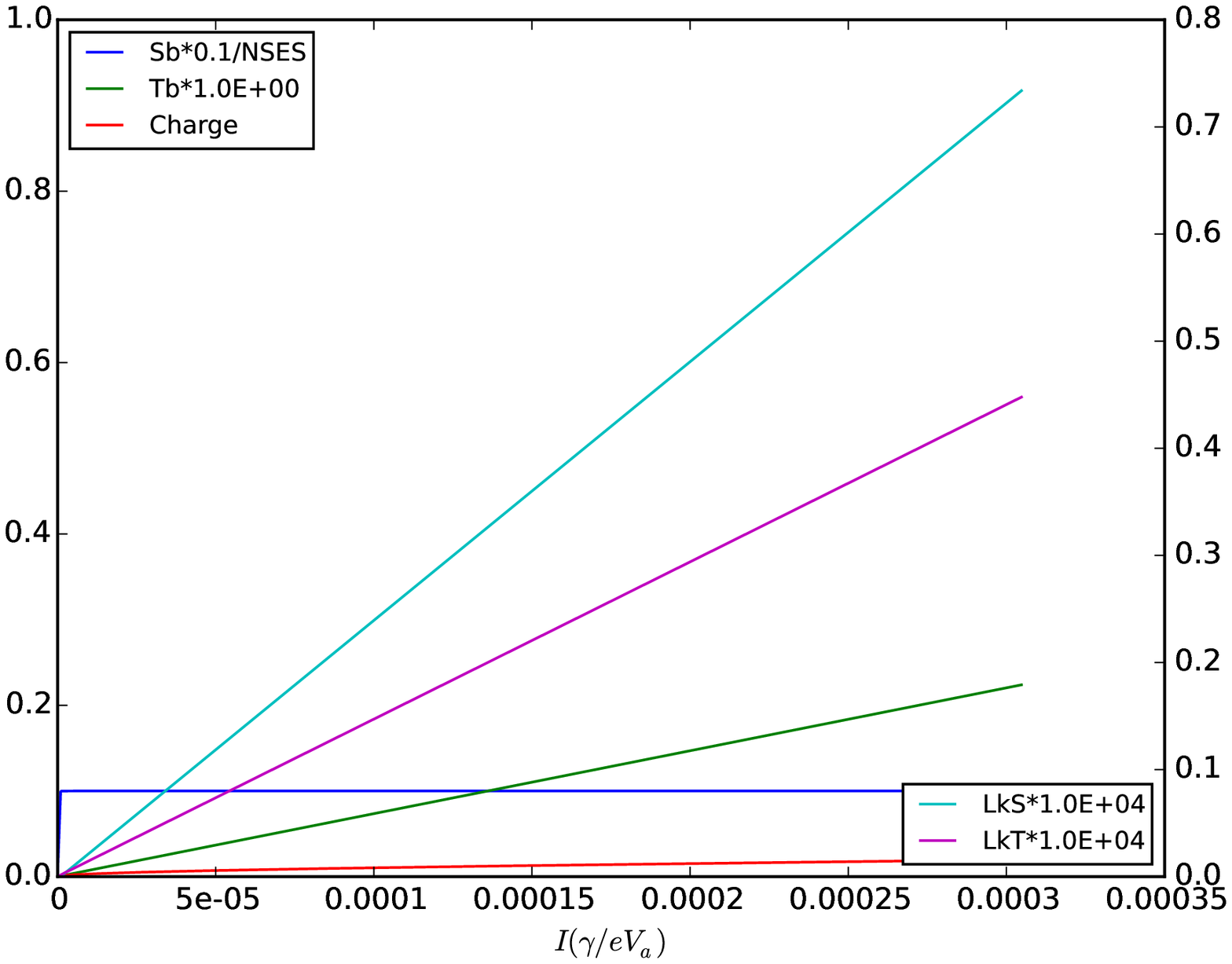}
    \label{fig:FIG.KTS.HighGapSinglet}
  } \\[5ex]
  \subfigure[\ Stimulated emission in triplet]{
    \includegraphics[scale=0.4]{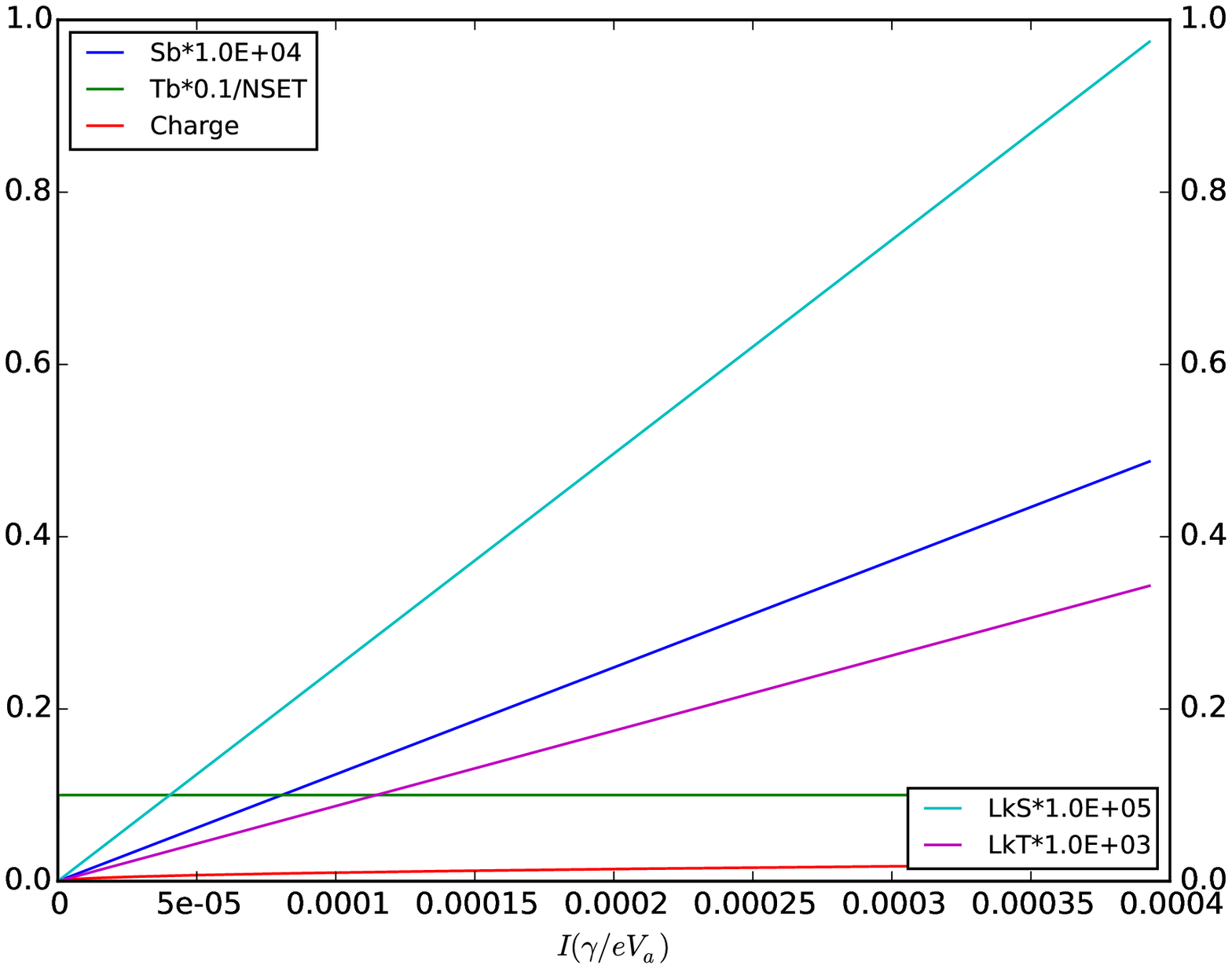}
    \label{fig:FIG.KTS.HighGapTriplet}
  }
  \subtable[b][\ Summary] {
   {\vbox to 1.8in{\hbox to 2.85in{\hspace{2mm}
    \begin{tabular}{lcc}
       & \hspace{20pt} $\ \Lk[S] / I$\hspace{10pt} & $ \Lk[T] / I$ \\
       \hline
      Both Spont.    & 0.0026 & 0.11 \\
      Singlet Stim.  & 0.24 & 0.15 \\
      Triplet Stim.  & 0.025 & 0.87 \\
    \end{tabular}
    }
    \vfill
    } }
   }
  \def\pars{KTS=100.0 Kisc=1.0 Krisc=7.14e-12}

%% file: FIG.KTS.LowGap.tex
  \subfigure[\ Spontaneous emission in singlet and triplet]{
    \includegraphics[scale=0.4]{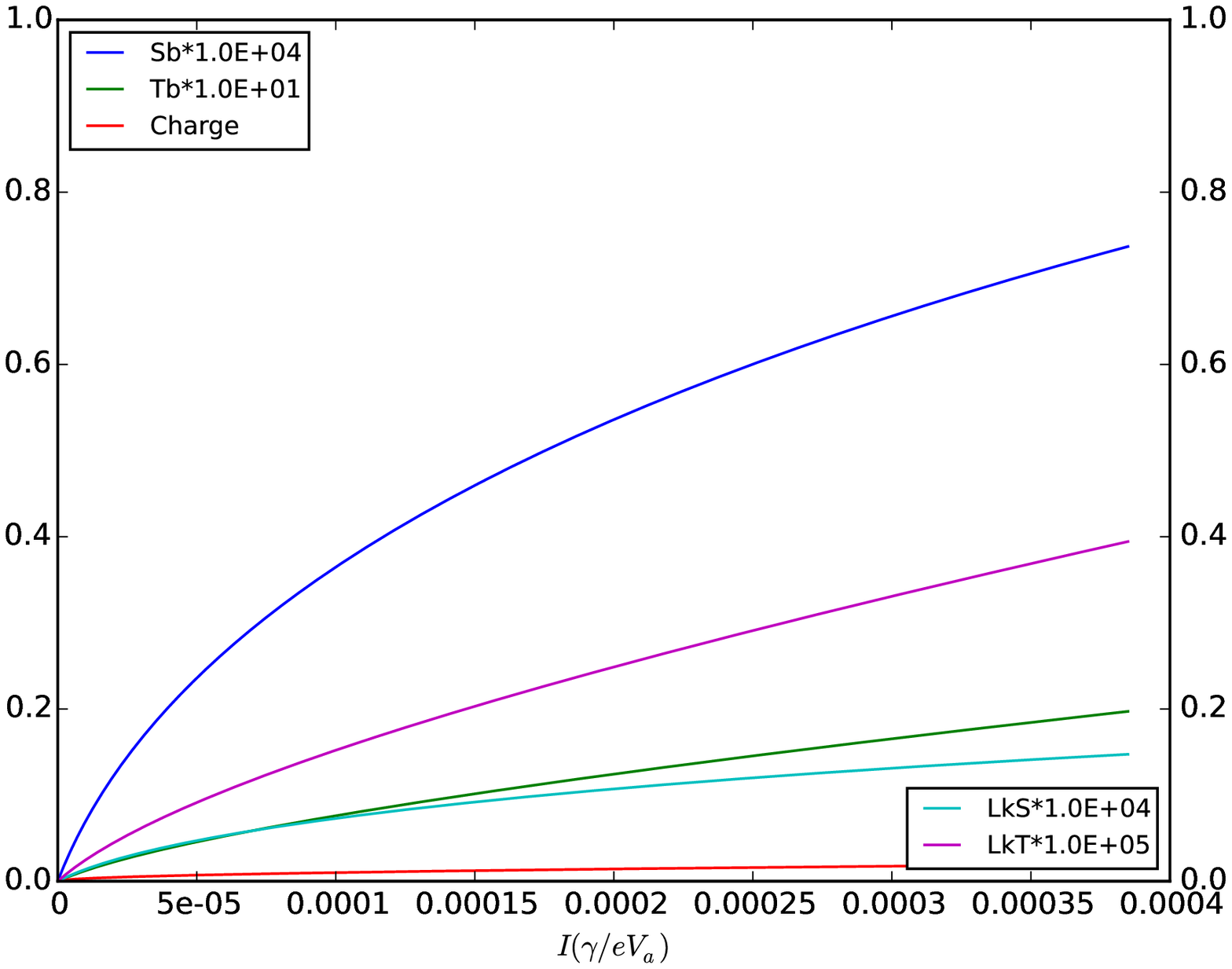}
    \label{fig:FIG.KTS.LowGapSpont}
  }
  \subfigure[\ Stimulated emission in singlet]{
    \includegraphics[scale=0.4]{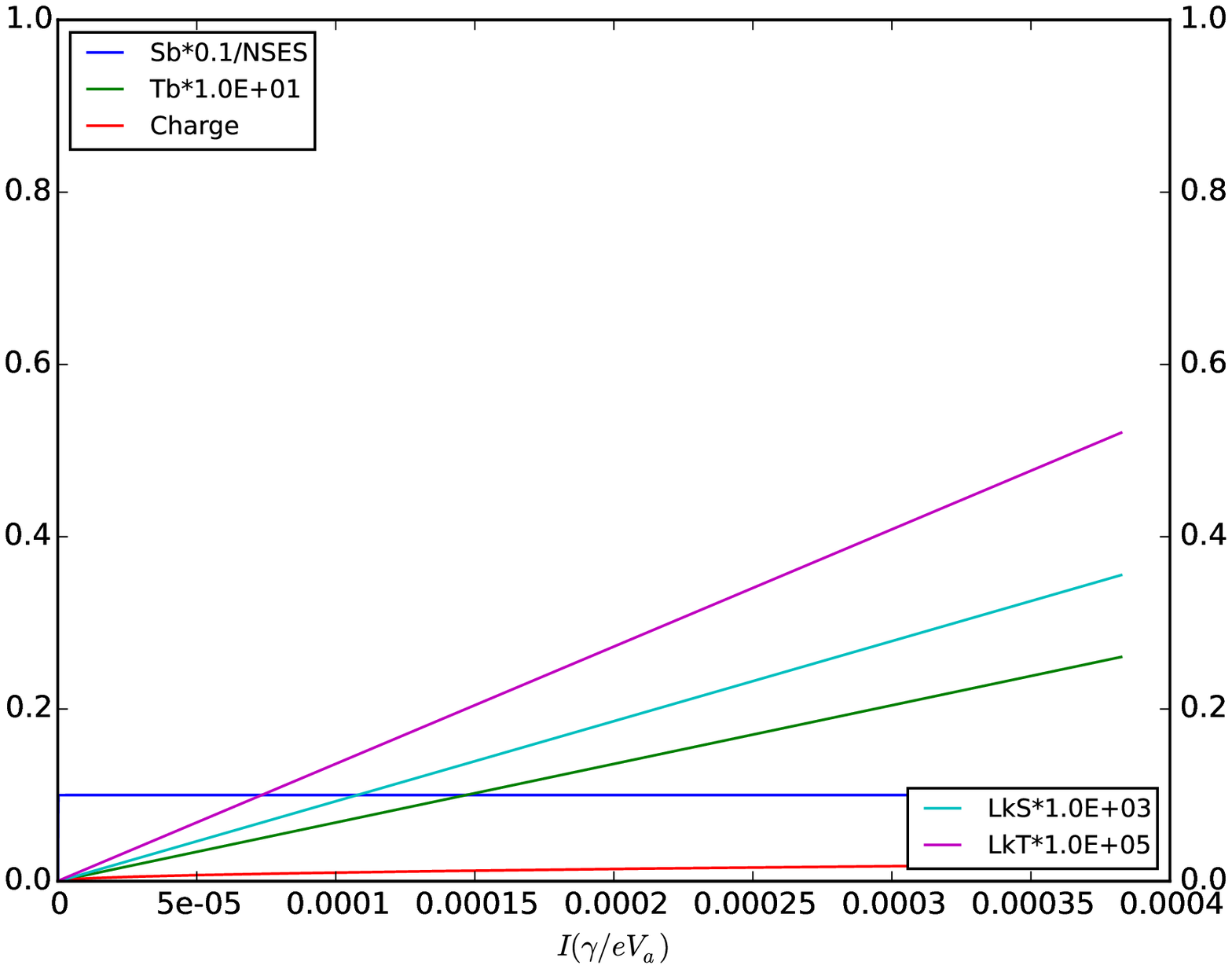}
    \label{fig:FIG.KTS.LowGapSinglet}
  } \\[5ex]
  \subfigure[\ Stimulated emission in triplet]{
    \includegraphics[scale=0.4]{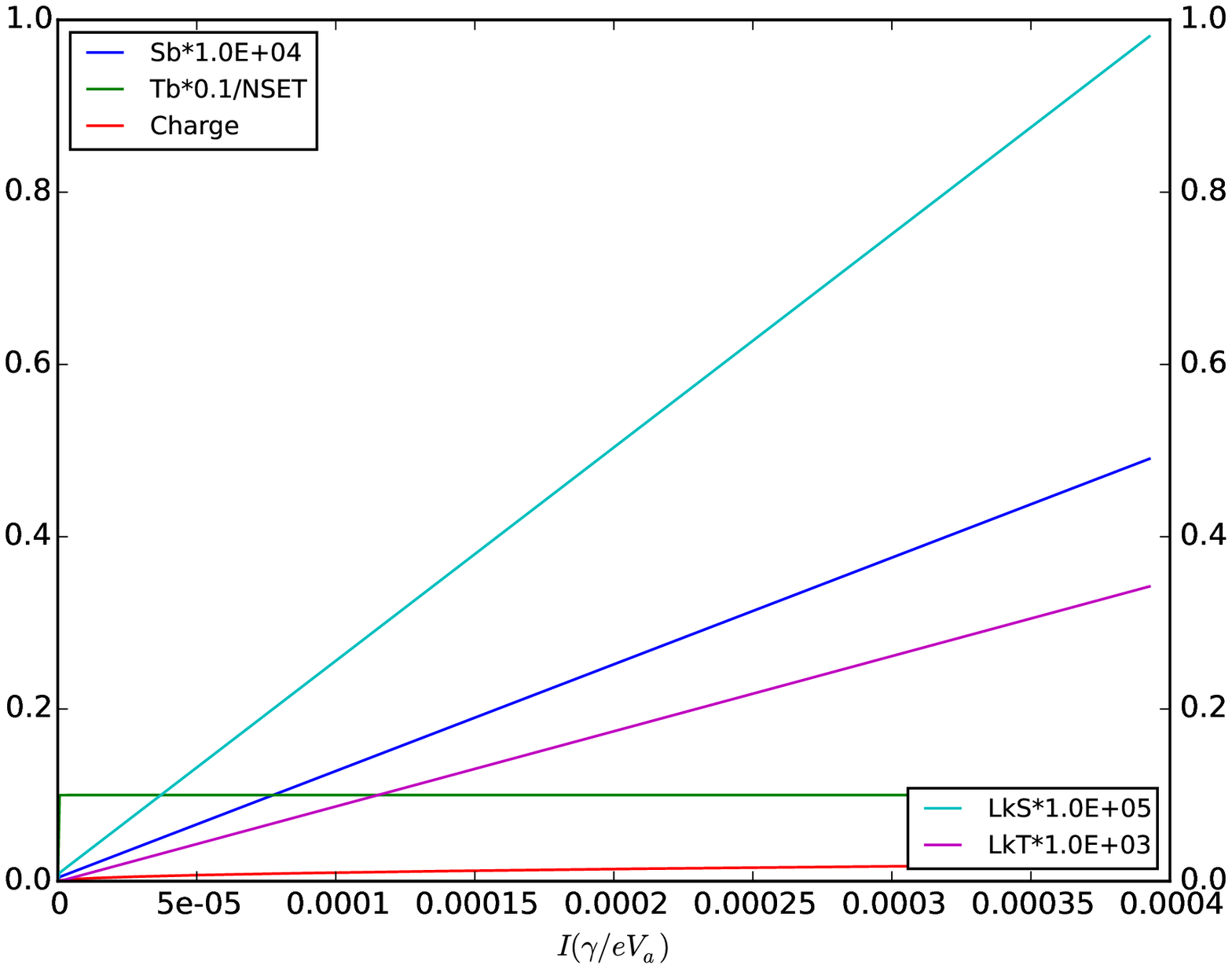}
    \label{fig:FIG.KTS.LowGapTriplet}
  }
  \subtable[b][\ Summary] {
   {\vbox to 1.8in{\hbox to 2.85in{\hspace{2mm}
    \begin{tabular}{lcc}
       & \hspace{20pt} $\ \Lk[S] / I$\hspace{10pt} & $ \Lk[T] / I$ \\
       \hline
      Both Spont.    & 0.038 & 0.01 \\
      Singlet Stim.  & 0.93 & 0.014 \\
      Triplet Stim.  & 0.025 & 0.87 \\
    \end{tabular}
    }
    \vfill
    } }
   }
  \def\pars{KTS=100.0 Kisc=1.0 Krisc=0.01}

%% file: FIG.KTTKTSKnT.FastISC.tex
  \subfigure[\ Spontaneous emission in singlet and triplet]{
    \includegraphics[scale=0.4]{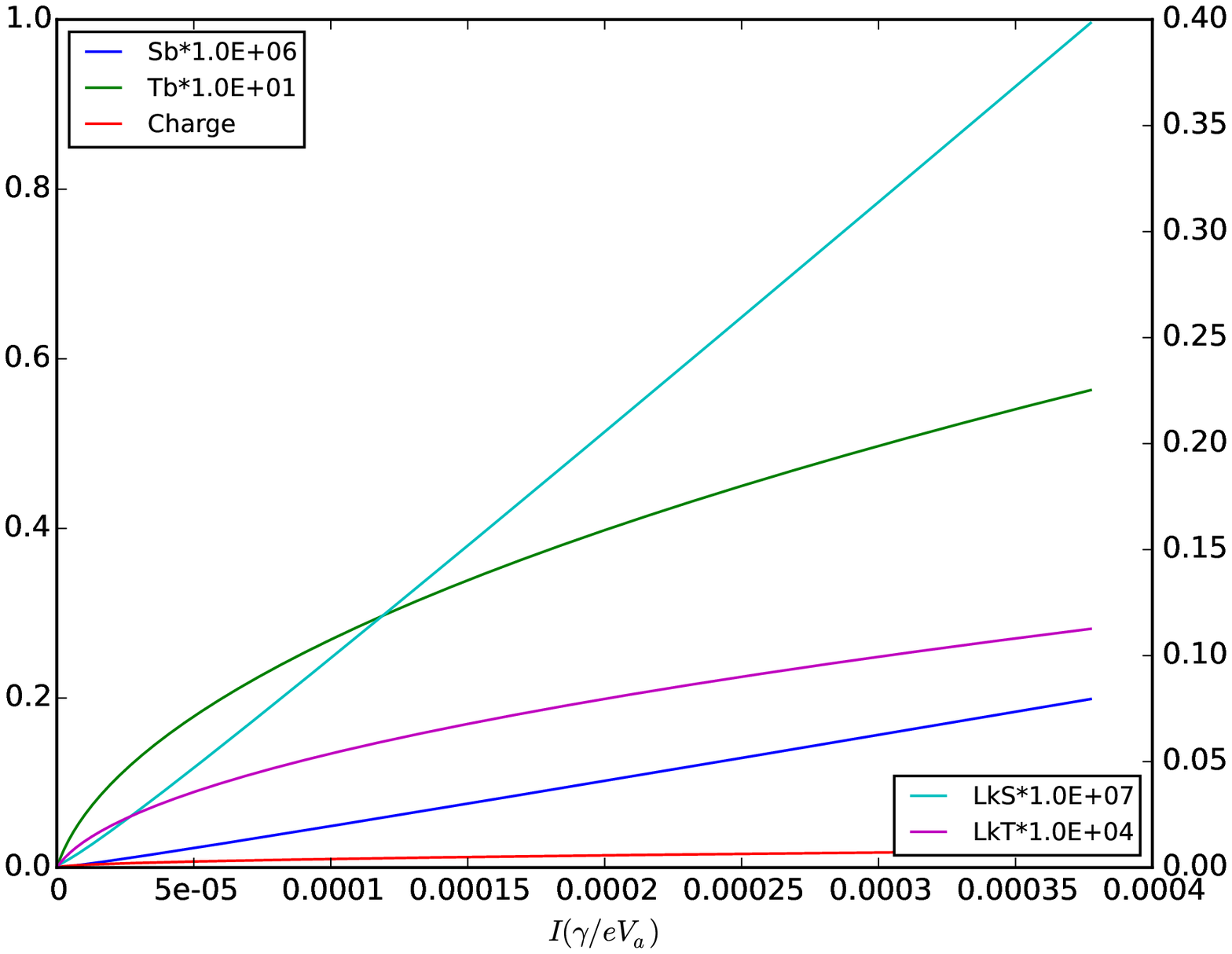}
    \label{fig:FIG.KTTKTSKnT.FastISCSpont}
  }
  \subfigure[\ Stimulated emission in singlet]{
    \includegraphics[scale=0.4]{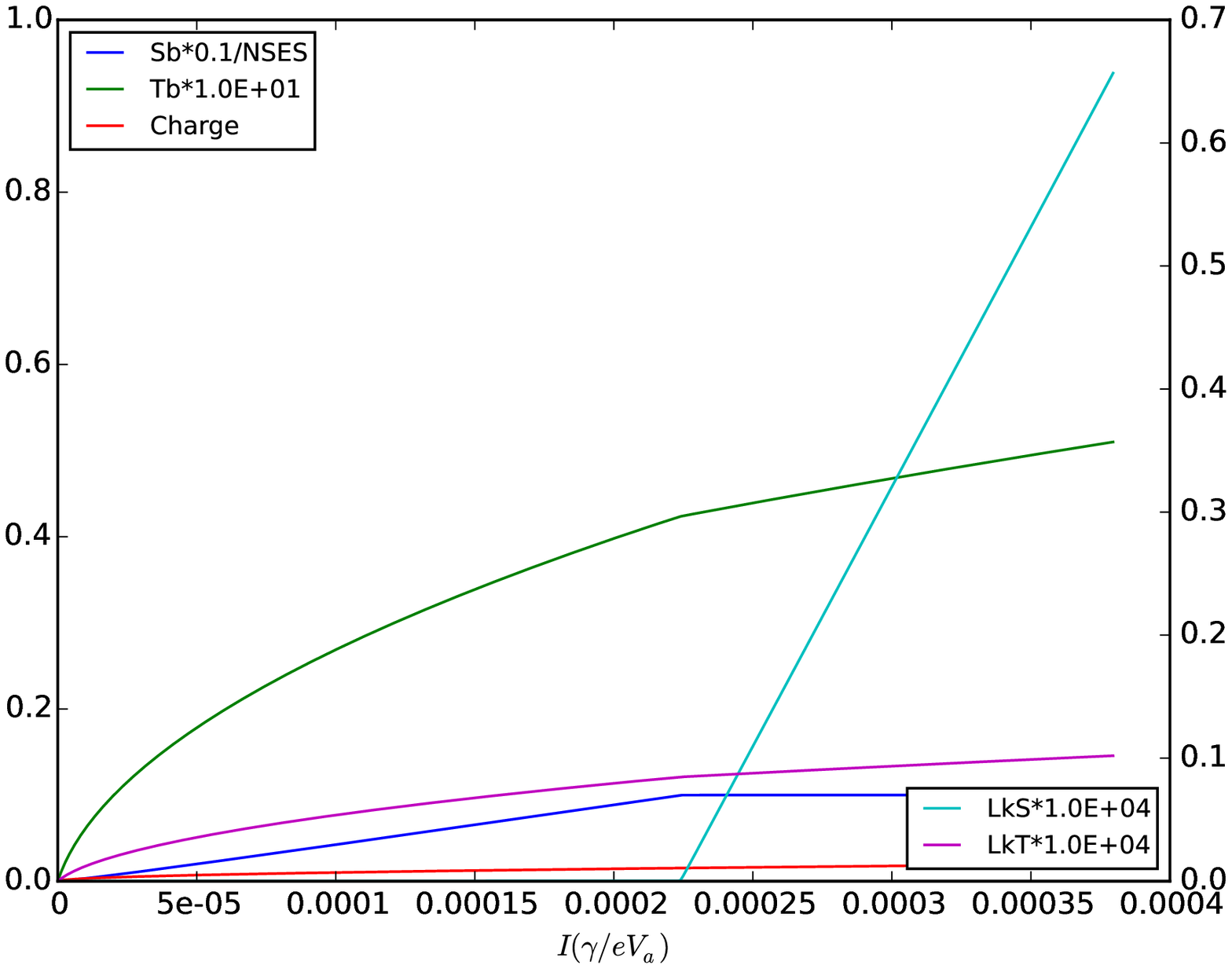}
    \label{fig:FIG.KTTKTSKnT.FastISCSinglet}
  } \\[5ex]
  \subfigure[\ Stimulated emission in triplet]{
    \includegraphics[scale=0.4]{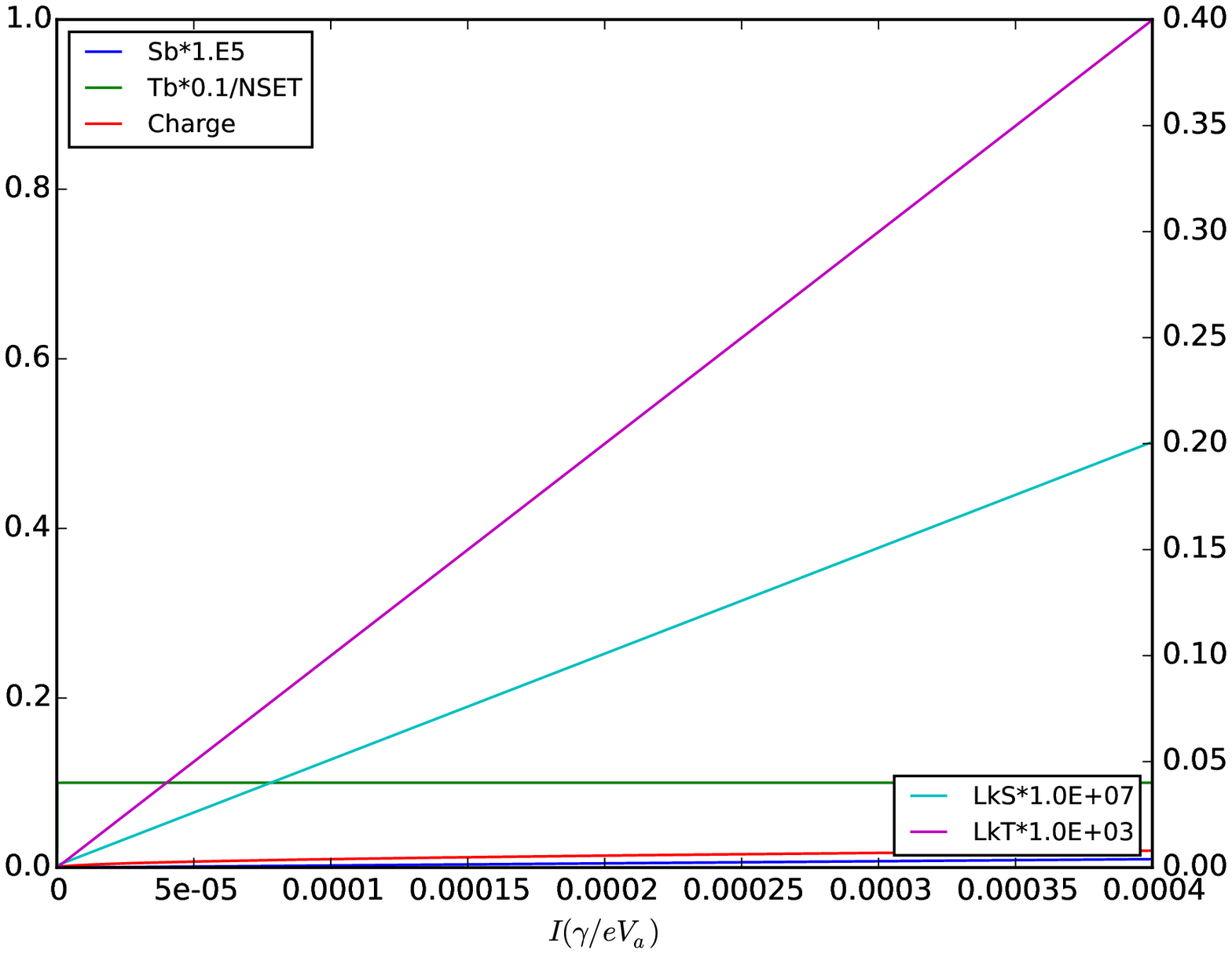}
    \label{fig:FIG.KTTKTSKnT.FastISCTriplet}
  }
  \subtable[b][\ Summary] {
   {\vbox to 1.8in{\hbox to 2.85in{\hspace{2mm}
    \begin{tabular}{lcc}
       & \hspace{20pt} $\ \Lk[S] / I$\hspace{10pt} & $ \Lk[T] / I$ \\
       \hline
      Both Spont.    & 0.00011 & 0.03 \\
      Singlet Stim.  & 0.17 & 0.027 \\
      Triplet Stim.  & 5e-05 & 1 \\
    \end{tabular}
    }
    \vfill
    } }
   }
  \def\pars{KTT=0.1 KTS=100.0 KnT=0.001 Kisc=1000.0 Krisc=0.0}

%% file: FIG.KTTKTSKnT.HighGap.tex
  \subfigure[\ Spontaneous emission in singlet and triplet]{
    \includegraphics[scale=0.4]{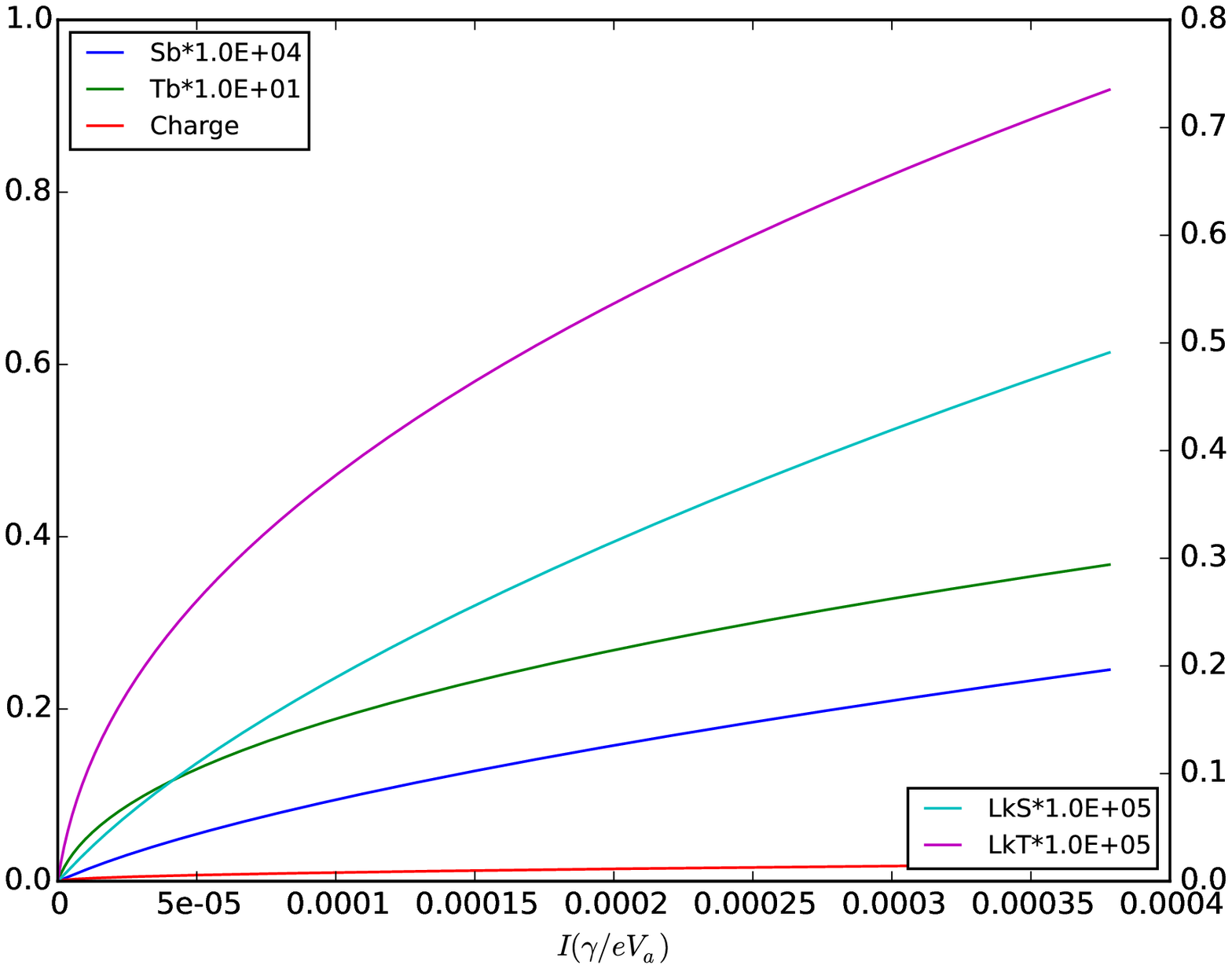}
    \label{fig:FIG.KTTKTSKnT.HighGapSpont}
  }
  \subfigure[\ Stimulated emission in singlet]{
    \includegraphics[scale=0.4]{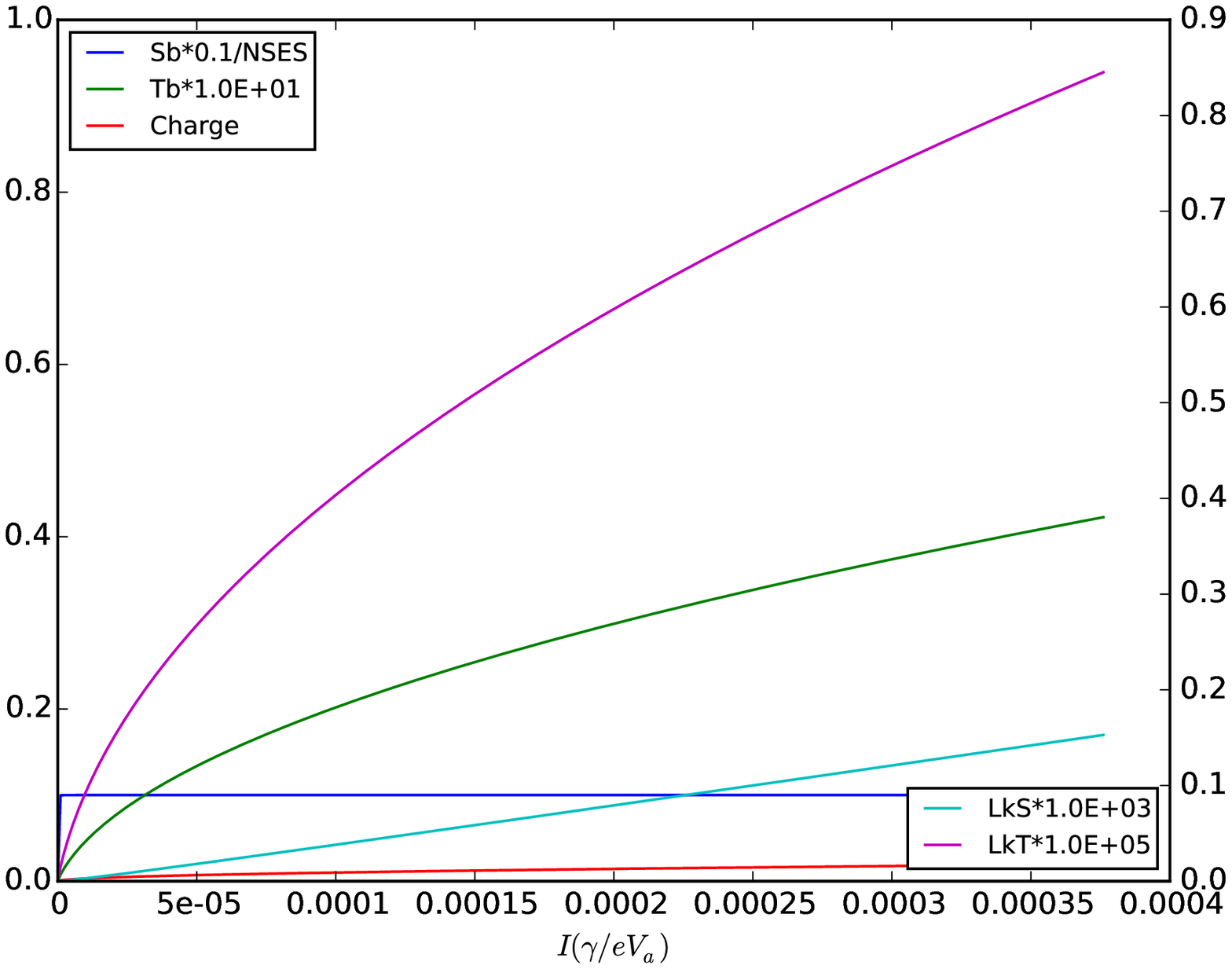}
    \label{fig:FIG.KTTKTSKnT.HighGapSinglet}
  } \\[5ex]
  \subfigure[\ Stimulated emission in triplet]{
    \includegraphics[scale=0.4]{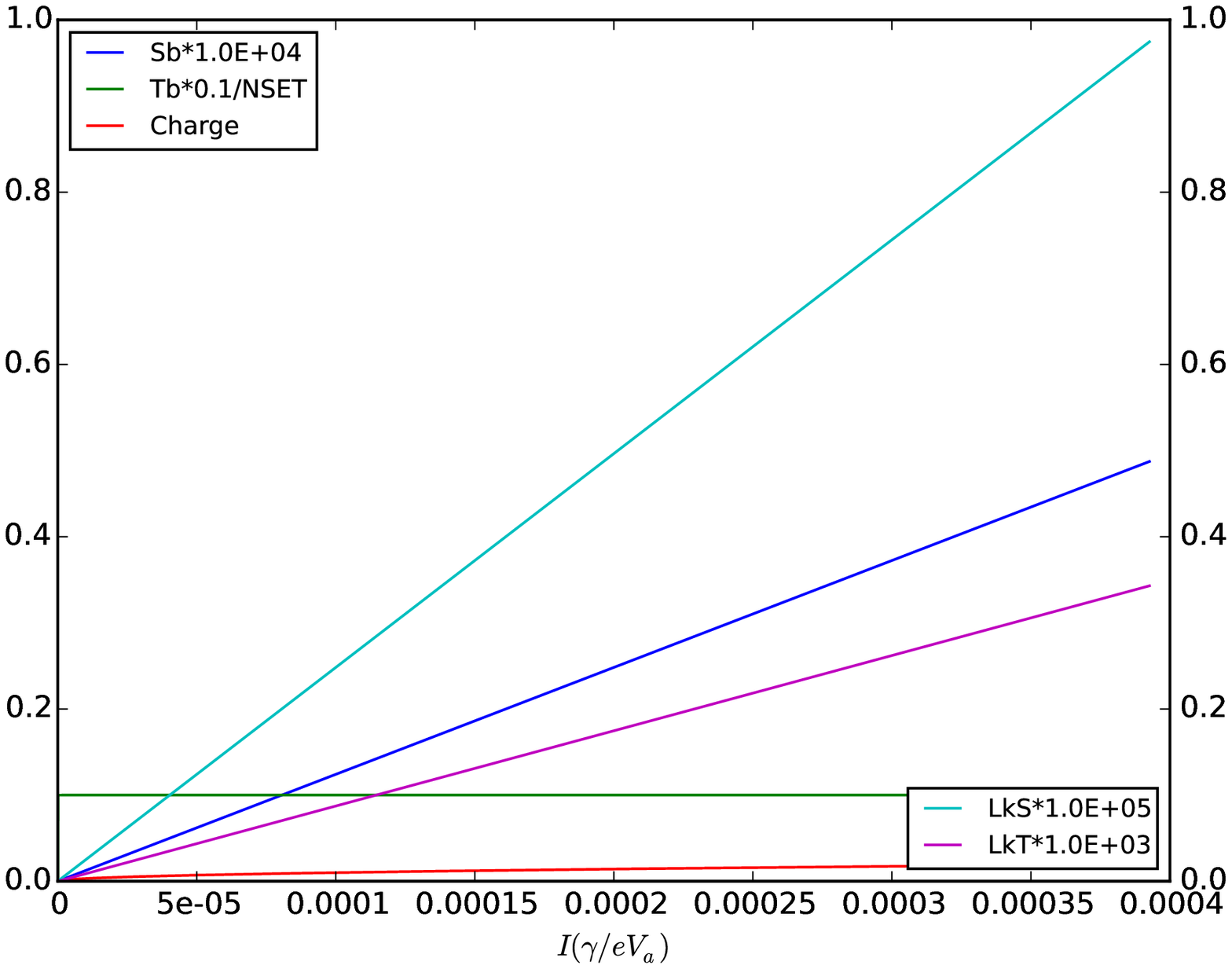}
    \label{fig:FIG.KTTKTSKnT.HighGapTriplet}
  }
  \subtable[b][\ Summary] {
   {\vbox to 1.8in{\hbox to 2.85in{\hspace{2mm}
    \begin{tabular}{lcc}
       & \hspace{20pt} $\ \Lk[S] / I$\hspace{10pt} & $ \Lk[T] / I$ \\
       \hline
      Both Spont.    & 0.027 & 0.019 \\
      Singlet Stim.  & 0.41 & 0.022 \\
      Triplet Stim.  & 0.025 & 0.87 \\
    \end{tabular}
    }
    \vfill
    } }
   }
  \def\pars{KTT=0.1 KTS=100.0 KnT=0.001 Kisc=1.0 Krisc=7.14e-12}

%% file: FIG.KTTKTSKnT.LowGap.tex
  \subfigure[\ Spontaneous emission in singlet and triplet]{
    \includegraphics[scale=0.4]{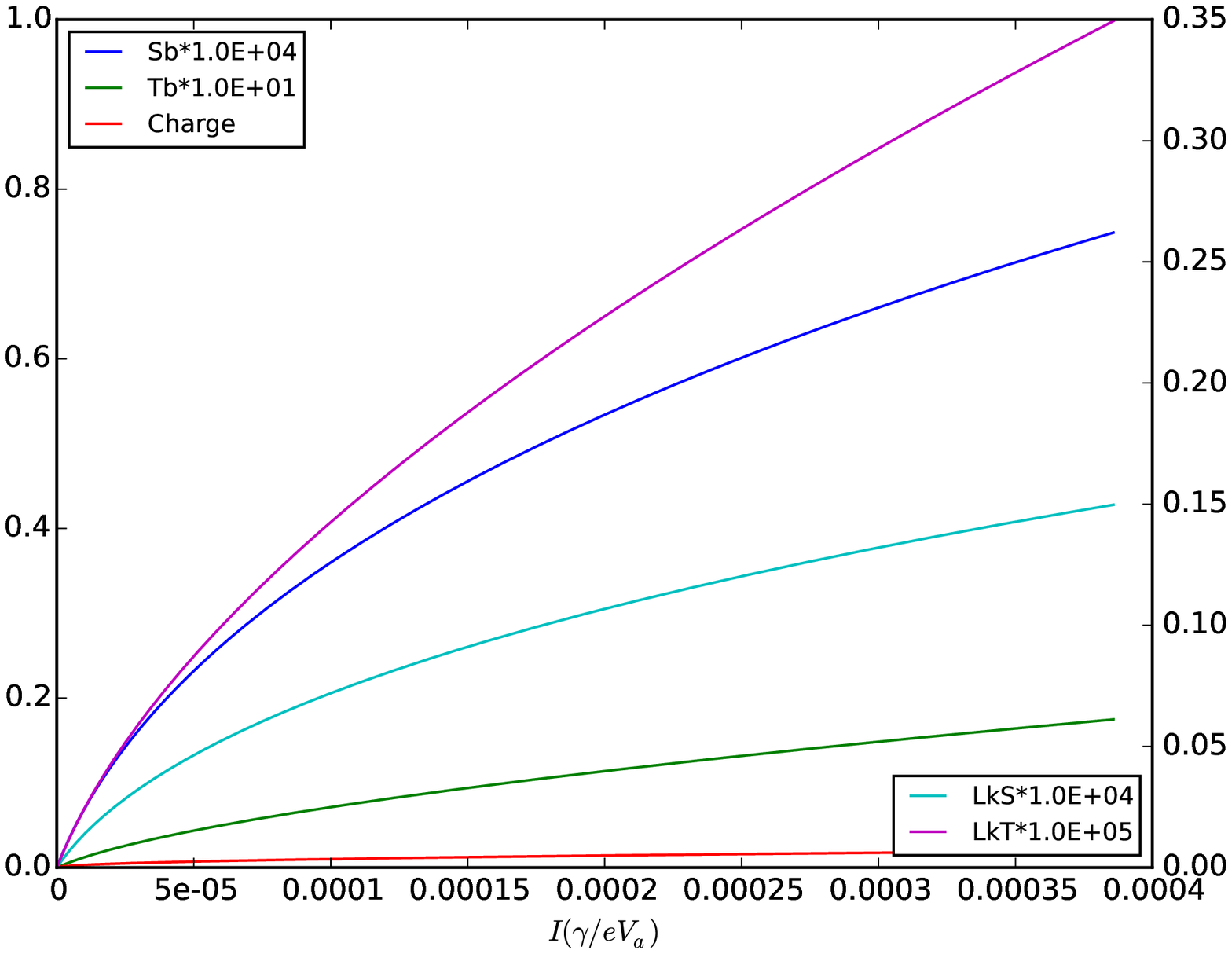}
    \label{fig:FIG.KTTKTSKnT.LowGapSpont}
  }
  \subfigure[\ Stimulated emission in singlet]{
    \includegraphics[scale=0.4]{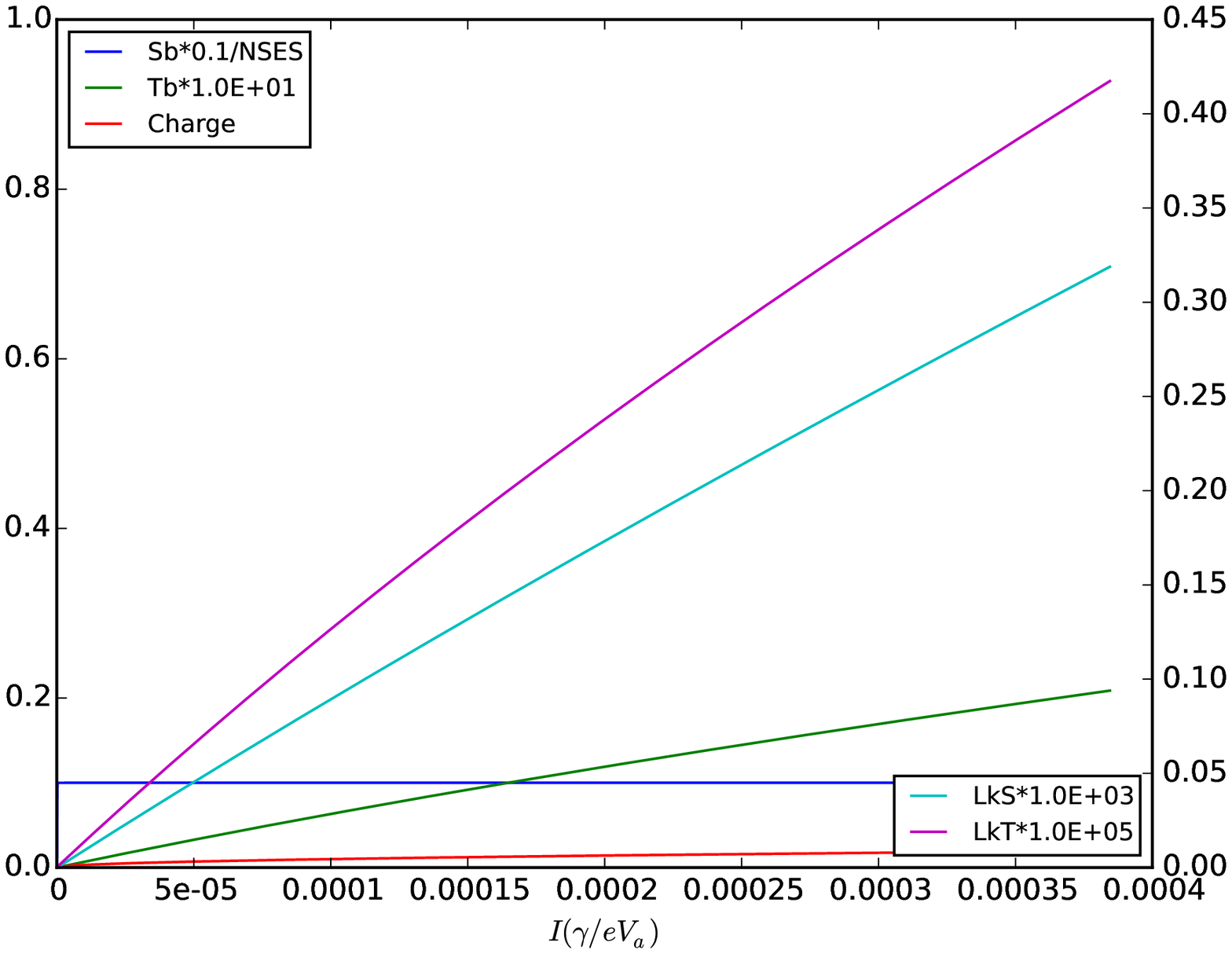}
    \label{fig:FIG.KTTKTSKnT.LowGapSinglet}
  } \\[5ex]
  \subfigure[\ Stimulated emission in triplet]{
    \includegraphics[scale=0.4]{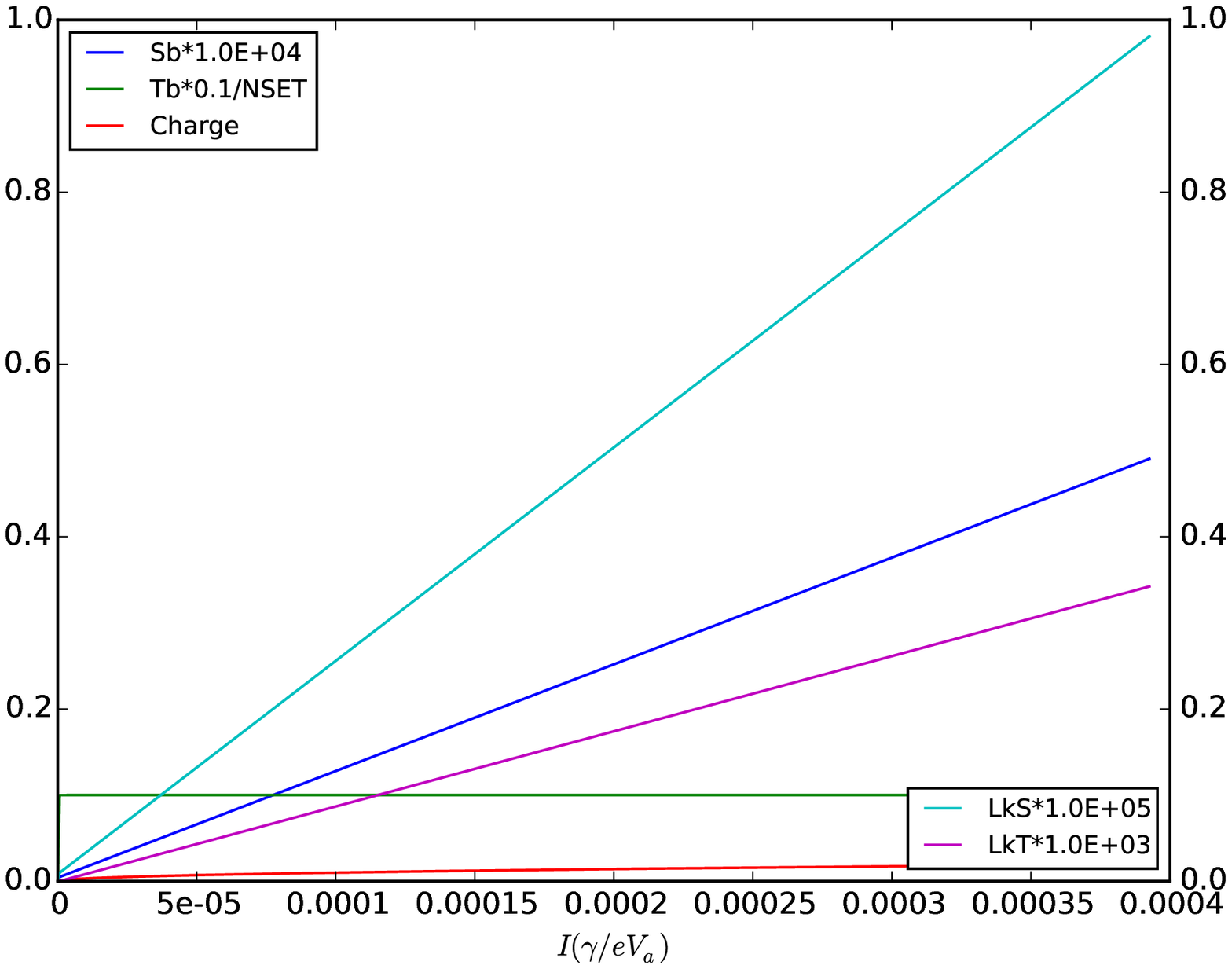}
    \label{fig:FIG.KTTKTSKnT.LowGapTriplet}
  }
  \subtable[b][\ Summary] {
   {\vbox to 1.8in{\hbox to 2.85in{\hspace{2mm}
    \begin{tabular}{lcc}
       & \hspace{20pt} $\ \Lk[S] / I$\hspace{10pt} & $ \Lk[T] / I$ \\
       \hline
      Both Spont.    & 0.038 & 0.009 \\
      Singlet Stim.  & 0.92 & 0.011 \\
      Triplet Stim.  & 0.025 & 0.87 \\
    \end{tabular}
    }
    \vfill
    } }
   }
  \def\pars{KTT=0.1 KTS=100.0 KnT=0.001 Kisc=1.0 Krisc=0.01}